\documentclass[multphys,vecphys]{svmult}

\usepackage{makeidx}     
\usepackage{graphicx}    
\usepackage{multicol}    
\usepackage[]{natbib}    
\usepackage[]{psfig}    
\bibpunct[]{(}{)}{;}{a}{}{,}  
\def\ga{\;\rlap{\lower 2.5pt\hbox{$\sim$}}\raise 1.5pt\hbox{$>$}\;}
\def\la{\;\rlap{\lower 2.5pt\hbox{$\sim$}}\raise 1.5pt\hbox{$<$}\;}

\makeindex             

\begin{document}

\title*{Physics of Stellar Coronae}
\titlerunning{Stellar Coronae}
\author{Manuel G\"udel}
\institute{Paul Scherrer Institut, W\"urenlingen and Villigen, CH-5232 
           Villigen PSI, Switzerland  
\texttt{guedel@astro.phys.ethz.ch}}
%
%
\maketitle

\section{Introduction}\label{intro}

For the plasma physicist, the solar corona offers an outstanding example of a space plasma,
and surely one that deserves a lifetime of study. Not only can we observe the solar corona
on scales of a few hundred kilometers and monitor its changes in the course of seconds
to minutes, we also have a wide range of detailed diagnostics at our disposal that provide
immediate access to the prevalent physical processes. 

Yet, solar physics offers a rich field of unsolved plasma-physical problems. How is the coronal
plasma continuously heated to $> 10^6$~K? How and where are high-energy particles episodically
accelerated? What is the internal dynamics of plasma in magnetic loops? What is the initial
trigger of a coronal flare? How does the corona link to the solar wind, and how and where is
the latter accelerated? How is plasma transported into the solar corona? 

Why, then, study 
stellar coronae that remain spatially unresolved in X-rays and are only marginally resolved at 
radio wavelengths, objects that require exposure times of several hours before approximate
measurements of the  ensemble of plasma structures can be obtained?

\begin{table}[t!] 
\caption{Symbols and units used throughout the text} 
\label{symbols}        
\begin{tabular}{ll} 
\hline\noalign{\smallskip}
Symbol, acronym	          	    & Explanation	       \\
\noalign{\smallskip}\hline\noalign{\smallskip}
$R_*$ or $R$                                & Stellar radius [cm]		      \\    
$R_{\odot}$             	            & Solar radius [$7\times 10^{10}$~cm]		      \\    
$c$                                         & Speed of light [$3\times 10^{10}$~cm~s$^{-1}$]		      \\    
$k$                                         & Boltzmann constant [$1.38\times 10^{-16}$~erg~cm$^{3}~$K$^{-1}$]		      \\    
$m_e$                                       & Mass of electron [$9.1\times 10^{-28}$~g]		      \\    
$d$             	                    & Distance [pc]		      \\    
$p$             	                    & Pressure [dyne~cm$^{-2}$]		      \\    
$L$             	                    & Coronal loop semi-length [cm]		      \\    
$T$             	        	    & Coronal electron temperature [K]	  \\	
$T_b$             	        	    & Radio brightness temperature [K]	  \\	
$n, N$                                      & non-thermal electron density ($N$: integrated in energy)       \\
$N$                                         & Number of flares       \\
$n_e$             	        	    & Electron density [cm$^{-3}$]	  \\	
$n_{\rm H}$             	            & Hydrogen density [cm$^{-3}$]	  \\	
$B$             	        	    & Magnetic field strength [G]	  \\	
$f$             	        	    & Surface filling factor [\%]	  \\	
$\Gamma$             	        	    & Loop area expansion factor (apex to base)	  \\	
$E$             	                    & Energy  [erg]		      \\    
$L_R$             	                    & Radio luminosity 	[erg~s$^{-1}$~Hz$^{-1}$]		      \\    
$L_X$             	                    & X-ray luminosity 	[erg~s$^{-1}$]		      \\    
$L_{\rm bol}$             	            & Stellar bolometric luminosity 	[erg~s$^{-1}$]		      \\    
$\Lambda = \Lambda_0T^{\chi}$               & Cooling function  [erg~s$^{-1}$~cm$^{3}$]   \\    
$\tau$                                      & (Decay) time scales, also: optical depth   \\
$\nu$                                       & Radio frequency    \\
$\nu_p$, $\omega_p$                         & Plasma frequency, angular  plasma frequency \\
$\nu_c$, $\Omega_c$                         & Gyrofrequency, angular  gyrofrequency \\
$\kappa$                                    & Absorption coefficient    \\
$\gamma$                                    & Lorentz factor \\
$L_B$                                       & Magnetic scale height \\
HRD          	                	    & Hertzsprung-Russell Diagram			 \\   
EM             	                	    & Emission Measure  		     \\    
$Q$, DEM             	        	    & Differential Emission Measure  Distribution		      \\    
EMD             	        	    & (discretized, binned) Emission Measure Distribution      \\    
\noalign{\smallskip}\hline
\end{tabular}
\end{table}

There are many reasons. In the context of the {\it solar-stellar connection,} stellar X-ray astronomy 
has introduced a range of stellar rotation periods, gravities, masses, and ages into the debate 
on the magnetic dynamo. Coronal magnetic structures and heating  mechanisms may vary together 
with variations of these parameters. Parameter studies could provide valuable insight for 
constraining relevant theory. Different topologies and sizes of magnetic field structures lead 
to different wind mass-loss rates, and this will regulate the stellar spin-down rates differently. 

Including stars into the big picture of coronal research has also widened our view of coronal plasma physics.
While solar coronal plasma resides typically at $(1-5)\times 10^6$~K with temporary excursions to $\approx 20$~MK
during large flares, much higher temperatures are found  on some active stars, with steady plasma temperatures
of several tens of MK and flare peaks beyond 100~MK. Energy release in stellar flares involves up to $10^5$ times
more thermal energy than in solar flares, and pressures that are not encountered in the solar corona.

The present chapter provides a ``stellar astronomer's view'' of magnetic coronae. It cannot replace 
the knowledge of detailed physical processes in the solar corona, nor do we expect to find entirely
new physical concepts from  the study of stellar coronae without guidance by solar physics. However, 
as I hope to show in the following sections, stellar astronomy has provided some unexpected and 
systematic trends that may well help understand systematics of coronal behavior across a large range 
of stellar parameters. In this sense, the goal of this chapter is to present an overview of the
basic, observed stellar-coronal phenomena and their interpretation analogous to solar physics, rather 
than the derivation of the basic plasma-physical mechanisms themselves. For a deeper understanding of 
the latter, I must refer the reader to appropriate textbooks and lectures on solar coronal physics.

\section{Stellar coronae - Defining the theme} 

We consider coronae to be the ensemble of closed magnetic structures above 
the stellar photosphere and chromosphere together with their plasma content, regardless of
whether the latter is the thermal bulk plasma or a non-thermal population of accelerated particles.
Although often narrowed down to some specific energy ranges, coronal emission is 
intrinsically a multi-wavelength phenomenon revealing itself from the meter-wave radio range
to gamma rays. The most important wavelength regions from which we have learned {\it diagnostically} on 
{\it stellar} coronae include the radio (decimetric to centimetric) range and the X-ray domain.
The former is sensitive to accelerated 
electrons in magnetic fields, and that has provided the only direct means of imaging 
stellar coronal structure, through very long baseline interferometry. 

The soft X-ray (0.1--10~keV) 
diagnostic power has been instrumental for our understanding of physical processes in the hot, magnetically
trapped coronal plasma, and  the recent advent of high-resolution X-ray spectroscopy with the {\it Chandra} 
and {\it XMM-Newton} X-ray observatories is now accessing physical parameters of coronal plasma  
directly. The adjacent extreme ultraviolet (EUV) range contains diagnostic relevant for the same 
temperature range as X-rays.

\section{The coronal Hertzsprung-Russell diagram}

\begin{figure} 
\hbox{
\resizebox{0.49\hsize}{!}{\includegraphics{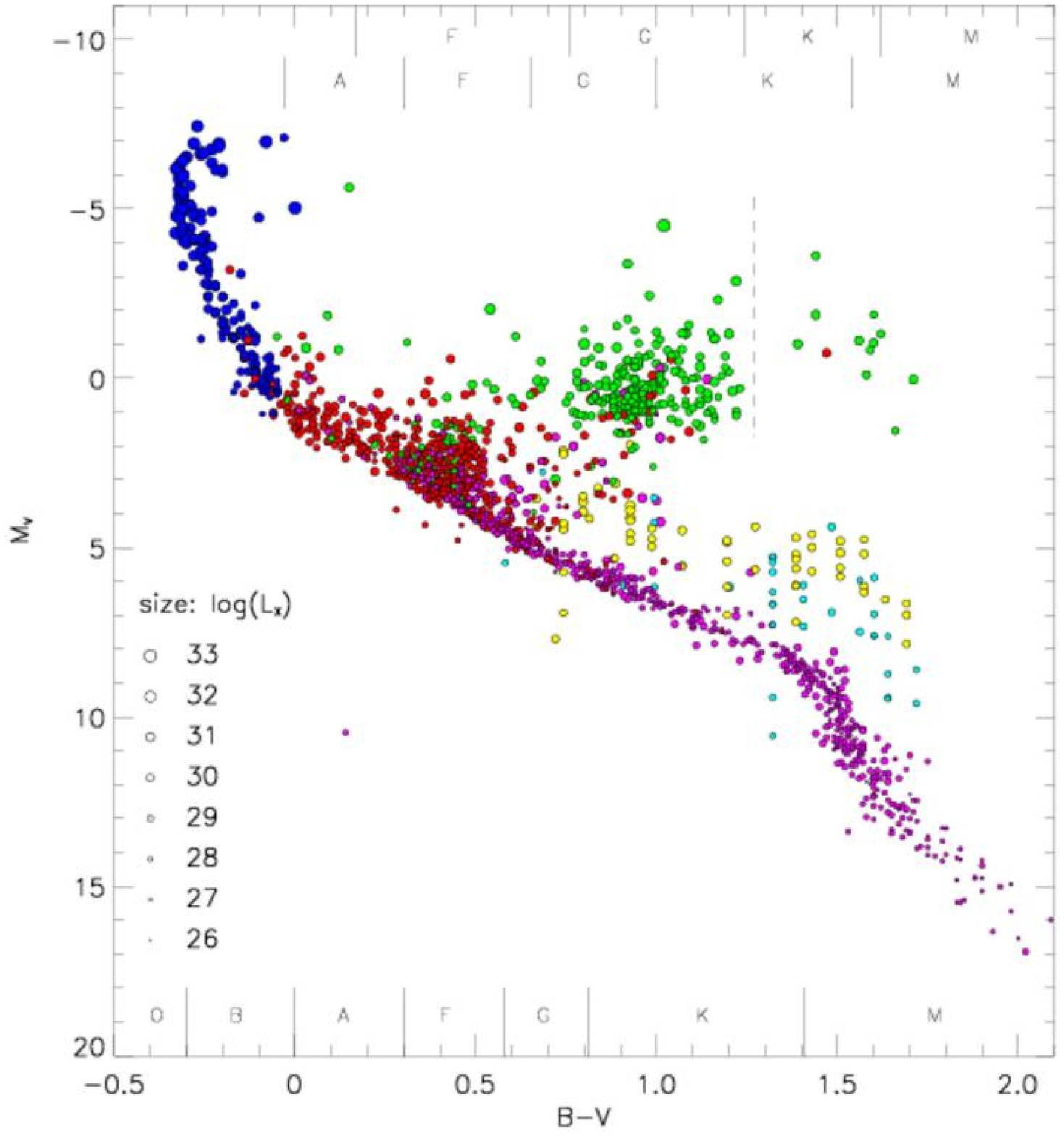}}    
\resizebox{0.51\hsize}{!}{\includegraphics{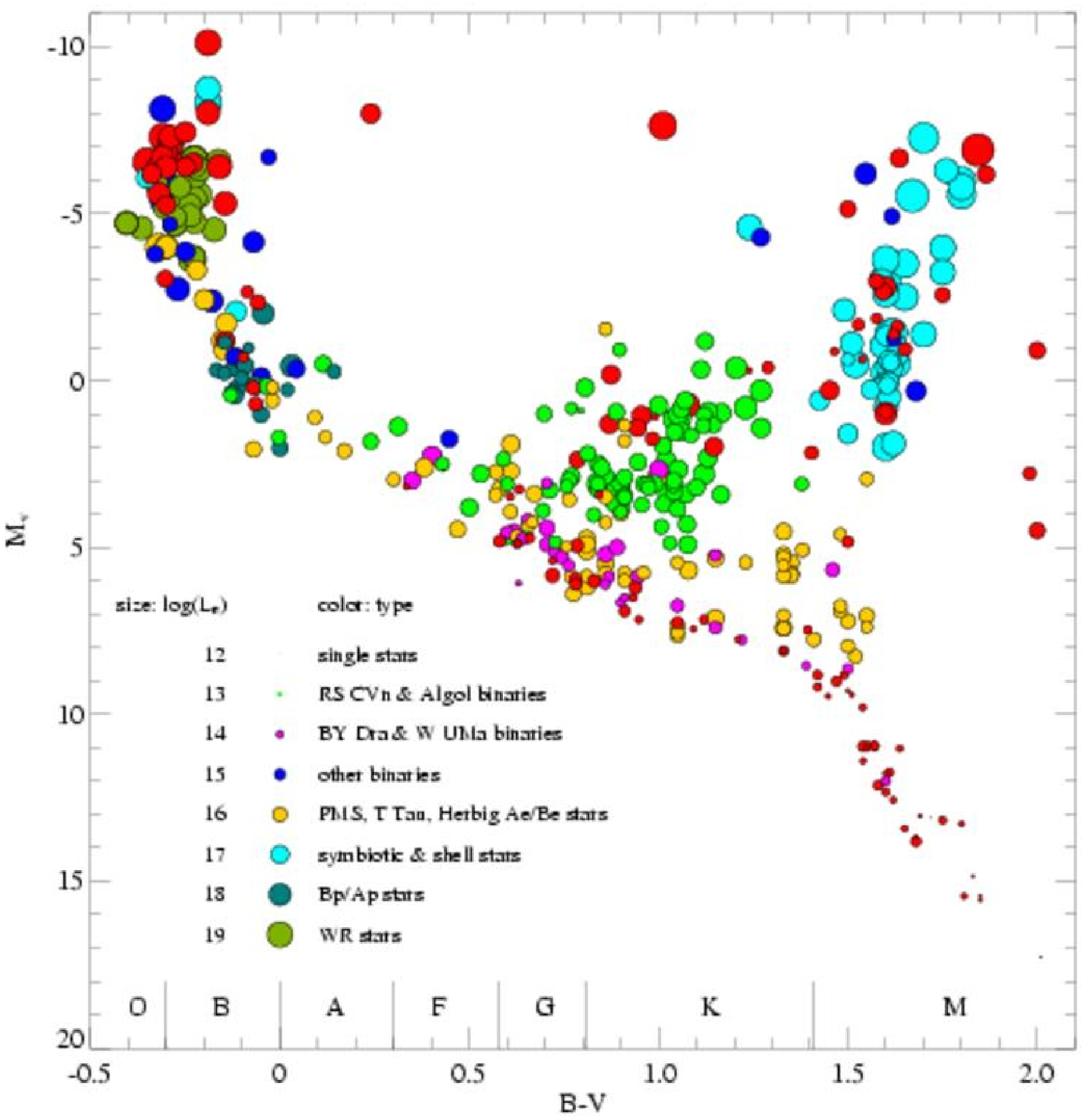}}
}
\caption{{\bf Left:} Hertzsprung-Russell diagram based on about 2000 X-ray detected stars extracted from  
survey catalogs (see \citealt{guedel04b} for references).  The size of the circles
characterizes log\,$L_\mathrm{X}$ as indicated in the panel at lower left. The ranges for the spectral classes
are given at the top (upper row for supergiants, lower row for giants), and at the bottom of the figure (for 
main-sequence stars). -- {\bf Right:} Similar, but for 440 radio stars detected between 1--10~GHz 
(after  \citealt{guedel02d}).}\label{hrd}
\end{figure}

Before discussing specific physical problems in stellar coronal physics, I will briefly 
review the phenomenology of stellar radio (here: 1--20~GHz) and X-ray emission and summarize 
stellar classes that are prolific coronal emitters. Fig.~\ref{hrd} presents Hertzsprung-Russell 
diagrams (HRD) of detected X-ray (left) and radio (right) stars. 
They show all basic features that we know from an optical HRD.  Although the samples 
used for these figures are in no way ``complete'' (in volume or brightness), the main sequence 
is clearly evident, and so is the giant branch. The cool half of the subgiant and giant area 
is dominated by the X-ray and radio-strong sample of RS CVn and Algol-type close binaries.  
The top right part of the diagram, comprising cool giants, is almost devoid of X-ray detections, 
however (although well populated by radio emitters). The so-called corona vs. 
wind dividing line  (dashed in Fig.~\ref{hrd}a; after \citealt{linsky79}) 
separates coronal giants and supergiants to its left from stars with massive winds
to its right. It is unknown whether the wind giants possess magnetically
structured coronae at the base of their winds -- the X-rays may simply be absorbed 
by the overlying wind material. Additionally,
a very prominent population of (presumably) coronal radio and X-ray sources just above 
the main sequence is made up of various classes of pre-main sequence stars, such as classical 
and weak-lined T Tauri stars.  

Moving toward A-type stars on the HRD, one expects, and
finds, a  significant drop of coronal emission owing to the absence of magnetic dynamo action
in these stars. However, this is also the region of the chemically peculiar Ap/Bp stars that 
possess strong magnetic fields and many of which are now known to be non-thermal  radio 
sources as well. Finally, the very luminous radio and X-ray emissions from O and B stars are 
believed to originate in non-magnetic stellar winds - we will not further discuss these stars.

\section{Nonflaring radio emission from stellar coronae}

``Quiescent'' (non-flaring) radio emission at levels of 
$10^{12} - 10^{16}$~erg~s$^{-1}$~Hz$^{-1}$ from magnetically active stars 
was entirely unanticipated but constitutes an important achievement 
in stellar radio astronomy: There simply is no solar counterpart!
Quiescent emission  can be  defined by the absence of impulsive,
rapidly variable flare-like events. Common characteristics of quiescent 
emission are (i) slow variations on time scales of hours and days, (ii) broad-band 
radio spectra,  (iii) brightness temperatures 
in excess of coronal temperatures measured in X-rays, 
and usually (iv) low polarization degrees. 

\subsection{Bremsstrahlung}\label{brems}

The Sun emits steady, full-disk, optically thick thermal radio emission at chromospheric and
transition region levels of a few $10^4$~K. However, such emission cannot be detected 
with present-day facilities, except for radiation from the very nearest stars, or giants 
subtending  a large solid angle. Using the Rayleigh-Jeans approximation for the flux density $S$
\begin{equation}\label{rayleigh}
S = {2kT\tau\nu^2\over c^2}{\pi R^2\over d^2} \approx 0.049\left({T\over 10^6~{\rm K}}\right)\left({\nu\over 1~{\rm GHz}}\right)^2
       \left({R\over R_{\odot}}\right)^2\left({1~{\rm pc}\over d}\right)^2\tau~{\rm mJy} 
\end{equation}
($R$ = source radius, $T$ = electron temperature, $d$ = stellar distance, $k$ = Boltzmann constant,
$\nu$ = observing frequency, $\tau$ = optical depth), we find for optically thick chromospheric 
emission (with $\tau = 1, T = 1.5\times 10^4$~K, $\nu = 8.4$~GHz)
\begin{equation}
S \approx {0.05\over d_{\rm pc}^2}\left({R\over R_{\odot}}\right)^2~{\rm mJy}.
\end{equation}

Optically thin free-free emission from the hot, X-ray emitting plasma in coronal loops
can be estimated as follows: The radio optical depth is
\begin{equation} 
 \tau = \int\kappa{\rm d}l \approx {0.16\over \nu^2T^{3/2}}\int n_e^2{\rm d}l
\end{equation}
and the X-ray volume emission measure EM, using a filling factor $f$ in the approximation of a
small coronal height,
\begin{equation}
{\rm EM_X} = 2R^2 \pi f \int n_e^2{\rm d}l.
\end{equation}
(the factor of two accounts for the visible and invisible hemispheres, assuming that the 
EM is uniformly distributed). For plasma between 1--20~MK, the EM can be estimated
from the X-ray luminosity $L_{\rm X}$ with a conversion factor $\Lambda(T)$ (\citealt{mewe85}, 
$\Lambda$ larger for lower $T$)
\begin{equation}\label{lxEM}
L_{\rm X} = \Lambda(T){\rm EM_X} \approx (1.5-6)\times 10^{-23} {\rm EM_X} \quad {\rm [erg~s^{-1}]}
\end{equation}
Thus,  we find (using $T_{\rm MK} = T/[10^6~{\rm K}]$)  
\begin{equation}
S = 2.6\times 10^{-52}{L_{\rm X}\over \Lambda(T)T_{\rm MK}^{1/2}fd_{\rm pc}^2}\ {\rm mJy} \approx
(4-17)\times 10^{-30}{L_{\rm X}\over T_{\rm MK}^{1/2} f d_{\rm pc}^2}\quad {\rm mJy}.
\end{equation}
Coronal bremsstrahlung contributions are presently out of reach for almost all stars.

\subsection{Gyroresonance emission}\label{gyrores}

Because active stars show high coronal temperatures and large magnetic filling factors 
that prevent magnetic fields from strongly diverging with increasing 
height, the radio optical depth can become significant at coronal levels
owing to  gyroresonance absorption. This type of emission is observed above solar 
sunspots. 
The optical depth for the $s$th harmonic is (after \citealt{whites94})
\begin{eqnarray}
&&\tau(s,\nu) = {\pi^3\over 4}{\nu_p L_B\over \nu_c} {s^2(2s -2)! \over 2^{2s-2}s![(s-1)!]^2}
         \left({s^2\over 2\mu}\right)^{s-1}\nonumber \\
	 &=& 1.45{L_B\over R_*}{R_*\over R_{\odot}}{n_e\over \nu_{\rm GHz}} 
	 (1.3\times 10^{-4}, 4.1\times 10^{-6}, 1.8\times 10^{-7}) 
	 \left( {T\over 20~{\rm MK}}\right)^{s-1}
\end{eqnarray}
where the three coefficients in the parenthesis are for the harmonics $s = 3, 4,$ and 5, respectively,
of the gyrofrequency, $\nu_{\rm GHz}$ is the observing frequency in GHz, $n_e$ is the electron density of the
emitting hot plasma at temperature $T$, and $\mu = m_ec^2/kT$. The magnetic scale height, $L_B$,
is not precisely known but will be assumed to be 0.3$R_*$ (see arguments in \citealt{whites94}).
Using typical X-ray derived temperatures ($[1-3]\times 10^7$) and X-ray emission measures EM = $n_e^2V$ (to
estimate electron densities in the emitting volume), one finds that $\tau$ invariably reaches unity 
at $s = 3, 4,$ or 5 for an observing frequency of 15~GHz while $\tau < 1$ for larger $s$. 
The highest harmonic that is still optically thick is relevant.

\begin{figure}[t!]
\centerline{\psfig{file=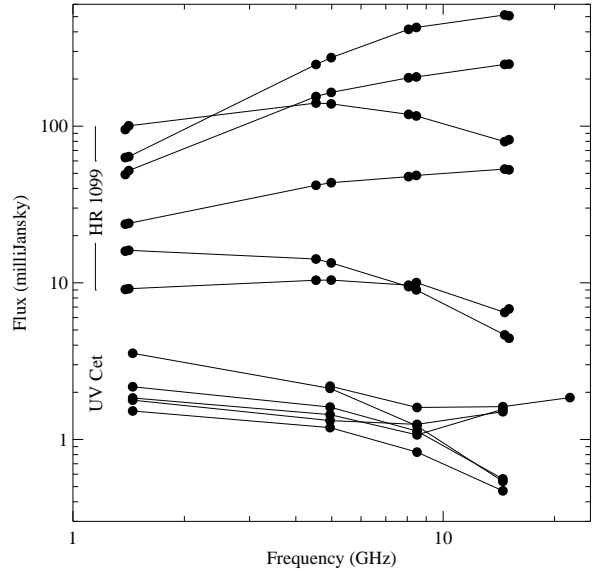,width=8truecm}}
\caption{Radio spectra of the RS CVn binary HR~1099  (upper set) and of the dMe dwarf  
UV Cet (lower set) at different flux levels. The gently bent spectra are indicative
of gyrosynchrotron emission, and the high-frequency part of U-shaped spectra for 
UV Cet has been interpreted as a gyroresonance component (HR~1099 spectra: courtesy
of S.~M. White).\label{spec}}
\end{figure}

The measured flux from the optically thick layer is then again
given by (\ref{rayleigh}). Observations have shown the following:
\begin{itemize}

\item The observed fluxes at 15~GHz from M dwarfs are not compatible with the above
      prediction if the emitting layer of hot plasma covers the entire stellar surface.
      The observed flux is typically much smaller. The microwave spectra are falling,
      not compatible with optically thick radiation (\citealt{whites94} - see Fig.~\ref{spec}). 
      
\item There are exceptions in which a rising spectrum from 5 to 15~GHz (Fig.~\ref{spec})  
      could be explained by gyroresonance emission, while the lower-frequency
      spectrum cannot \citep{guedel89b}. The same stars may show this 
      feature only temporarily.
      
\end{itemize}
These results imply that the hot plasma in general is not coincident with the strong-magnetic 
field regions in the corona. It must reside in lower-$B$  regions in between or above the
 cooler regions, possibly implying rather extended coronal structures seen in X-rays. This 
 plasma could be induced by flares in which magnetic loops reconnect to larger structures 
 in which the low-density gas will not produce appreciable gyroresonance emission \citep{whites94}.
The cooler plasma also usually observed in active stars might be trapped in the strong fields 
but its gyroresonance emission is also negligible.

Lower-frequency radio emission cannot be due to gyroresonance absorption: 
The radius of the optically thick layer would still
at $s = 3-5$, but it would then be located at  $ > 3R_*$  (for dMe stars). If we
extrapolate the   the corresponding magnetic fields of  more than 100~G down to 
photospheric levels, we would find photospheric field strengths much in excess of
those observed \citep{gary81}.

\subsection{Gyrosynchrotron  emission}\label{gyrosynch}

We could, however, allow  for much higher $T_{\mathrm{eff}}$. The optically 
thick layer would in that case shift to harmonics above 10, the range of gyrosynchrotron radiation.
The optically thick source sizes are then more reasonable for M dwarfs, with
$R \approx R_*$ \citep{linsky83}, and the optically thin emission may still be strong enough
for detection. However, for a thermal plasma, the spectral power drops like $\nu^{-8}$ at 
high  frequencies, in contrast to  observed microwave spectra that  show $\nu^{-(0.3...1)}$ 
for  magnetically  active stars  (Fig.~\ref{spec}). 

Instead, the electron population could be  {\it non-thermal}. Electron energy distributions
in cosmic sources are often found to follow a power law, and this also holds for
solar and stellar microwave flares \citep{kundu85}:
\begin{equation}\label{npowerlaw}
n(\gamma)=K(\gamma -1)^{-\delta}
\end{equation}
There is wide support for this model from estimates of the brightness temperature (e.g., based 
on the stellar radius, or from resolved interferometric images). The important question then is 
how  these coronae are continuously replenished with high-energy electrons.

\begin{figure} 
\vskip -2truecm\hbox{
\hskip -0.3truecm\resizebox{0.32\textwidth}{!}{\includegraphics{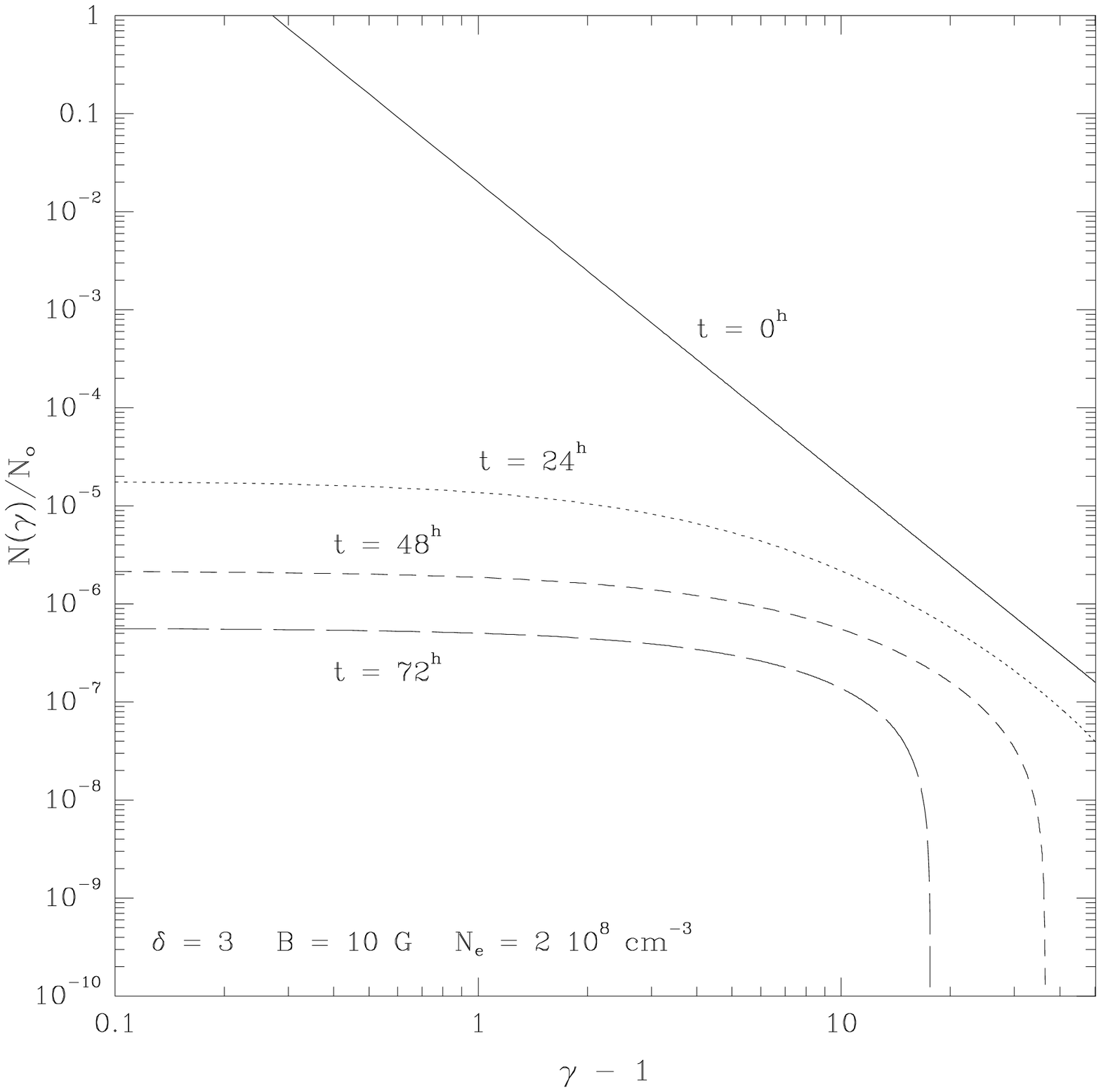}}
\hskip -1truecm\resizebox{0.42\textwidth}{!}{\includegraphics{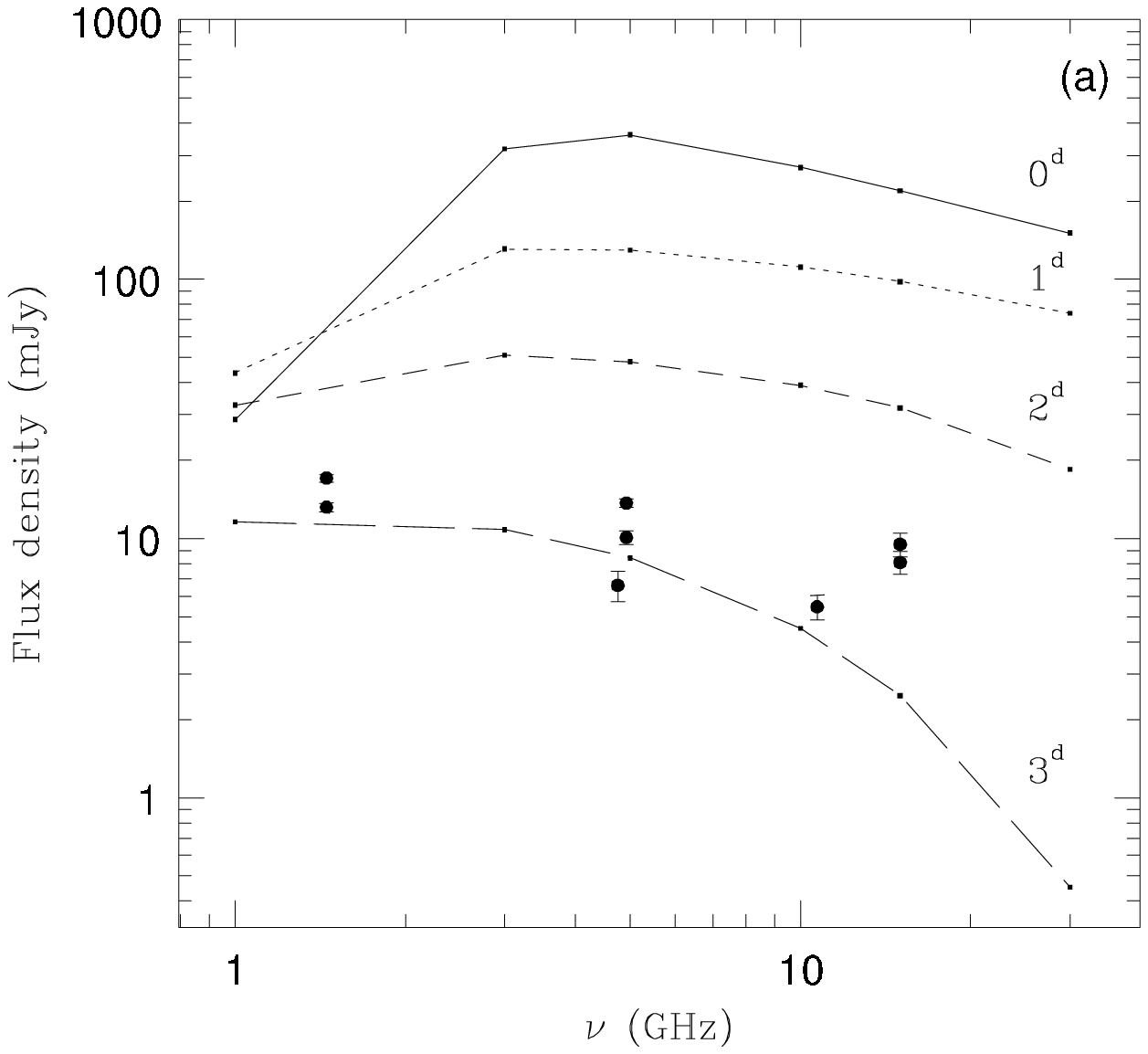}}
\hskip -1truecm\resizebox{0.42\textwidth}{!}{\includegraphics{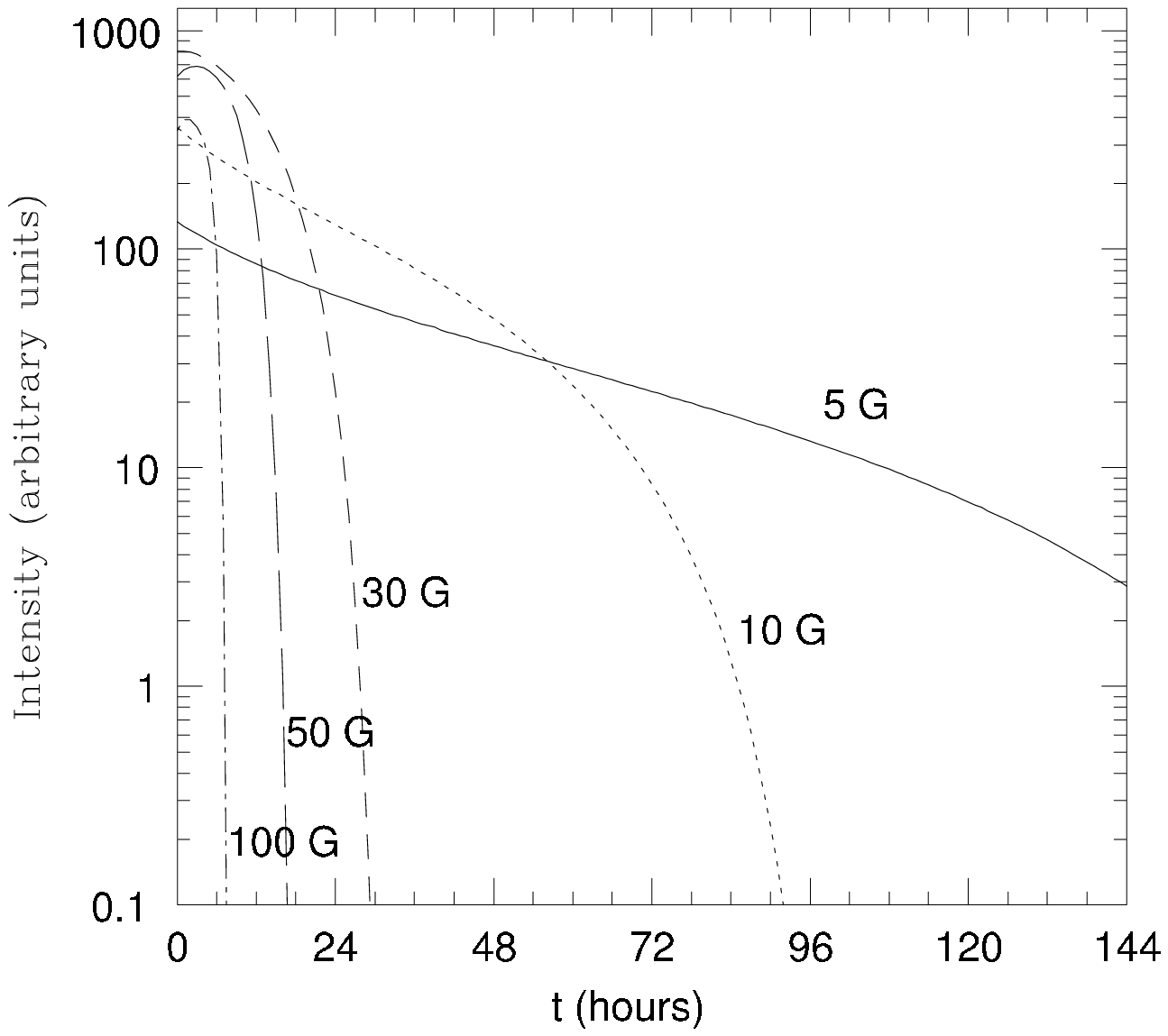}}
}
\vskip -1truecm\caption{Time-dependent spectral model of an RS CVn magnetosphere after injection of
a power-law electron population. {\it Left:} Evolution of the electron energy distribution
due to collisional and synchrotron losses, starting from a power law. {\it Middle:}
Evolution of the microwave spectrum with time, for a magnetic field of 
$B = 10$~G and an initial electron power-law index of $\delta = 2$. The electron
density in the thermal power-law is $n_e = 2\times 10^8$~cm$^{-3}$. {\it Right:}
Time evolution of the radiation intensity at 5~GHz, for different magnetic field strength.
(From \citealt{chiuderidrago93}, courtesy of E. Franciosini.)
}\label{rscvnmodel}
\end{figure}

The radio-spectral time development implied by this model can be analytically calculated
for the case of a short injection of electrons into the corona \citep{chiuderidrago93}.
The outline of the derivation is as follows: For trapped electrons with nonthermal 
number density $n$, an equation of continuity applies in energy,
\begin{equation}
{\partial n(\gamma,t) \over \partial t} + {\partial \over \partial \gamma} \left[ n(\gamma, t)
             {d\gamma \over dt} \right] = 0
\end{equation}
with the solution
\begin{equation}
n(\gamma,t)d\gamma = n(\gamma_0)d\gamma_0.
\end{equation}
Here, $n(\gamma_0)d\gamma_0$ is the initial distribution.
We now need an expression for the energy loss rate of an electron, $d\gamma/dt$. In an extended
magnetosphere,  the most relevant loss mechanisms are synchrotron loss and Coulomb-collisional	
loss, given by, respectively,  
\begin{equation}\label{collloss}
-{\dot{\gamma}}_{\mathrm{coll}} 
     = 5\times 10^{-13}n_e\quad \mathrm{[s^{-1}]}, \quad\quad \tau_{\mathrm{coll}}   
     = 2\times 10^{12}{\gamma\over n_e} \quad \mathrm{[s]}
\end{equation}
\begin{equation}\label{synchloss}
-\dot{{\gamma}}_{\mathrm{B}}  
     = 1.5\times 10^{-9}B^2\gamma^2\quad \mathrm{[s^{-1}]},\quad\quad \tau_{\mathrm{B}} = 
     {6.7\times 10^8\over B^2\gamma}\quad \mathrm{[s]}
\end{equation}
(where $n_e$ is the ambient thermal electron density: see, e.g., \citealt{petrosian85}; we have 
assumed a pitch angle of $\pi/3$). The total energy loss rate is thus
\begin{equation}\label{totloss}
{d\gamma\over dt} = \alpha + \beta\gamma^2
\end{equation}
with the appropriate coefficients from (\ref{collloss}) and (\ref{synchloss}). This equation has
an analytical solution for $n(\gamma,t)$ if the initial distribution is a power law as given by
(\ref{npowerlaw}), namely
\begin{equation}
n(\gamma,t) = K(1+\tan^2A\alpha t) A^{\delta} 
              {  [A\gamma(1+A\tan A\alpha t) - A + \tan A\alpha t]^{-\delta} \over
	         (1 - A\gamma\tan A\alpha t)^{2-\delta}   }
\end{equation}
where $A = (\beta/\alpha)^{1/2}$, and the initial power-law has been bounded by $\gamma_{0,1} < \gamma_0$
(typically a low energy, e.g., $\gamma_{0,1} = 1.1$) and $\gamma_{0,2} > \gamma_0$, typically a 
very high energy, formally including $\gamma_{0,2} = \infty$. The evolution of the initial 
power-law boundaries can be computed; after a finite time, 
all non-thermal electrons have thermalized. Some characteristic results of these calculations
are shown in Fig.~\ref{rscvnmodel}.

Figure~\ref{rscvnmodel}a shows the flattening with time of the electron energy distribution due to
the collisional losses at lower energies, and a cut-off due to synchrotron losses
at higher energies. Figure~\ref{rscvnmodel}b illustrates the intensity of radiation after calculation of
the emissivity and the absorption coefficient, together with an observed spectrum of the
RS CVn binary UX Ari. In the early spectrum, the optically thick portion is well visible
on the low-frequency side. As the electron energy decays, the spectrum becomes optically thin,
and once the high-energy electron population is depleted, the spectrum begins to fall
off steeply as there are no electrons left emitting at the observing frequency.

The time development of the emitted intensity as a function of magnetic field strength
(Fig.~\ref{rscvnmodel}c) shows that the initial radiation originates in fields of higher strength (the ``core''), 
but  the responsible electron population decays rapidly due to synchrotron losses. Eventually, 
radiation  from the slowly-decaying population in the weaker, more extended magnetic fields (the ``halo'')
begins to dominate on longer time scales. 		  
 
A power-law index of $\delta = 2-3$ (as also used above) appears to usually fit active-stellar 
microwave spectra such as those shown in Fig.~\ref{spec} quite well (\citealt{white89, umana98}).
This suggests very hard electron distributions, similar to those seen in gradual solar flares
\citep{cliver86}. But such distributions are observed during so-called quiescence - an indication
that quiescent radio emission is due to a flare-like process? We will return to this hypothesis in
later sections.

\section{Thermal X-ray emission from stellar coronae}

\subsection{High-resolution X-ray spectroscopy}

The high-resolution X-ray spectrometers on {\it XMM-Newton} and {\it Chandra} cover a large 
range of spectral lines that are temperature- and partly  density-sensitive. The spectra thus contain 
the features required for deriving X-ray emission measure distributions, abundances, coronal
densities, and opacities.
 
\begin{figure} 
\centerline{\resizebox{1.0\textwidth}{!}{\includegraphics{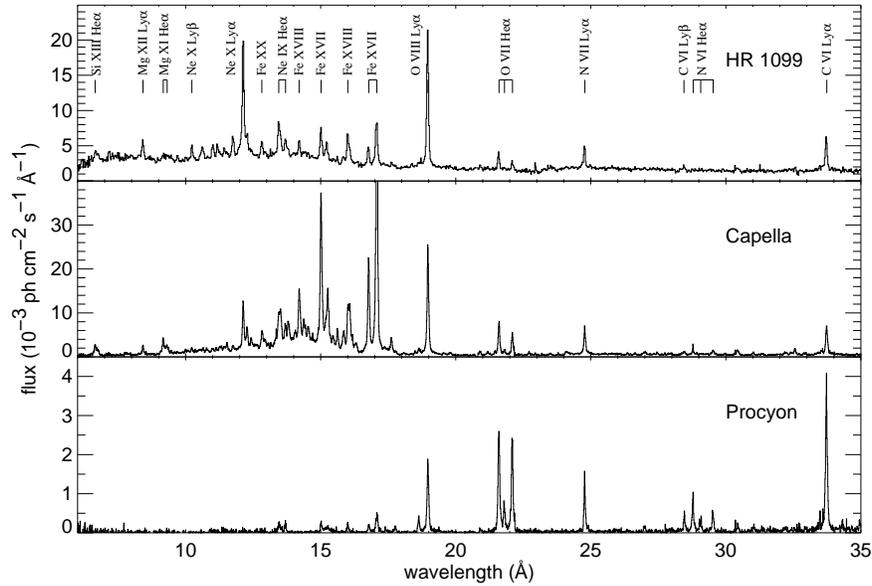}}}
\caption{Three high-resolution X-ray spectra of stars with largely differing activity levels: HR~1099,
Capella, and Procyon. Data from {\it XMM-Newton} RGS.}\label{spectra1}
\end{figure}

\begin{figure} 
\centerline{\resizebox{0.8\textwidth}{!}{\includegraphics{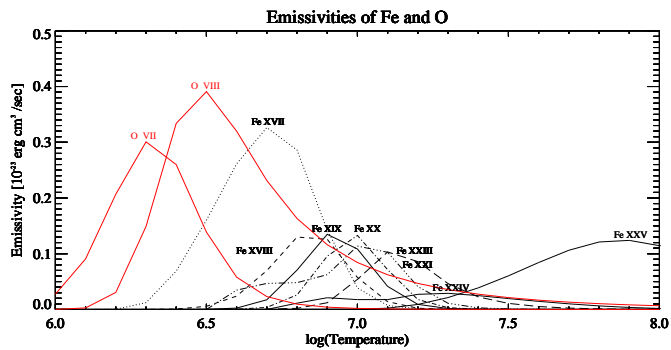}}}
\caption{Emissivities of several X-ray transitions, in particular of highly-ionized iron and
oxygen species (Fig. courtesy of A. Telleschi, after Telleschi et al. 2005).}\label{lineemissivity}
\end{figure}

Figure~\ref{spectra1}  shows examples of X-ray spectra. 
The stars cover the entire range of stellar activity: HR~1099 representing a very active RS CVn 
system, Capella an intermediately active binary, and Procyon an inactive F dwarf. The spectrum of HR~1099 
reveals a considerable amount of continuum and comparatively weak lines, which is a consequence of the very hot 
plasma in this corona ($T \approx 5-30$~MK). Note also the unusually strong Ne\,{\sc ix}/Fe\,{\sc xvii} and 
Ne\,{\sc x}/Fe\,{\sc xvii} flux ratios if compared to the other stellar spectra (these ratios are due
to an intrinsic compositional anomaly of the HR~1099 corona). The spectrum of Capella is dominated by 
Fe\,{\sc xvii} and Fe\,{\sc xviii} lines which are preferentially formed in this corona's plasma at 
$T \approx 6$~MK (Fig.~\ref{lineemissivity}). Procyon, in contrast, shows essentially no continuum and only very weak lines of Fe. 
Its spectrum is dominated by the H- and He-like transitions of C, N, and O formed  around 1--4~MK.  The 
flux ratios between H- to He-like transitions are also convenient  temperature indicators: The 
O\,{\sc viii}~$\lambda 18.97$/O\,{\sc vii}~$\lambda 21.6$ flux ratio, for example, is very large 
for HR~1099 but drops  below unity for Procyon.

\subsection{Thermal coronal components}\label{thermalcomp}

The large range of temperatures measured in stellar coronae has been a challenge for 
theoretical interpretation. Whereas much of the solar coronal plasma can be well described 
by a  component of a few million degrees, magnetically active stars
have consistently shown a wide distribution of electron temperatures, reaching 
values as high as 50~MK outside obvious flares.

The flux $\phi_j$ observed in a spectral line from a given atomic transition can be written as
\begin{equation}\label{fluxdem}
\phi_j = {1\over 4\pi d^2}   \int AG_j(T) {n_en_H dV\over d{\rm ln}T}d{\rm ln}T
\end{equation}
where $d$ denotes the distance, and $G_j(T)$ is the ``line cooling function'' (luminosity 
per unit EM; Fig.~\ref{lineemissivity}) that contains 
the atomic physics of the transition as well as the ionization  fraction for  the
ionization stage in question, and $A$ is the  abundance of the element with respect
to some basic tabulation used for $G_j$. For a fully ionized plasma
with cosmic abundances, the hydrogen density  $n_H \approx 0.85n_e$. The expression
\begin{equation}\label{demdef}
Q(T)  = {n_en_HdV\over d{\rm ln}T}
\end{equation}
defines the {\it differential emission measure distribution} (DEM). I will use this definition
throughout but note that some authors define $Q^{\prime}(T) = n_en_HdV/dT$  which is 
smaller by one power of $T$. For a plane-parallel  atmosphere with surface area $S$,
Eq.~(\ref{demdef}) implies 
\begin{equation}
Q(T) = n_en_H SH(T), \quad\quad H(T) = \left|{1\over T}{dT\over ds}\right|^{-1}
\end{equation} 
where  $H$ is the {\it temperature scale height.}

\subsection{Observational results}\label{demresults}

\begin{figure} 
\hbox{
\resizebox{0.535\textwidth}{!}{\includegraphics{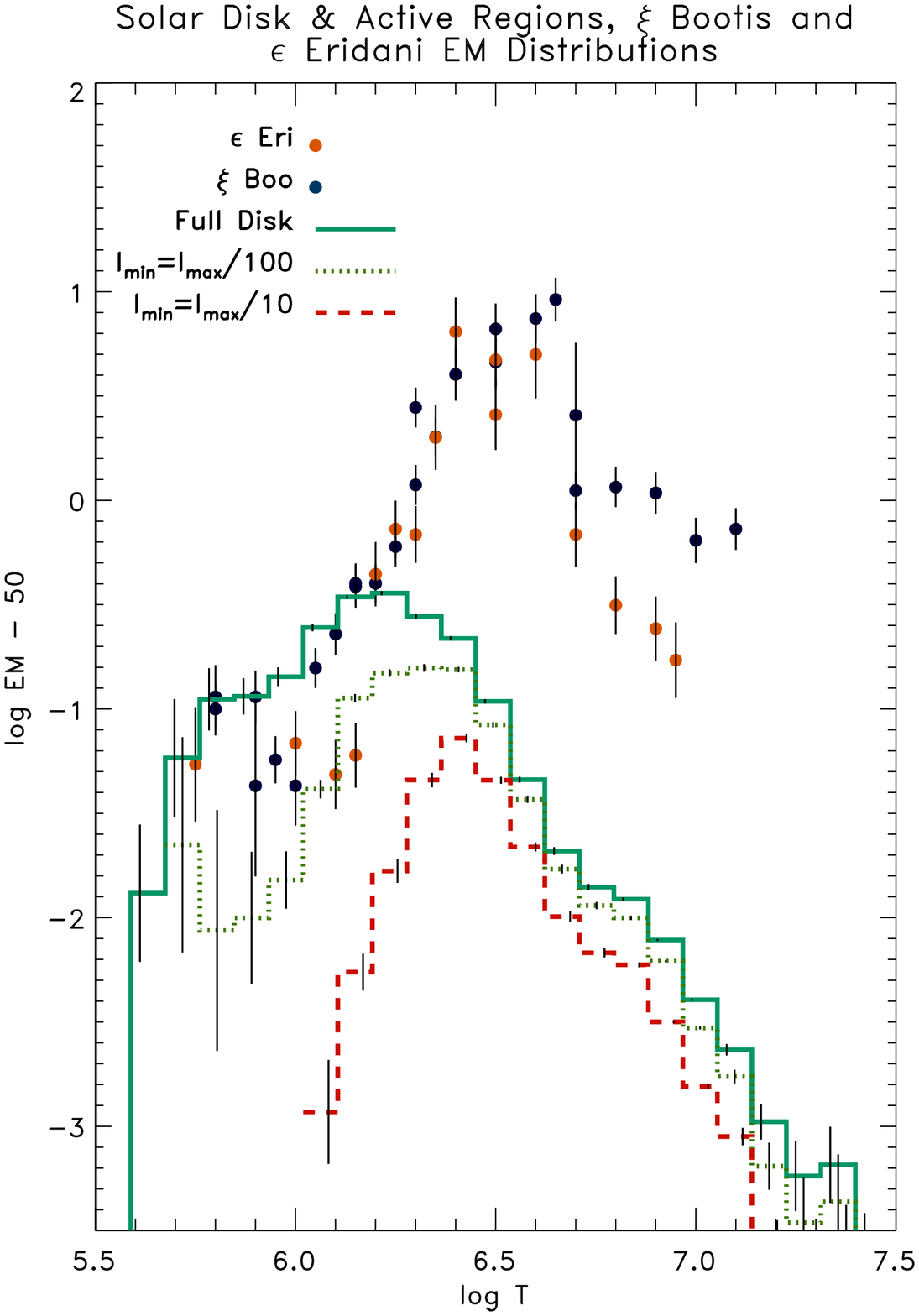}}
\vbox{
\resizebox{0.465\textwidth}{!}{\includegraphics{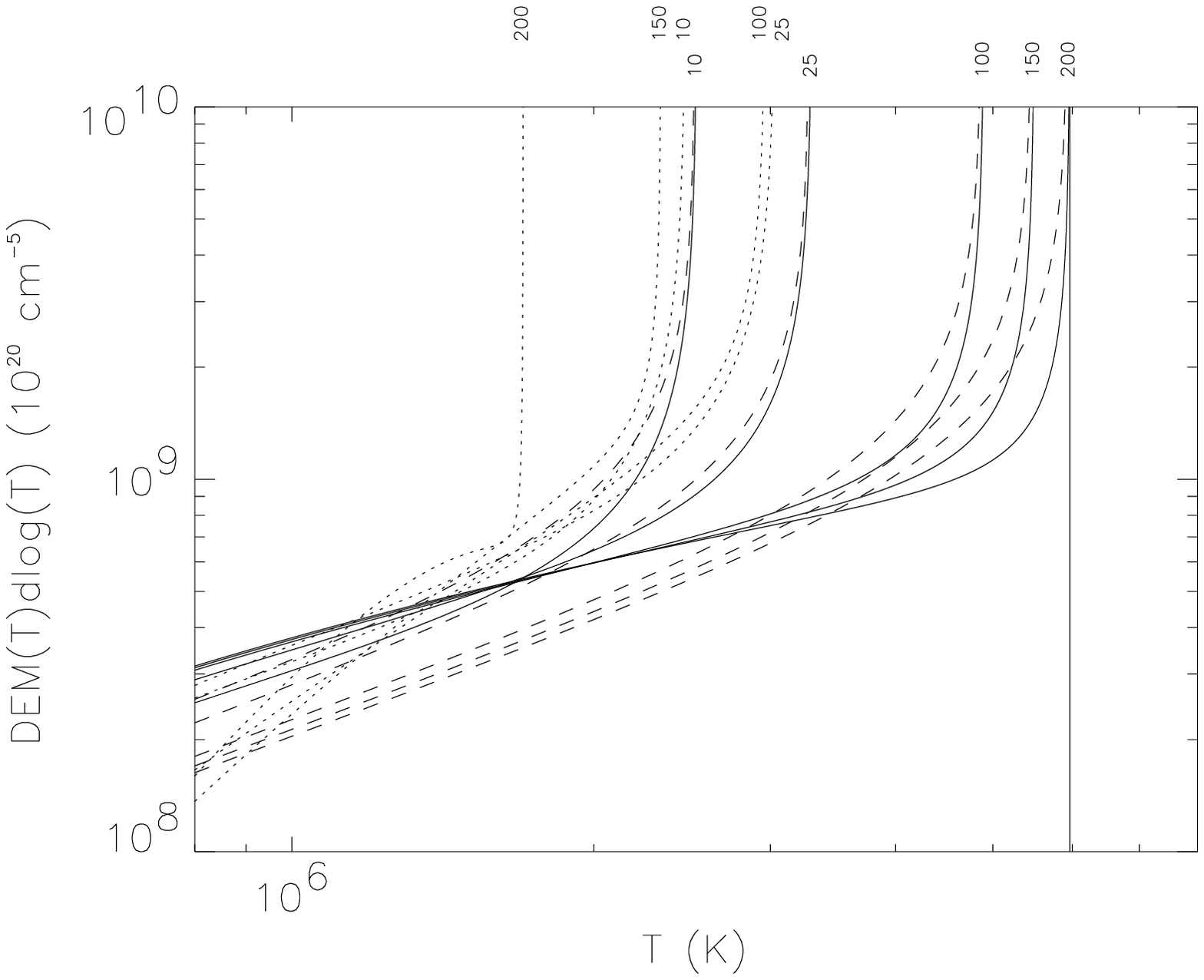}}\\
\resizebox{0.465\textwidth}{!}{\includegraphics{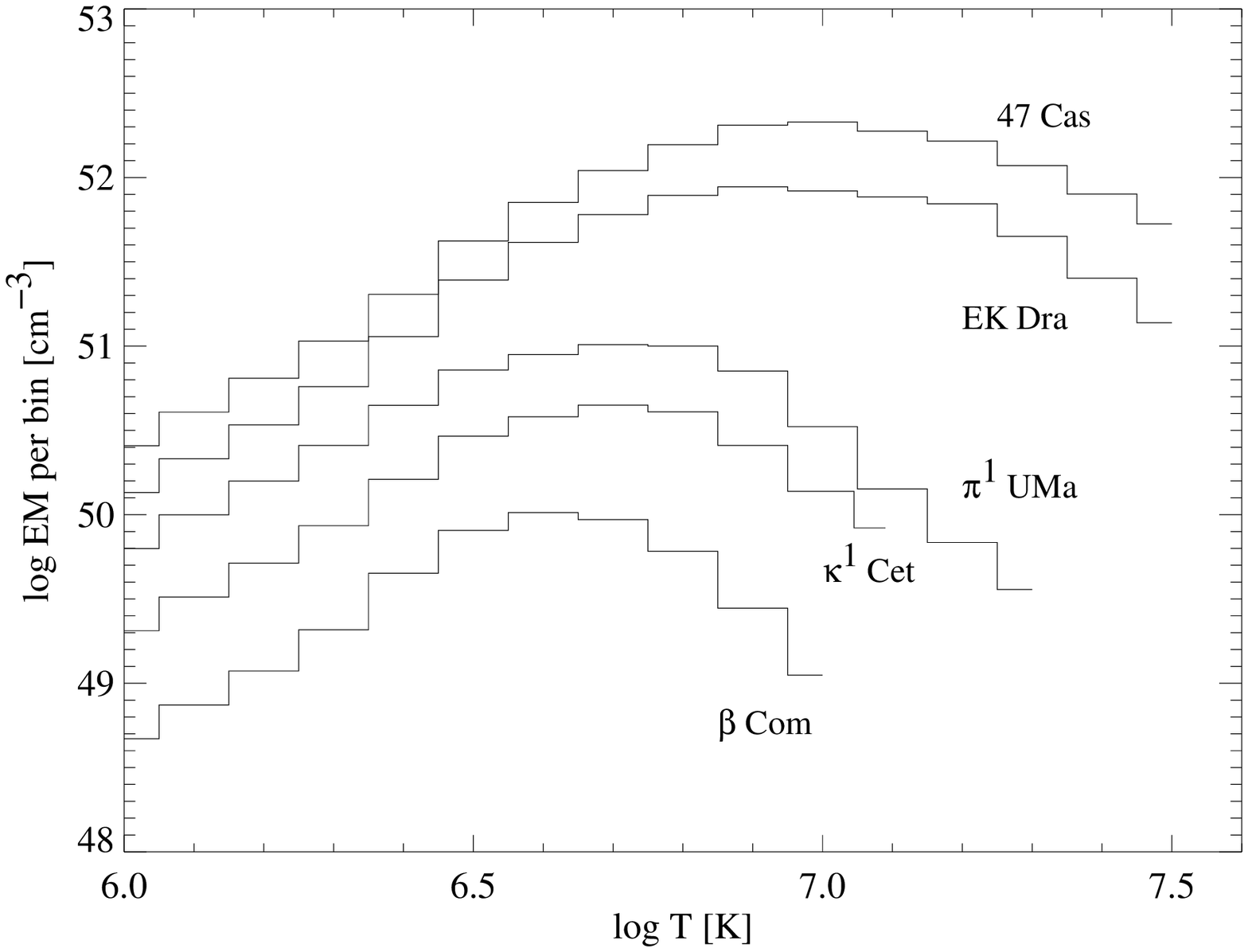}}
}
}
\caption{{\it Left:} Emission measure distributions  of two intermediately active stars 
(bullets) and the Sun (histograms; EM is in units of $10^{50}$~cm$^{-3}$). The histograms 
refer to full-disk solar EMDs derived from {\it Yohkoh} images at solar maximum, 
including also two versions for different lower cutoffs for the intensities in {\it Yohkoh} images. 
(figure courtesy of J. Drake, after \citealt{drake00}). 
-- {\it Upper right:} Calculated differential emission measures of individual
static loops. The solid curves refer to uniform heating along the loop and some
fixed footpoint heating flux, for different loop half-lengths labeled above the 
figure panel in megameters. The dashed curves illustrate the analytical
solutions presented by \citet{rosner78} for uniform heating. The dotted
lines show solutions assuming a heating scale height of $2\times 10^9$~cm (figure
courtesy of K. Schrijver, after \citealt{schrijver02}).
-- {\it Lower right:} Examples of discrete stellar emission measure distributions 
of solar analogs (figure courtesy of A. Telleschi).}
\label{demsolarstellar}
\end{figure}

``Discrete'' (binned) emission measure distributions (EMDs) reflecting the full 
DEMs can be obtained by various inversion algorithms from the observed spectra.
Most EMDs have generally been found to be singly or doubly peaked 
and confined on either side approximately by power laws (e.g.,  \citealt{schrijver95}). 
Interestingly, EMDs are often very steep on the low-$T$ side, and this is particularly 
true for the more active stars. For example, the slope of the stellar EMD in Fig.~\ref{demsolarstellar}a 
follows approximately  $Q \propto T^3$ on the low-$T$ side. 
 
It is notable that 
the complete EMD shifts to higher temperatures with increasing stellar activity 
as seen in Fig.~\ref{demsolarstellar}ac, often leaving very 
little EM at modest temperatures and correspondingly weak
spectral lines from ions of  C, N, and O. 

\begin{figure} 
\centerline{\resizebox{0.8\textwidth}{!}{\includegraphics{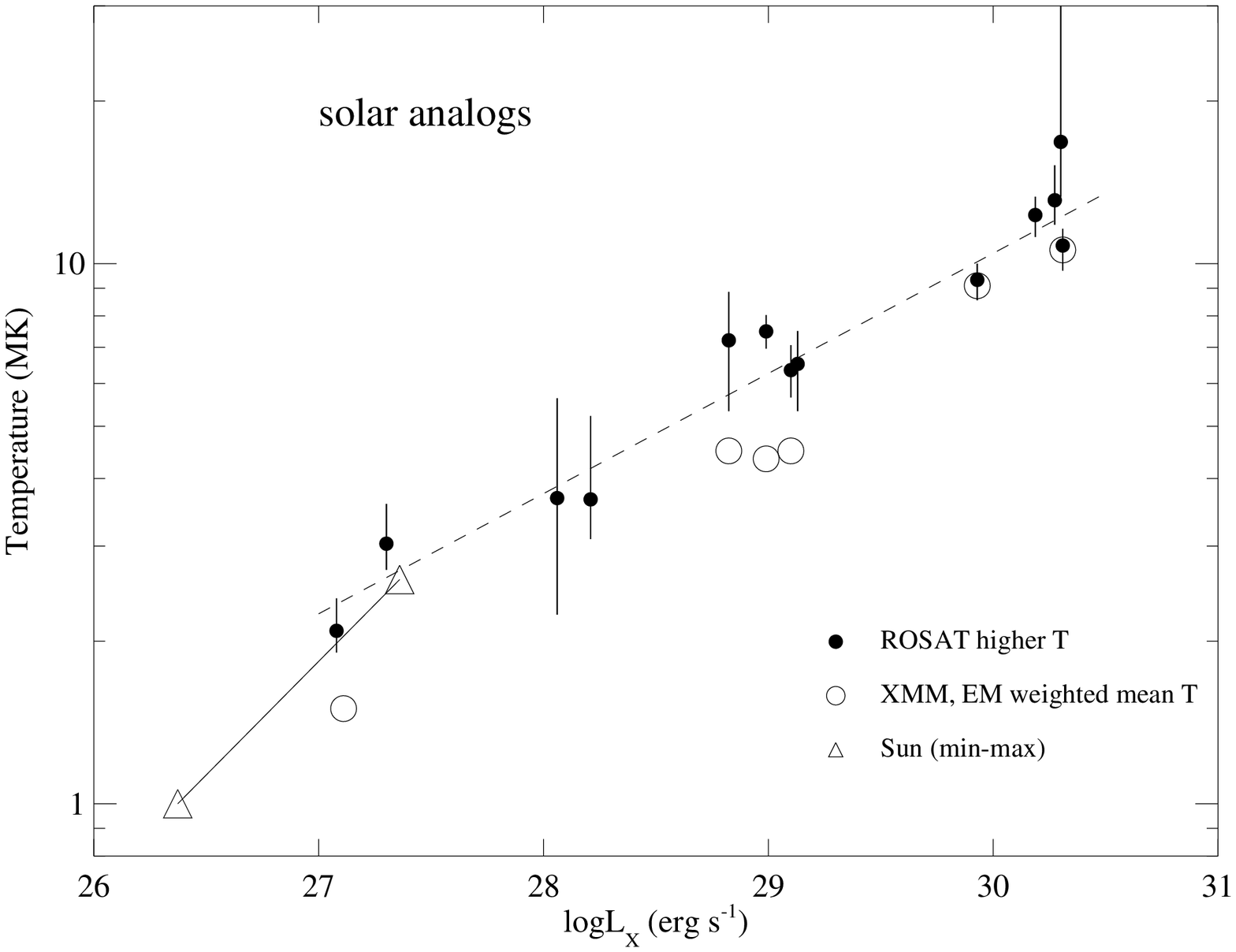}}}
\caption{Coronal temperature vs. X-ray luminosity for solar analogs. For details and
references, see \citet{guedel04b}.\label{temperaturelx}}
\end{figure}

As a consequence, a relatively tight
correlation between the characteristic coronal temperature (e.g., derived from a DEM)   
and the normalized coronal luminosity $L_X/L_{\rm bol}$ is found:
{\it Stars at higher activity levels support hotter coronae}, with the most
active stellar coronae reaching characteristic temperatures of
several tens of MK.  An example of solar 
analogs is shown together with the Sun itself during its activity maximum 
and minimum in  Fig.~\ref{temperaturelx}. Here,  
\begin{eqnarray}\label{TLx}
L_X \propto T^{4.5\pm 0.3} \\
EM  \propto T^{5.4\pm 0.6} 
\end{eqnarray}
where $L_X$ denotes  the total X-ray luminosity (but EM and $T$ refer to the 
``hotter'' component in standard 2-$T$ fits to {\it ROSAT} data). 

\subsection{Interpretation of differential emission measure distributions}\label{deminterpret}

Equations~(\ref{fluxdem}) and (\ref{demdef}) introduce the  DEM as the basic interface between 
the stellar X-ray observation and the model interpretation of the thermal source. It contains 
information on the plasma temperature and the density-weighted plasma mass that emits X-rays 
at any given temperature. Although a DEM is often a  highly degenerate description of a complex 
real corona, it provides important constraints on heating theories and on the range of coronal 
structures that it may describe. {\it Solar} DEMs can, similarly to the stellar cases, often be approximated
by two power laws $Q(T)  \propto T^s$  on either side of their peaks (Fig.~\ref{demsolarstellar}). 
The Sun has indeed given considerable guidance in physically interpreting the observed stellar DEMs. 

\subsubsection{The DEM of a static loop}\label{demstatic}

The DEM  of the plasma contained in a static magnetic loop follows from the hydrostatic 
equilibrium (see, e.g., derivation by \citealt{rosner78}). Under the conditions of negligible 
gravity, i.e., constant pressure in the entire loop, and negligible thermal conduction at the footpoints,
\begin{equation}\label{staticem}
Q(T) \propto pT^{3/4-\chi/2+\alpha}  {1 \over \left( 1 - \left[T/T_{\rm a}\right]^{2-\chi+\beta}\right)^{1/2}}
\end{equation}
\citep{bray91} where $T_{\rm a}$ is the loop apex temperature,  and $\alpha$ and $\beta$ are power-law 
indices of, respectively,  the loop cross section area $S$ and the heating power $q$ as a function of $T$:
$S(T) = S_0T^{\alpha}$,  $q(T) = q_0T^{\beta}$, and 
$\chi$ is the exponent in the cooling function over the relevant temperature range: 
$\Lambda(T) \propto T^{\chi}$. If $T$ is not close to $T_{\rm a}$ and the loops have constant cross
section ($\alpha = 0$), we have 
\begin{equation} 
Q(T) \propto T^{3/4 -\chi/2},
\end{equation}
i.e., under typical coronal conditions for non-flaring loops ($T < 10$~MK, $\chi \approx -0.5$), the DEM 
slope is near unity. If strong thermal conduction is included at the footpoints, then
the slope changes to +3/2 if not too close to $T_{\rm a}$ \citep{vdoord97}.
The single-loop DEM sharply increases at $T \approx T_{\rm a}$ (Fig.~\ref{demsolarstellar}). 

Loop expansion ($\alpha > 0$) obviously steepens the DEM. Increased heating at the loop footpoints 
(instead of uniform heating) makes the $T$ range narrower and will also increase the slope 
of the DEM  (see  numerical calculations of various loop examples by \citealt{schrijver02} and 
\citealt{aschwanden02}, Fig.~\ref{demsolarstellar}b).

\subsubsection{The DEM of flaring structures}\label{demflaring}

\citet{antiochos80} discussed DEMs of solar flaring loops 
that cool by i) static conduction (without flows), ii) evaporative conduction (including flows), and
iii) radiation. The inferred DEMs scale, in the above order, like
\begin{equation}
Q_{\rm cond} \propto  T^{1.5}, \quad\quad  Q_{\rm evap} \propto T^{0.5}, 
     \quad\quad Q_{\rm rad} \propto T^{-\chi + 1}. 
\end{equation}
Since $\chi \approx 0\pm 0.5$ in the range typically of interest for stellar flares
($5-50$~MK), all above DEMs are relatively flat (slope $1\pm 0.5$). If multiple loops with equal slope 
but different peak $T$ contribute, then the slope up to the first DEM peak can only become smaller.
Non-constant loop cross sections have a very limited influence on the DEM slopes. 

Stellar flare observations are often not of sufficient quality to derive temperature and EM characteristics for 
many different time bins. An interesting diagnostic was presented by \citet{mewe97} who calculated
the time-integrated (average) DEM of a flare that decays quasi-statically. They find
\begin{equation}\label{demflareint}
Q \propto T^{19/8}
\end{equation}
up to a maximum $T$ that is equal to the temperature at the start of the decay phase.

Systems of episodically flaring loops were 
computed semi-analytically by \cite{cargill94}, using analytic approximations for conductive and 
radiative decay phases of the flares. Here, the DEM is defined not by the internal loop structure 
but by the time evolution of a flaring plasma (assumed to be isothermal). Cargill argued that 
for radiative cooling, the (statistical) contribution of a flaring loop to the DEM is, to 
zeroth order, inversely proportional to the radiative decay time, which implies
\begin{equation}
Q(T) \propto T^{-\chi+1}
\end{equation}
up to a maximum $T_m$, and a factor of $T^{1/2}$ less if subsonic draining of the cooling loop is 
allowed. Simulations with a uniform distribution of small flares within a limited energy range agree with
these rough predictions, indicating a time-averaged DEM that is relatively flat below $10^6$~K but 
steep ($Q[T] \propto T^4$) up to a few MK, a range in which the cooling function drops rapidly.

Let us next assume - in analogy to solar flares - that the occurrence rate of stellar flares 
is distributed in energy as a power law with an index $\alpha$  ($dN/dE \propto E^{-\alpha}$) 
Then, an analytic expression can be derived for the 
time-averaged DEM of such a flare ensemble, i.e., a ``flare-heated corona'' \citep{guedel03a}. 
We present a brief outline of the derivation.

Observationally, the flare peak temperature is correlated with the peak EM
both in solar \citep{feldman95} and stellar flares \citep{guedel04b}:
\begin{equation}\label{e:feldman}
\mathrm{EM}_0  = aT_0^b \quad {\rm [cm^{-3}]}
\end{equation}
with $b\approx 4.3\pm 0.35$ in the range of $T = 10-100$~MK (see Fig.~\ref{tem_flare} further below).
The X-ray luminosity $L_X$ of a plasma due to bremsstrahlung and line emission can be expressed as 
\begin{equation}\label{e:luminosity}
L_X \approx \mathrm{EM}~\Lambda(T) = g~\mathrm{EM}~T^{-\phi}    
\end{equation}
with $\phi \approx 0.3$ over the  above temperature range (for broad-band X-ray losses). 

From (\ref{e:feldman}) and (\ref{e:luminosity}) we obtain a relation between
the flare peak temperature $T_0$ and its peak luminosity $L_{X,0}$,
\begin{equation}\label{e:peakt}
T_0 = \left({L_{X,0}\over ag}\right)^{1/(b-\phi)}.
\end{equation} 
We will investigate the general case in which $\tau$ varies with the
flare energy, namely
\begin{equation}
\tau = \tau_{\rm 0} E^{\beta}
\end{equation}
(where $\beta \ge 0$ is assumed, and $\tau_{\rm 0}$ is a constant adjusted to the larger 
detected  flares) in which case
\begin{equation}\label{e:peakdistribtau}
{dN\over dL_{X,0}} ={dN\over dE}{dE\over dL_{X,0}} = k^{\prime}L_{X,0}^{-(\alpha-\beta)/(1-\beta)}.
\end{equation}
where the constant $k^{\prime} = k\tau_{\rm 0}^{(1-\alpha)/(1-\beta)}/(1-\beta) > 0$ as 
long as $\beta <1$ (which can be reasonably assumed).
Since we neglect the short rise time of the flare, our flare light curves are
described  by their  exponential decay at $t \ge 0$,
\begin{equation}\label{e:decay}
L_X(t) = L_{X,0}e^{-t/\tau}.
\end{equation}
From hydrodynamic modeling, theory and observations, it is known  
that during the flare decays $T\propto n_e^{\zeta}$, where $n_e$ is the plasma density
\citep{reale93}.
The parameter $\zeta$ is usually found between 0.5 and 2.

Integrating the above equations for all emission contributions from the decaying 
and cooling flare plasma over the entire distribution of flares leads to two expressions 
valid for different regimes:
\begin{equation} 
Q(T) \propto \left\{ 
   \begin{array}{ll}\label{demflare}
        T^{2/\zeta}                                                  & \mbox{\quad for\quad  $T < T_m$ } \\
        T^{-(b-\phi)(\alpha-2\beta)/(1-\beta) +2b - \phi}            & \mbox{\quad for\quad  $T > T_m$ } 
   \end{array} 
   \right. 
\end{equation}
Here, $T_m$ (a free parameter) is the temperature of the DEM peak. It is controlled by 
the lower cutoff in the power-law energy distribution of flares (for details, see 
\citealt{guedel03a}). 

\subsection{Discussion and summary on the temperature structure}\label{sumdem}

There is, as of yet, no clear explanation for the observed DEMs in active stars.
The most notable feature is the often steep slope on the low-$T$ side. The steepness is
not easy to explain with standard static loop models, although strong footpoint heating
may be a way out. Another explanation are magnetic loops
with an expanding cross section from the base to the apex. The larger amount of plasma at
high temperatures near the apex evidently steepens the DEM. Solar imaging does not
prefer this type of loop, however.
 
Alternatively, the steep slope may be evidence for continual flaring; 
equation~(\ref{demflare}) predicts slopes between 1 and 4, similar to what 
is often found in magnetically active stars. There is other evidence for this model
to have some merit - we will encounter it again in the subsequent sections.
      
It is also well established that more luminous stars (of a given size) reveal hotter coronae (Fig.~\ref{temperaturelx}).
Again, the cause of this relation is unclear. Possibly, 
increased magnetic activity leads to more numerous interactions between 
adjacent magnetic field structures. The heating efficiency thus increases. 
In particular, we expect a higher rate of large flares. The increased flare rate produces 
higher X-ray luminosity because chromospheric evaporation produces more EM; at the 
same time, the plasma is heated to higher temperatures in larger flares 
\citep{guedel97}.

\subsection{Electron densities in stellar coronae}\label{coronaldensities}

Coronal electron densities control radiative losses from the 
coronal plasma; observationally, they can in principle also be used in conjunction with 
EMs to derive approximate coronal source volumes. The spectroscopic derivation of coronal 
densities is subtle, however.  Two principal methods  are available.

{\it Densities from Fe line ratios:}
The emissivities of many transitions of Fe ions in the EUV range 
are sensitive to densities  in the range of interest to  coronal research 
\citep{mewe85}. The different
density dependencies of different lines of the same Fe ion then also make their 
{\it line-flux ratios}, which (apart from blends) are easy to measure, useful 
diagnostics for the electron density.

\begin{figure} 
\hbox{
\resizebox{0.45\textwidth}{!}{\includegraphics{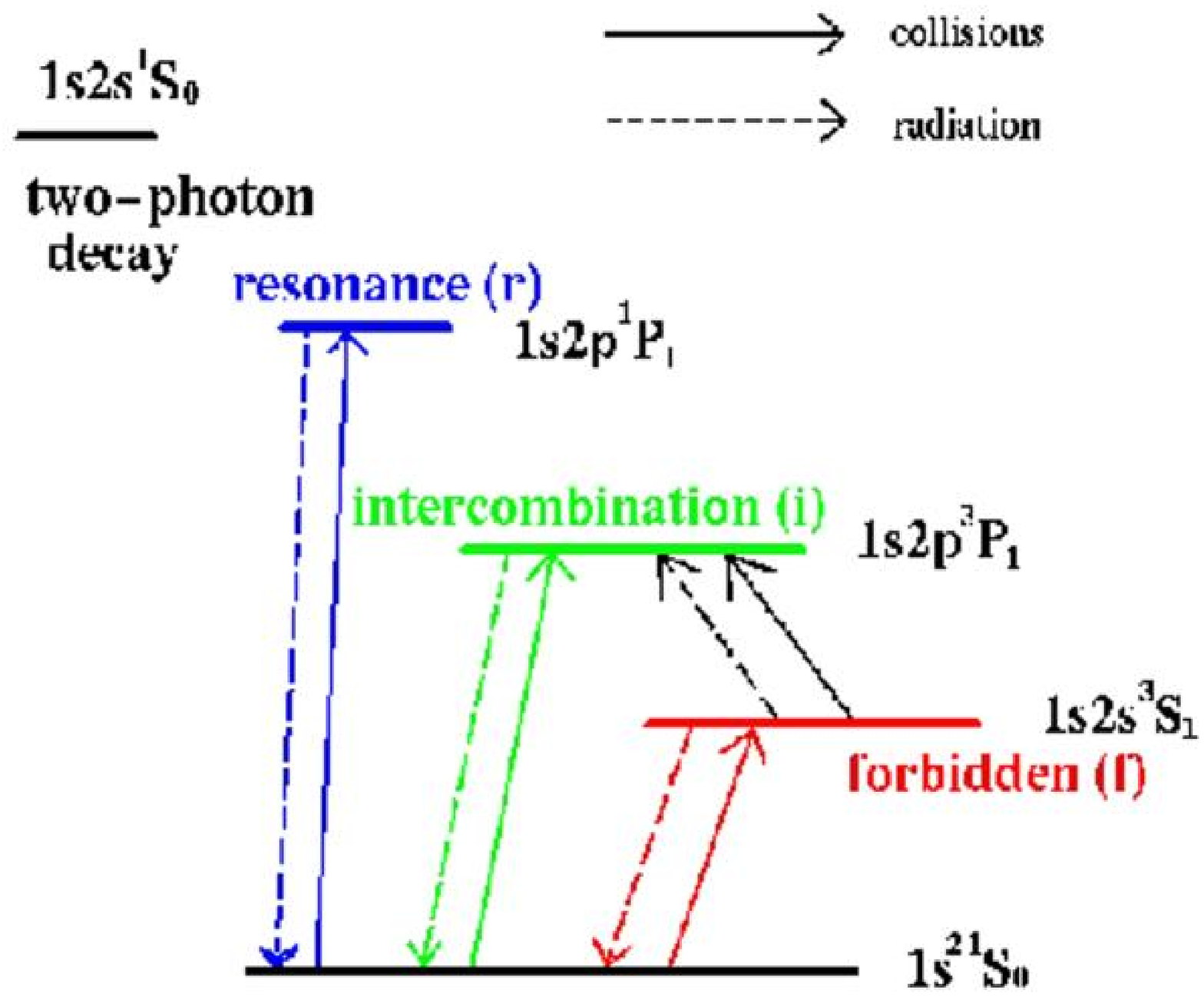}}
\resizebox{0.55\textwidth}{!}{\includegraphics{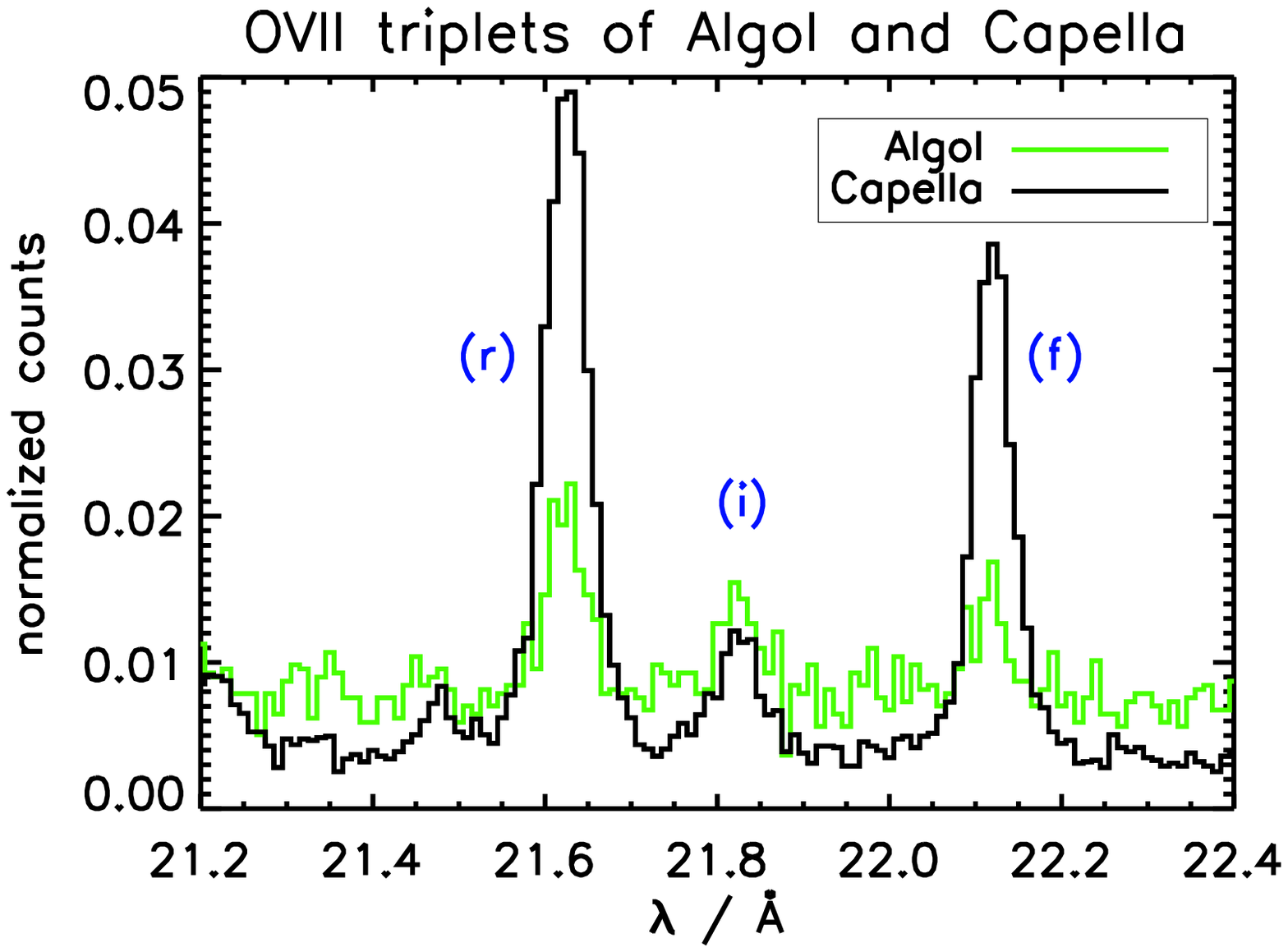}}
}
\vskip 0.4truecm\caption{{\it Left:} Term diagram for transitions in He-like triplets. The resonance, intercombination,
       and forbidden transitions are marked. The transition from $^3S_1$ to $^3P_1$ re-distributes electrons from the
       upper level of the forbidden transition to the upper level of the intercombination transition, thus making 
       the $f/i$ line-flux ratio density sensitive. In the presence of a strong UV field, however, the same 
       transition can be induced by radiation as well.
 {\it Right:} He-like triplet of O\,{\sc vii} for Capella (black) and Algol (green).
       The resonance (r), intercombination (i), and forbidden (f) lines are marked.
       The $f/i$ flux ratio of Algol is suppressed probably due to the strong UV radiation field
       of the primary B star (data from {\it Chandra}; both figures courtesy of J.-U. Ness).}\label{oxygentriplet}
\end{figure}

{\it Densities from He-like line triplets:} The He-like triplets of C\,{\sc v}, N\,{\sc vi}, 
O\,{\sc vii}, Ne\,{\sc ix}, Mg\,{\sc xi}, and Si\,{\sc xiii}  provide another interesting 
density diagnostic for stellar coronae. Two examples are shown in Fig.~\ref{oxygentriplet} (right). 
The spectra show, in order of increasing wavelength, the resonance, the intercombination, and 
the forbidden line of the O\,{\sc vii} triplet. The corresponding transitions are depicted in the left panel 
of the Figure.
The ratio between the fluxes in the forbidden line and the intercombination line is sensitive to density
\citep{gabriel69} for the following reason: if the electron collision rate is sufficiently high, 
ions in the upper level of the forbidden transition, $1s2s\ ^3S_1$, do not return to the ground 
level, $1s^2\ ^1S_0$, instead the ions are collisionally excited to the upper levels of the 
intercombination transitions, $1s2p\ ^3P_{1,2}$, from where they decay radiatively to the ground 
state. They thus enhance the flux  in the 
intercombination line and weaken the flux in the forbidden line. The measured ratio $R = f/i$ of the 
forbidden to the intercombination line flux can be written as
\begin{equation}
R = {R_0 \over 1 + n_e/N_c} = {f\over i}
\end{equation}
where $R_0$ is the limiting flux ratio at low densities and $N_c$ is the critical
density at which $R$ drops to $R_0/2$. 
The  parameters $R_0$ and $N_c$ are slightly dependent on the electron temperature in
the emitting source. A few useful
parameters are collected in Table~\ref{lineratios}. A systematic problem with He-like triplets 
is that the critical density $N_c$ increases with the formation temperature of 
the ion, i.e., higher-$Z$ ions measure only high densities at high $T$, while
the lower-density analysis based on C\,{\sc v}, N\,{\sc vi}, O\,{\sc vii}, and Ne\,{\sc ix} is 
applicable only to cool plasma.

\begin{table}
\caption{Density-sensitive He-like triplets$^a$}
\label{lineratios}     
\begin{tabular}{llllll}
\hline\noalign{\smallskip}
Ion       & \quad $\lambda(r,i,f)$ (\AA)&   \quad $R_0$          & \quad $N_c$              & \quad log\,$n_e$ range$^b$  &  \quad  $T$ range$^c$ (MK) \\
\noalign{\smallskip}\hline\noalign{\smallskip}
C\,{\sc v}       & \quad 40.28/40.71/41.46      &  \quad 11.4		 & \quad $6\times 10^8$     &  \quad 7.7--10	     &  \quad  0.5--2	   \\
N\,{\sc vi}      & \quad 28.79/29.07/29.53      &  \quad  5.3		 & \quad $5.3\times 10^9$   &  \quad 8.7--10.7       &   \quad 0.7--3	   \\
O\,{\sc vii}     & \quad 21.60/21.80/22.10      &  \quad  3.74  	 & \quad $3.5\times 10^{10}$&  \quad 9.5--11.5       &   \quad 1.0--4.0    \\ 
Ne\,{\sc ix}     & \quad 13.45/13.55/13.70      &   \quad 3.08           & \quad $8.3\times 10^{11}$&  \quad 11.0--13.0      &   \quad 2.0--8.0    \\ 
Mg\,{\sc xi}     & \quad 9.17/9.23/9.31         &   \quad 2.66$^d$       & \quad $1.0\times 10^{13}$&  \quad 12.0--14.0      &   \quad 3.3--13     \\
Si\,{\sc xiii}   & \quad 6.65/6.68/6.74         &  \quad  2.33$^d$       & \quad $8.6\times 10^{13}$&  \quad 13.0--15.0      &   \quad 5.0--20     \\
\noalign{\smallskip}\hline
\multicolumn{6}{l}{$^a$data derived from \citet{porquet01} at maximum formation temperature of ion} \\
\multicolumn{6}{l}{$^b$range where $R$ is within approximately [0.1,0.9] times $R_0$ } \\
\multicolumn{6}{l}{$^c$range of 0.5--2 times maximum formation temperature of ion} \\
\multicolumn{6}{l}{$^d$for  measurement with {\it Chandra} HETGS-MEG spectral resolution} \\
\end{tabular}
\end{table}

A review of the literature shows a rather unexpected
segregation of coronal densities into two realms at different temperatures. 
The cool coronal plasma measured by C, N, and O lines in inactive stars is typically found 
at low, solar-like  densities of order $10^9$~cm$^{-3}$ -- $10^{10}$~cm$^{-3}$. In {\it active} stars, 
the cooler components may show elevated densities up to several times $10^{10}$~cm$^{-3}$, but it is the 
hotter plasma component that apparently reveals extreme values up to $>10^{13}$~cm$^{-3}$ (e.g.,
\citealt{dupree93, schrijver95}). A basic concern with these latter measurements is that
most of the reported  densities are only slightly above the low-density limits for the
respective ratios, and upper limits  have equally been reported, sometimes resulting in conflicting
statements for different line ratios in the same spectrum \citep{ness04, phillips01, osten03}.
Several authors  have concluded that the extremely
high densities found in some active stars are spurious and perhaps not representative of 
coronal features. 
The observational situation is clearly unsatisfactory at the time of  writing. 
The resolution of these contradictions requires a careful reconsideration of atomic physics issues.

\section{The structure of stellar coronae}\label{coronalstructure}

The magnetic structure of stellar coronae is one of the central topics in the stellar coronal
research discipline. The extent and predominant locations of magnetic structures
currently hold the key to our understanding of the internal magnetic dynamo. 
All X-ray inferences of coronal magnetic structure in stars other than the Sun are so 
far indirect, while direct imaging, although at modest resolution, is available
at radio wavelengths. 

\subsection{Magnetic loop models}

\begin{figure} 
\centerline{
\hbox{
\resizebox{0.5\textwidth}{!}{\includegraphics{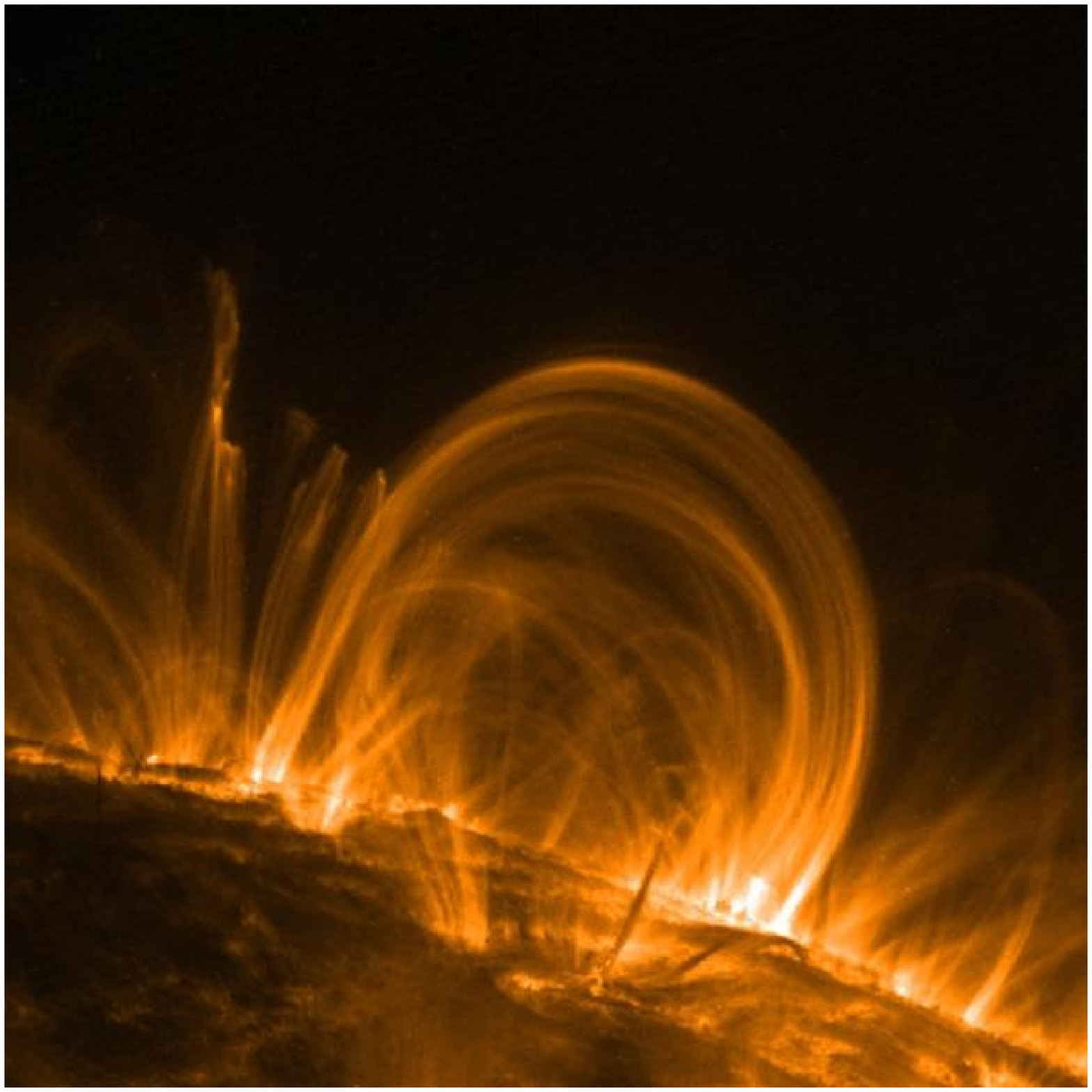}}
\resizebox{0.5\textwidth}{!}{\includegraphics{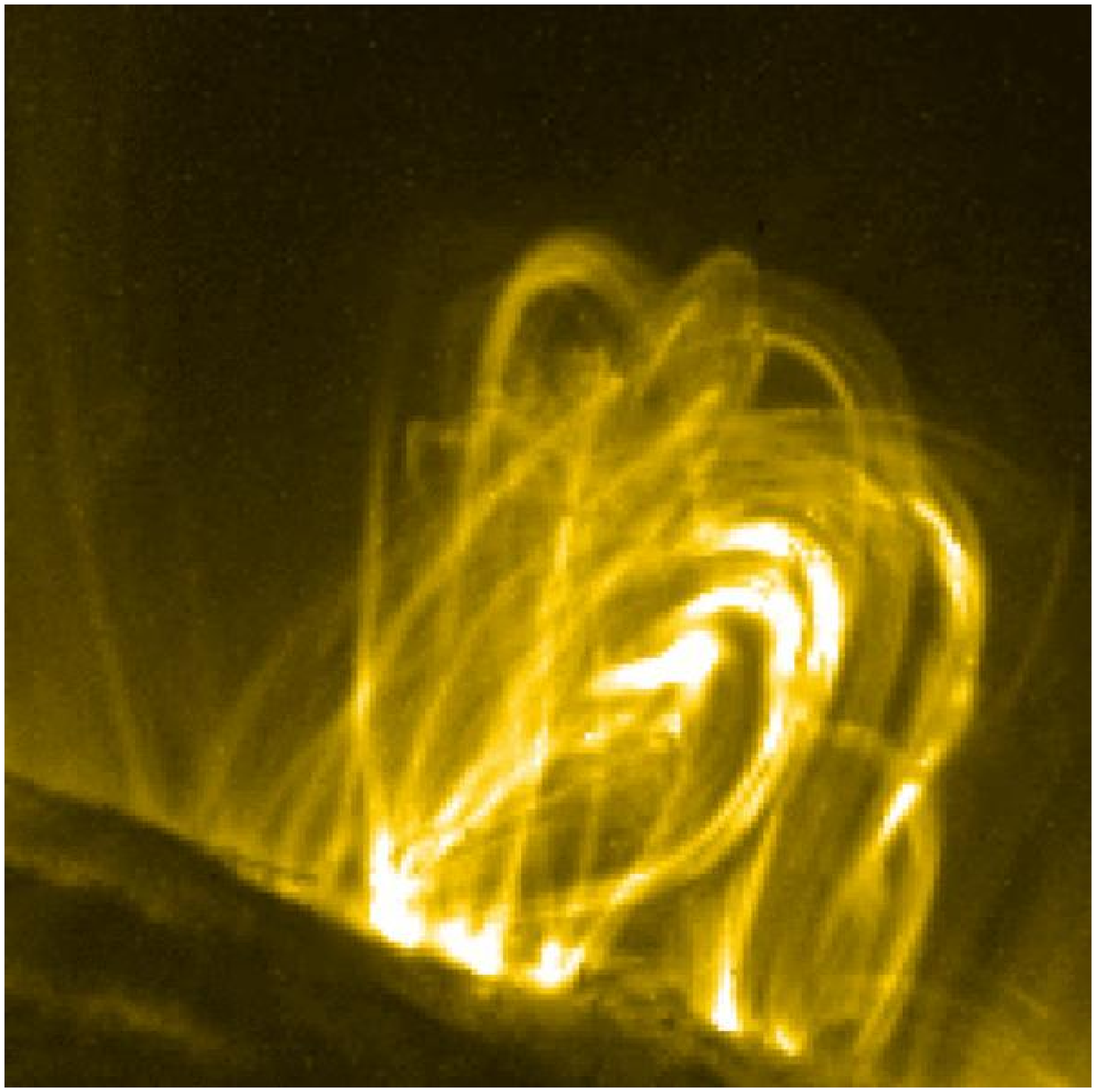}}
}
}
\caption{{\it Left:} Example of a solar coronal loop system observed by {\it TRACE}. {\it Right:}
Flaring loop system (observation by {\it TRACE} at 171\AA). Although these images show the emission from
relatively cool coronal plasma, they illustrate the possible complexity of magnetic fields.}\label{traceloop}
\end{figure}

Closed magnetic loops are the fundamental ``building blocks'' of the solar corona. When interpreting stellar
coronae of any kind, we assume that this concept applies as well, although caution is
in order. Even in the solar case, loops come in a wide variety of shapes and sizes (Fig.~\ref{traceloop}) and appear to imply heating 
mechanisms and heating locations that are poorly understood - see, for example, \citet{aschwanden00}. Nevertheless, 
simplified loop models offer an important starting point for coronal structure studies and possibly for coronal heating diagnostics.
A short summary of some elementary properties follows.

\citet{rosner78} (RTV) have modeled hydrostatic loops with
constant pressure (i.e., the loop height is smaller than the pressure scale height). They also assumed
constant cross section,  uniform heating, and absence of gravity, and found two scaling laws relating the loop semi-length $L$ (in cm),
the volumetric heating rate $\epsilon$ (in erg~cm$^{-3}$s$^{-1}$), the pressure $p$ (in dynes~cm$^{-2}$), 
and the loop apex temperature  $T_{\rm a}$ (in K),
\begin{equation}\label{RTV}
T_{\rm a} = 1400(pL)^{1/3}; \quad\quad \epsilon = 9.8\times 10^4p^{7/6}L^{-5/6}. 
\end{equation}
\citet{serio81} extended these scaling laws to loops exceeding the pressure scale height $s_p$, whereby, however,
the limiting height at which the loops grow unstable is $(2-3)s_p$:
\begin{equation}
T_{\rm a} = 1400(pL)^{1/3}e^{-0.04L(2/s_H + 1/s_p)}; \quad\quad \epsilon = 10^5p^{7/6}L^{-5/6}e^{0.5L(1/s_H - 1/s_p)}.
\end{equation}
Here, $s_H$ is the heat deposition scale height. For loops with 
an area expansion factor  $\Gamma > 1$, \citet{vesecky79}  found numerical solutions that approximately follow the scaling 
laws \citep{schrijver89} 
\begin{equation}
T_{\rm a} \approx 1400\Gamma^{-0.1}(pL)^{1/3}; \quad\quad T_{\rm a} = 60\Gamma^{-0.1}L^{4/7}\epsilon^{2/7}.
\end{equation}
There are serious disagreements between some {\it solar}-loop observations and the RTV formalism so 
long as simplified quasi-static heating laws are assumed, the loops being more isothermal than
predicted by the models. There is, however, only limited understanding 
of possible remedies, such as heating that is strongly concentrated at the loop footpoints,
or dynamical processes in the loops (see, for example, a summary of this debate in
\citealt{schrijver02}).

\subsection{Coronal structure from loop models}\label{loopmodels}

When we interpret stellar coronal spectra, we assume, to first order, that some physical
loop parameters map on our measured quantities, such as temperature and EM (and possibly
density), in a straightforward way. In the simplest approach, we assume that the
observed luminosity $L_X$ is produced  by
an ensemble of identical coronal loops with characteristic half-length $L$, surface filling
factor $f$, and an apex temperature $T$ used for the entire loop; then, on using (\ref{RTV}) 
and identifying $L_X =  \epsilon V$, we obtain
\begin{equation}
L \approx 6\times 10^{16} \left({R_*\over R_{\odot}}\right)^2 {f\over L_X}T^{3.5}\quad {\rm [cm]}.
\end{equation}
This relation can only hold if $L$ is smaller than the pressure scale height. As an example, for
an active solar analog ($R = R_{\odot}$, $L_X = 10^{30}$~erg~s$^{-1}$, $T = 10^7$~K) 
we obtain $L \approx 2\times 10^{11}f$~cm. The coronal volume is approximately $V \approx 8R^2fL$
and the electron density $n_e = (EM/V)^{1/2}$. Further, $L_X \approx 2\times 10^{-23}$~EM~erg~s$^{-1}$.
For the solar analog, thus, we find $n_e f \approx 2.5\times 10^9$~cm$^{-3}$.
The luminous, hot plasma component in magnetically active stars therefore seems to
invariably require either very large, moderate-pressure loops with a large filling factor,
or solar-sized high-pressure compact loops with a very small ($\la 1$\%) filling factor.

While the above interpretational work identifies spect\-ral-fit parameters such as $T$
or EMs with parameters  of theoretical loop models, a physically more appealing
approach involves full hydrostatic models whose calculated emission spectra are directly
fitted to the observations.

Such studies \citep{stern86, giampapa96, maggio97, ventura98} have found that the cooler
component at $\approx 1-2$~MK requires loops of small length ($L \ll R_*$) but high pressure
($p>p_{\odot}$), whereas the high-$T$ component at $\approx 5-10$~MK must be confined 
by very compact loops with  extremely high base pressures 
(up to hundreds of dynes cm$^{-2}$) and small  ($<1$\%) filling factors. 
These parameters are suspiciously ``flare-like'' - the observed hot plasma is perhaps 
indeed related to multiple, very compact flaring regions. 

The most essential conclusion from these exercises is perhaps  that,
within the framework of such simplistic models, the loop heating rate required for
magnetically active stars may exceed values for typical solar loops 
by  orders of magnitude, pointing toward some enhanced
heating process reminiscent of the energy deposition in flares. 
The compactness of the hot loops and the consequent high pressures also set these coronal
structures clearly apart from any non-flaring solar coronal features.

\subsection{Coronal structure from densities}\label{densityopacity}

Spectroscopically measured densities provide, in conjunction with EMs, important
estimates of emitting volumes. If the trend suggested from density-sensitive
line flux ratios holds, namely that for increasing temperature, the pressures become
progressively higher, then  progressively smaller volumes are a consequence.  The volume
required for a luminosity of $10^{30}$~erg~s$^{-1}$ at 10~MK is $V = L_X/(2\times 10^{-23}n_e^2)$~cm$^{-3}$
= $ 5\times 10^{26}$~cm$^{-3}$ (where the coefficient $2\times 10^ {-23}$ is from (\ref{lxEM}), 
appropriate for $T = 10$~MK), corresponding to a layer of only 80~m height around a solar-like star,
or 8~km for a filling factor of only 1\%! Such scales are much smaller than chromospheric scale height 
and therefore problematic. Still smaller filling factors must be assumed for a star of this kind.
Similarly, from the RTV loop scaling law (\ref{RTV}) - if applicable -,
a loop height $ h = 2L/\pi = 8.5\times 10^5 T^2/n_e \approx 80$~km is found, again an unreasonably
small size.  

The confinement of plasma at such  high densities in compact sources 
would also require coronal magnetic field strengths of order $B$ $>$ $(16\pi n_ekT)^{1/2}$ $\approx$~1~kG,
i.e., field strengths like those very close and just above (sun-)spots.
In that case, the typical magnetic dissipation time is
only a few seconds for $n_e \approx 10^{13}$~cm$^{-3}$ if the energy is derived 
from the same magnetic fields, suggesting that the small, bright loops light up 
only briefly. In other words, the stellar corona would be made up of numerous 
ephemeral loop sources that  cannot be treated as being in a quasi-static equilibrium
\citep{vdoord97}.

\subsection{X-ray coronal imaging: Overview}

X-ray images of stellar coronae have been derived from eclipses in binaries,
or from rotational modulation in rapidly rotating stars. We keep in mind that
any indirect imaging of this kind is highly biased by observational constraints 
(e.g., the volume that is subject to eclipses or self-eclipses, or the accessible temperature range) 
and by the amount and density of plasma trapped in the magnetic fields. X-ray imaging
captures strongly emitting plasma, {\it not} the entire magnetic-field structure.

\begin{figure} 
\centerline{\resizebox{0.75\textwidth}{!}{\includegraphics{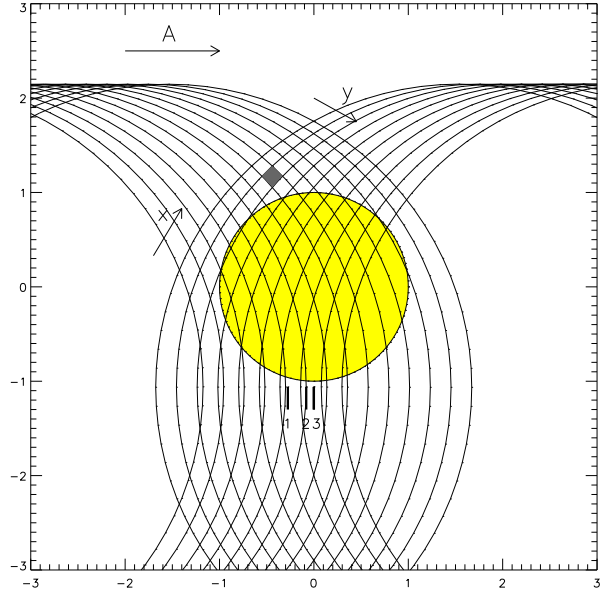}} }
\caption{Sketch showing the  geometry of an eclipsing binary (in this case, $\alpha$ CrB, figure
from \citealt{guedel03b}). The large circles
illustrate the limbs of the eclipsing star that moves from left to right in front of the eclipsed star 
(shown in yellow). The limbs projected at different times during ingress and egress
define a distorted 2-D  array $(x,y)$ of pixels (an example of a pixel is shown in gray).}\label{eclipsegeometry}
\end{figure}

The ``image'' to be reconstructed consists of volume elements at coordinates 
$(x,y,z)$  with {\it optically thin} fluxes $f(x,y,z)$ assumed to be constant in time. 
In the special case of negligible  stellar rotation during the observation, the 
problem can be reduced to a 2-D projection onto the plane of the sky, at the cost
of  positional information along the line of sight (Fig.~\ref{eclipsegeometry}). In general, 
one thus seeks the geometric brightness distribution $f(x,y,z) = f_{ijk}$ ($i,j,k$ being the
discrete number indices of the volume elements) from a binned, observed light curve 
$F_s = F(t_s)$ that undergoes a modulation due to an eclipse or due to rotation.

\subsection{Active-region modeling}\label{activeregions}

In the most basic approach, the emitting X-ray or radio corona can be modeled by
making use of a small 
number of simple, elementary building blocks that are essentially described by their size, 
their brightness, and their location. This approach is the 3-D equivalent
to standard surface spot modeling. 
Preferred building block shapes are radially directed, uniformly bright, optically thin,
radially truncated spherical cones with their apexes at the stellar center. Free parameters 
are their opening angles, their heights above the stellar surface, their radiances, and 
their central latitudes and longitudes. These parameters are then varied until the 
model  fits the observed light curve. 

A minimum solution can be found for a rotationally modulated star \citep{guedel95}.
If a rotationally modulated feature is {\it invisible} during a phase interval $\varphi$ of
the stellar rotation, then all sources contributing to this feature must be confined
to within a maximum volume, $V_{\rm max}$, given by
\begin{equation}\label{minimumvolume}
{V_{\rm max}\over R_*^3} = {\psi\over 3} - {(2\pi - \varphi)(1+{\rm sin}^2i) \over 6{\rm sin}i}
                 +{2{\rm cot}(\chi/2)\over 3{\rm tan}i}
\end{equation}
where ${\rm tan}\chi = {\rm tan}(\varphi/2){\rm cos}i$, 
      ${\rm sin}(\psi/2) = {\rm sin}(\varphi/2){\rm sin}i$ with $0 \le \psi/2\le \pi/2$,
      and $\chi$ and $\varphi/2$ lie in the same quadrant ($i$ is the stellar inclination,
$0 \le i \le \pi/2$). Together with the modulated fraction of the luminosity, lower limits to average 
electron densities in the modulated region follow directly.

\subsection{Maximum-entropy image reconstruction}\label{mem}
  
 Maximum entropy methods (MEM) are applicable both to rotationally modulated light curves
and to eclipse observations. The standard MEM  selects among all images $f_{ijk}$ 
(defined in units of counts per volume element) that are  compatible with the observation, 
the one that minimizes the Kullback contrast (``relative entropy'')
\begin{equation}\label{K}
K = \sum_{i,j,k} f_{ijk} \ln \frac{f_{ijk}}{f^a_{ijk}}
\end{equation}
with respect to an a priori image $f^a_{ijk}$, which is usually unity inside the allowed area or volume
and vanishes where no brightness is admitted. Minimizing $K$ thus introduces the least possible information 
while being compatible with the observation. The contrast $K$ is minimum if $f_{ijk}$ is proportional to 
$f_{ijk}^a$ and thus flat inside the field of view, and it is maximum if the whole flux is 
concentrated in a single pixel $(i,j,k)$. The compatibility with the observed count light 
curve is measured by $\chi^2$,
\begin{equation}
\chi^2 = \sum_s \frac{(F^*_s-F_s)^2}{F^*_s}
\label{chi2}
\end{equation}
where $F_s$ and $F^*_s$ are, respectively, the observed number of counts and the number of 
counts predicted from  $f_{ijk}$ and the eclipse geometry. Poisson statistics usually requires 
more than 15 counts per bin. Finally, normalization is enforced by means of the constraint
\begin{equation}
N = \frac{ \left( {\displaystyle f^{\rm tot} - \sum_{ijk} f_{ijk}}\right)^2  }{f^{\rm tot}}
\label{N}
\end{equation}
where $f^{\rm tot}$ is the sum of all fluxes in the model. 

The final algorithm minimizes the cost function
\begin{equation}
C = \chi^2 + \xi K + \eta N\, .
\label{F}
\end{equation}
The trade-off between the compatibility with the observation, normalization, and unbiasedness
is determined by the Lagrange multipliers $\xi$ and $\eta$ such that the reduced $\chi^2$ 
is $\la 1$, and normalization holds within a few percent. 

\subsection{Lucy/Withbroe image reconstruction}\label{lucy}

This method (after \citealt{lucy74} and \citealt{withbroe75} )1iteratively adjusts fluxes in a 
given set of volume elements based on the mismatch between the model and the observed light curves 
in all time bins to which the volume element contributes.  At any given time $t_s$ during the eclipse, 
the observed  flux $F_s$ is the sum of the fluxes $f_{ijk}$ from all volume elements that are unocculted:
\begin{equation}\label{withbroe1}
F(t_s) = {\displaystyle \sum_{i,j,k}} f_{ijk}\mathbf{m}_s(i,j,k)
\end{equation}
where $\mathbf{m}_s$ is the ``occultation matrix'' for the time $t_s$:  it puts, for any given time
$t_s$, a weight of unity to all visible volume elements and zero to all invisible elements 
(and intermediate values for partially occulted elements). Since $F_s$ is given, one  needs to 
solve (\ref{withbroe1}) for the flux distribution, which is done iteratively as follows:
\begin{equation}\label{withbroe2}
f_{ijk}^{n+1}= 
    f_{ijk}^{n}\frac{{\displaystyle \sum_s \frac{F_o(t_s)}{F^n_m(t_s)}}\mathbf{m}_s(i,j,k)}
                       {{\displaystyle \sum_s \mathbf{m}_s(i,j,k)}}
\end{equation}
where $F_o(t_s)$ and $F^n_m(t_s)$ are, respectively, the observed flux and the model flux (or counts)
in the bin at time $t_s$, both for the iteration step $n$. Initially, 
a plausible, smooth distribution of flux is assumed, e.g., constant brightness, or
some $r^{-p}$ radial dependence. 

\subsection{Backprojection and Clean image reconstruction}\label{clean}
 
If rotation can be neglected during an eclipse, for example in long-period detached binaries, then the 
limb of the eclipsing star is projected at regular time intervals onto the plane of the sky and therefore 
onto a specific part of the eclipsed corona, first during ingress, later during
egress \citep{guedel03b}. The two limb sets define a 2-D grid of distorted, curved pixels (Fig.~\ref{eclipsegeometry}). The brightness 
decrement during ingress or, respectively, the brightness increment during egress within a time step $[t_s, t_{s+1}]$ originates from 
within a region confined by the two respective limb projections at $t_s$ and $t_{s+1}$. Ingress and 
egress thus each define a 1-D image by backprojection from the light curve gradients onto the plane 
of the sky. The relevant reconstruction problem from multiple geometric projections is known in 
tomography. The limiting case of only two independent projections can be augmented by a CLEAN step, as follows. 
The pixel with the largest {\it sum} of projected fluxes  from ingress and egress is assumed to represent 
the location of a real source. A fraction, $g < 1$, of this source flux  is then subtracted 
from the two projections and saved on a clean map, and the process is iterated until all flux is 
transferred onto the latter.

\subsection{X-ray coronal structure inferred from eclipses}\label{imageeclipse}

\subsubsection{Extent of eclipsed coronal features} 
 
Some shallow X-ray eclipses in tidally interacting binary systems of the RS CVn, Algol, 
or BY Dra type have provided important information
on extended coronal structure. For example, Walter et al. (1983) concluded
that the coronae in the AR Lac binary components  are bi-modal in size, 
consisting of compact, high-pressure (i.e., 50--100 dynes cm$^{-2}$)
active regions with a scale height $<R_*$, while the subgiant K star is additionally surrounded 
by an extended (2.7$R_*$) low-pressure corona. 
Further, a hot component pervading the entire binary system was implied from the 
absence of an eclipse in the hard ME detector on {\it EXOSAT} \citep{white90}, and similar
conclusions have been drawn from detailed light-curve inversion analysis \citep{pres95}.

X-ray dips or any periodic modulation have often been absent in X-ray observations of
binaries. This again has been taken as evidence of a very extended (about 1$R_*$ in the case of 
Algol, \citealt{white86}) X-ray corona unless more compact structures sit at high latitudes where they
remain uneclipsed.

\begin{figure}[t!] 
\centerline{\resizebox{0.72\textwidth}{!}{\includegraphics{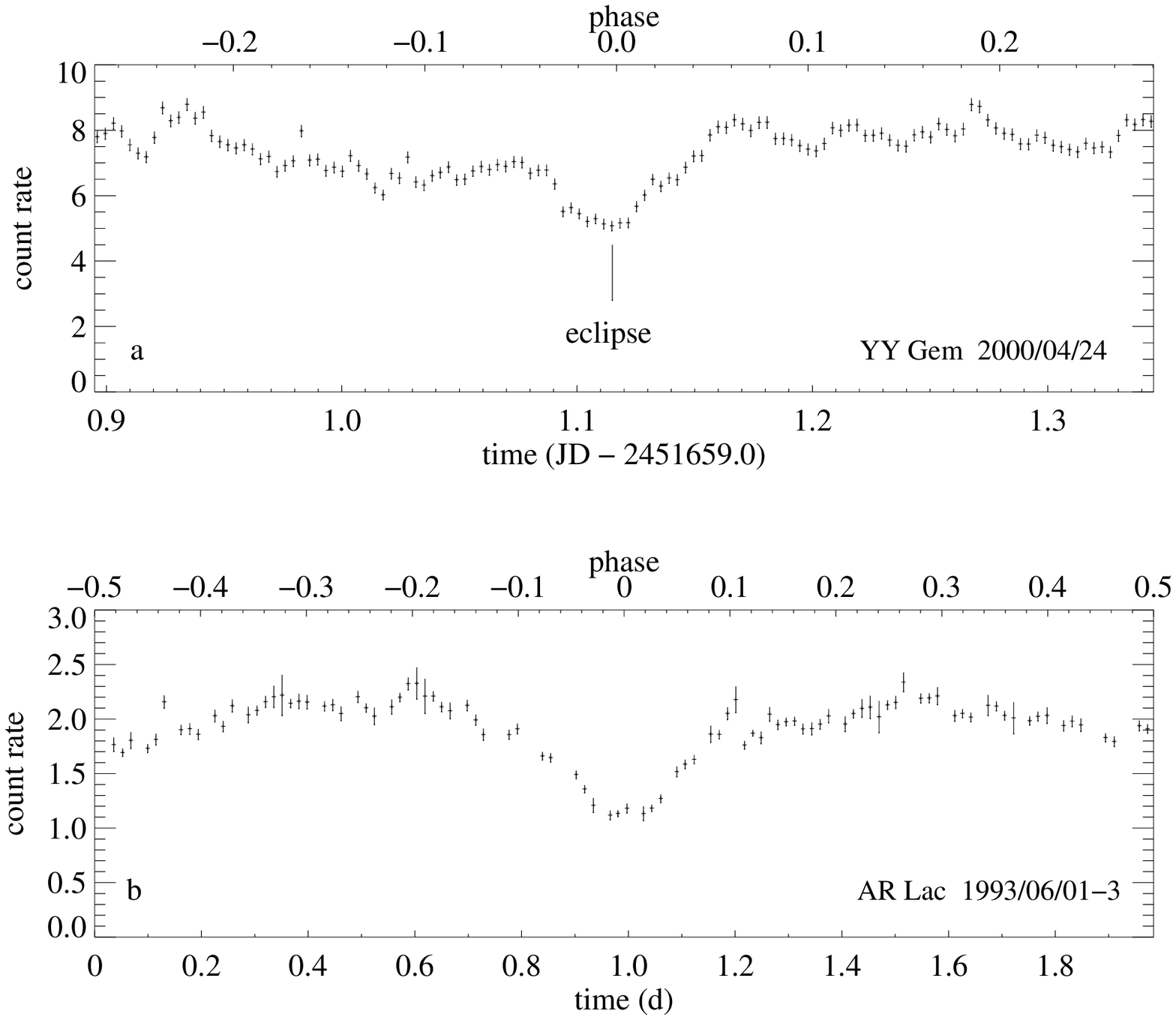}}}
\center{
\hbox{
\vbox{
\resizebox{0.5\textwidth}{!}{\includegraphics{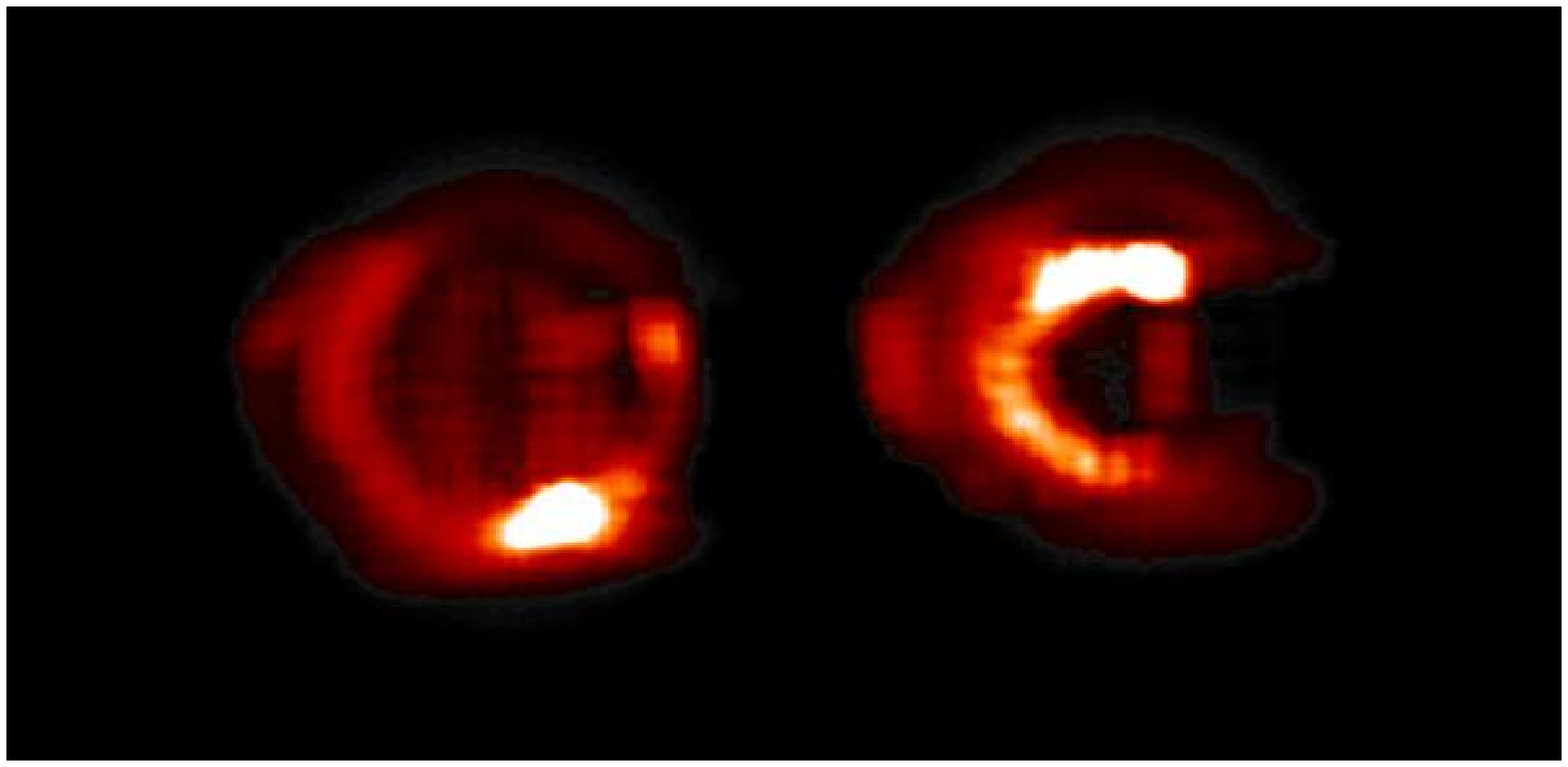}}\\
\resizebox{0.5\textwidth}{!}{\includegraphics{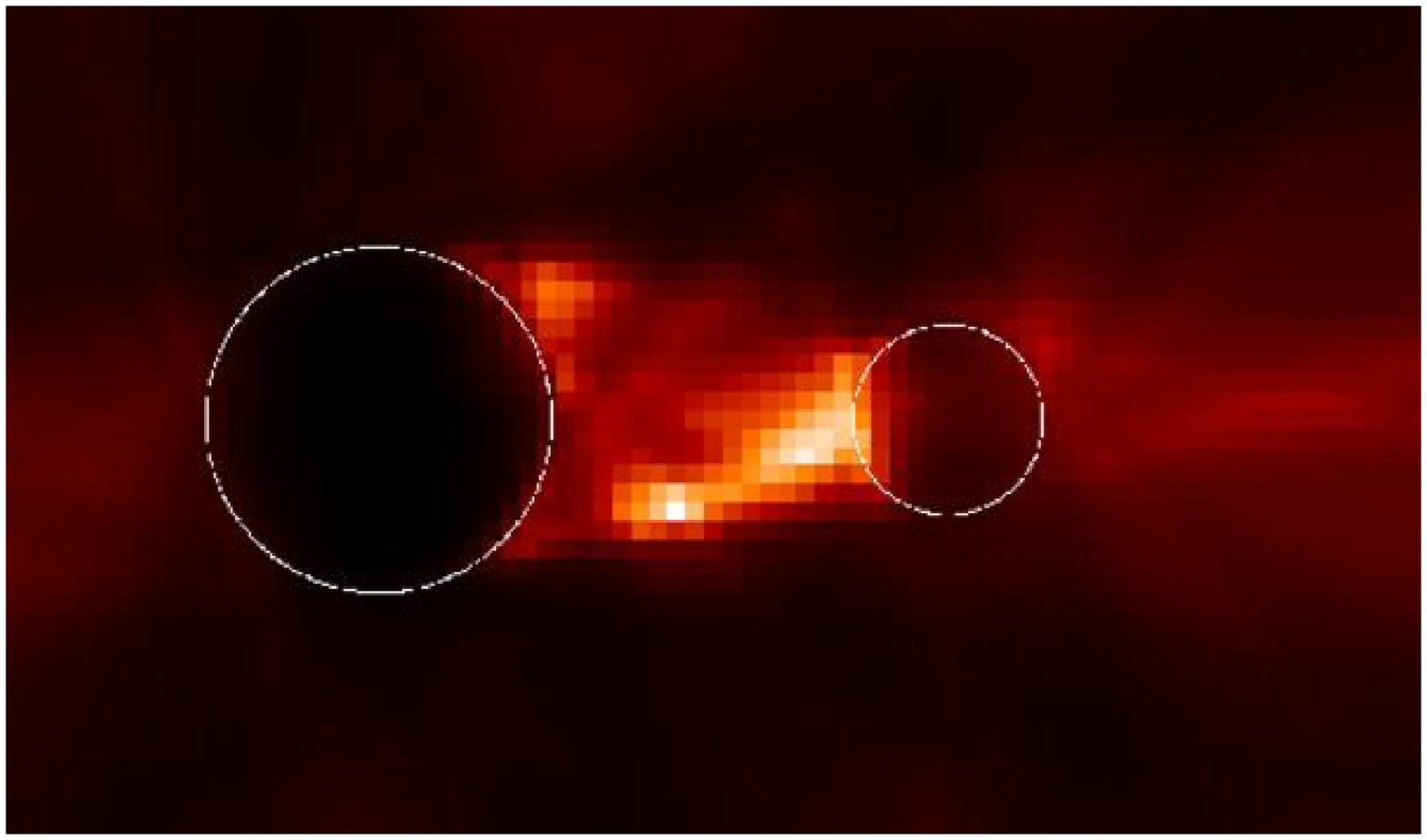}}
}
}
}
\vskip 0.2truecm\caption{Two examples of eclipses and the corresponding coronal image reconstructions. 
{\it From top to bottom:} Light curve of the YY Gem system (from \citealt{guedel01a}, observation with
{\it XMM-Newton} EPIC);
light curve of the AR Lac system (after \citealt{siarkowski96}, observation with {\it ASCA} SIS);
reconstructed image of the coronal structure of, respectively, YY Gem (at phase 0.375) and AR Lac (at quadrature).
The latter figure shows a solution with intrabinary emission. 
(The light curve of AR Lac is phase-folded; the actual observation started around phase 0; data and image
for AR Lac courtesy of M. Siarkowski.)
}\label{eclipsefig}
\end{figure}

\begin{figure}[t!] 
\centerline{
\hbox{
\resizebox{0.55\textwidth}{!}{\includegraphics{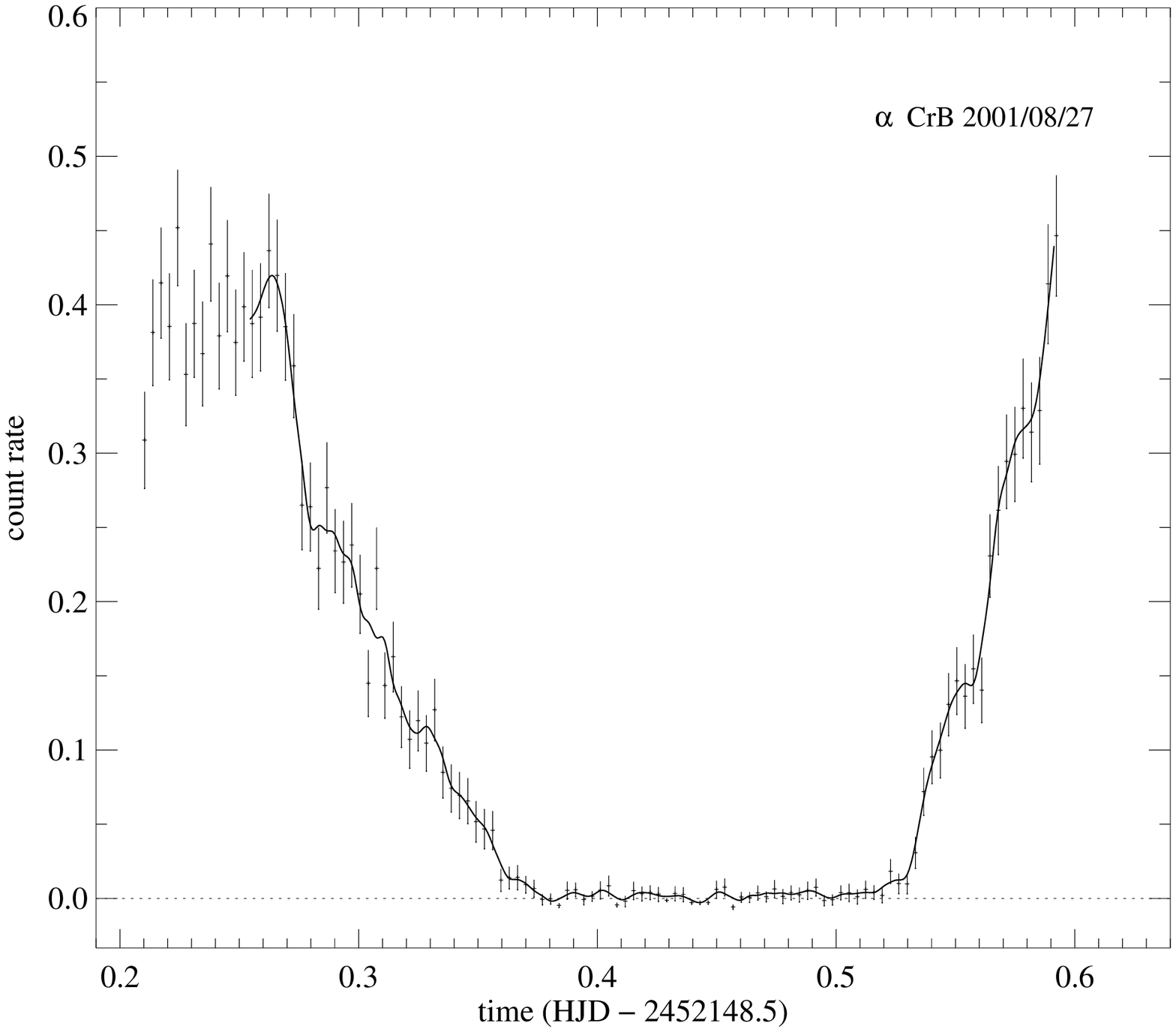}}
\resizebox{0.45\textwidth}{!}{\includegraphics{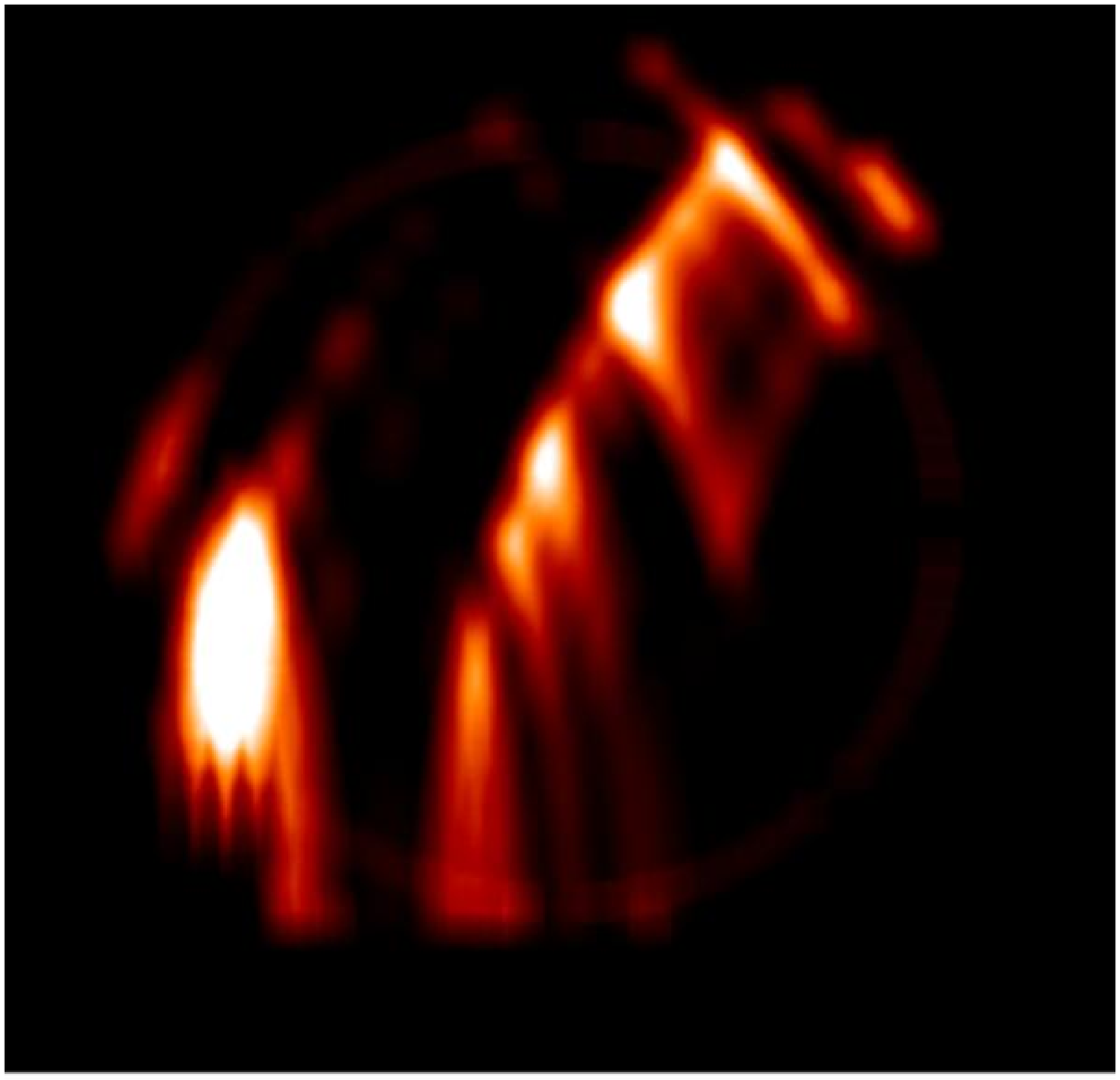}}
}
}
\vskip 0.2truecm\caption{Light curve and image reconstruction of the A+G binary
$\alpha$ CrB. The left panel shows the light curve from observations with {\it XMM-Newton}, 
the right panel illustrates the reconstructed X-ray brightness distribution  on the G star (after 
\citealt{guedel03b}).}\label{eclipsefig2}
\end{figure}

Among wide, non-interacting eclipsing stars, $\alpha$ CrB provides a particularly well-suited
example because its X-ray active, young solar analog (G5~V) is totally eclipsed every 17 days
by the optical primary, an A0~V star that is perfectly X-ray dark. Other parameters are
ideal as well, such as the non-central eclipse, the eclipse time-scale of a few hours, and the relatively
slow rotation period of the secondary. Image reconstructions from eclipse observations  
\citep{guedel03b} reveal patches of active 
regions across the face of  the G star; not much material is found significantly beyond its 
limb (Fig.~\ref{eclipsefig2}). The structures tend to be of modest size ($\approx 5\times 10^9$~cm), 
with large, X-ray faint areas in between, although the star's luminosity exceeds that of the active 
Sun by a factor of $\approx$30.  These observations imply moderately high densities in the 
emitting active regions, reaching a few $10^{10}$~cm$^{-3}$ in the brightest active regions.

\subsection{Eclipsed X-ray flares}\label{eclipseflare}

Eclipses of flares have contributed very valuable 
information on densities and the geometric size of flaring structures.  Only few reports
are available, among them the following:

\citet{choi98} described a full eclipse of an X-ray flare in progress in
the contact binary system VW Cep. During a narrow dip in the flare decay,
the X-ray flux returned essentially to the pre-flare level. Geometric
considerations then placed the flare near one of the poles of the primary star, with a size
scale of order $5.5\times 10^{10}$~cm or somewhat smaller than the secondary star. 
The authors consequently inferred an electron density of $5\times 10^{10}$~cm$^{-3}$. 
A polar location was also advocated for a flare on Algol observed across an eclipse by 
\citet{schmitt99}. The  flare emission was again eclipsed completely, and judged from the known system geometry, 
the flare was located above one of the poles, with a maximum source height of no more than approximately 
0.5$R_*$, implying a minimum electron density of $9.4\times 10^{10}$~cm$^{-3}$ if the volume filling 
factor was unity. 
A more moderate flare was observed during an eclipse in the Algol system by \citet{schmitt03}. 
In this case, the image reconstruction required an equatorial location, with a compact flare source of
height $h \approx 0.1R_*$. Most of the  source volume exceeded densities of
$10^{11}$~cm$^{-3}$, with the highest values at $\approx 2\times 10^{11}$~cm$^{-3}$. Because
the {\it quiescent} flux level was attained throughout the flare eclipse, the authors
argued that its source, in turn, must be concentrated near the polar region with a modest filling
factor of $f < 0.1$ and  electron densities of $\approx 3\times 10^{10}$~cm$^{-3}$.

\subsection{Radio Very Long Baseline Interferometry}\label{vlbi}

By interferometrically combining radio telescopes over large distances, angular resolutions of
as little as one milliarcsecond can be achieved in the microwave range.

VLBI techniques have been very demanding for single late-type dwarf stars, owing both
to low flux levels and small coronal sizes. Some  observations
with mas angular resolutions show unresolved quiescent or flaring sources, thus 
constraining the brightness temperature to  $T_b > 10^{10}$~K (e.g., \citealt{benz95}), whereas others
show evidence for extended coronae with coronal sizes up to several times the 
stellar size.

\begin{figure}[t!]
\hbox{
\resizebox{0.46\textwidth}{!}{\includegraphics{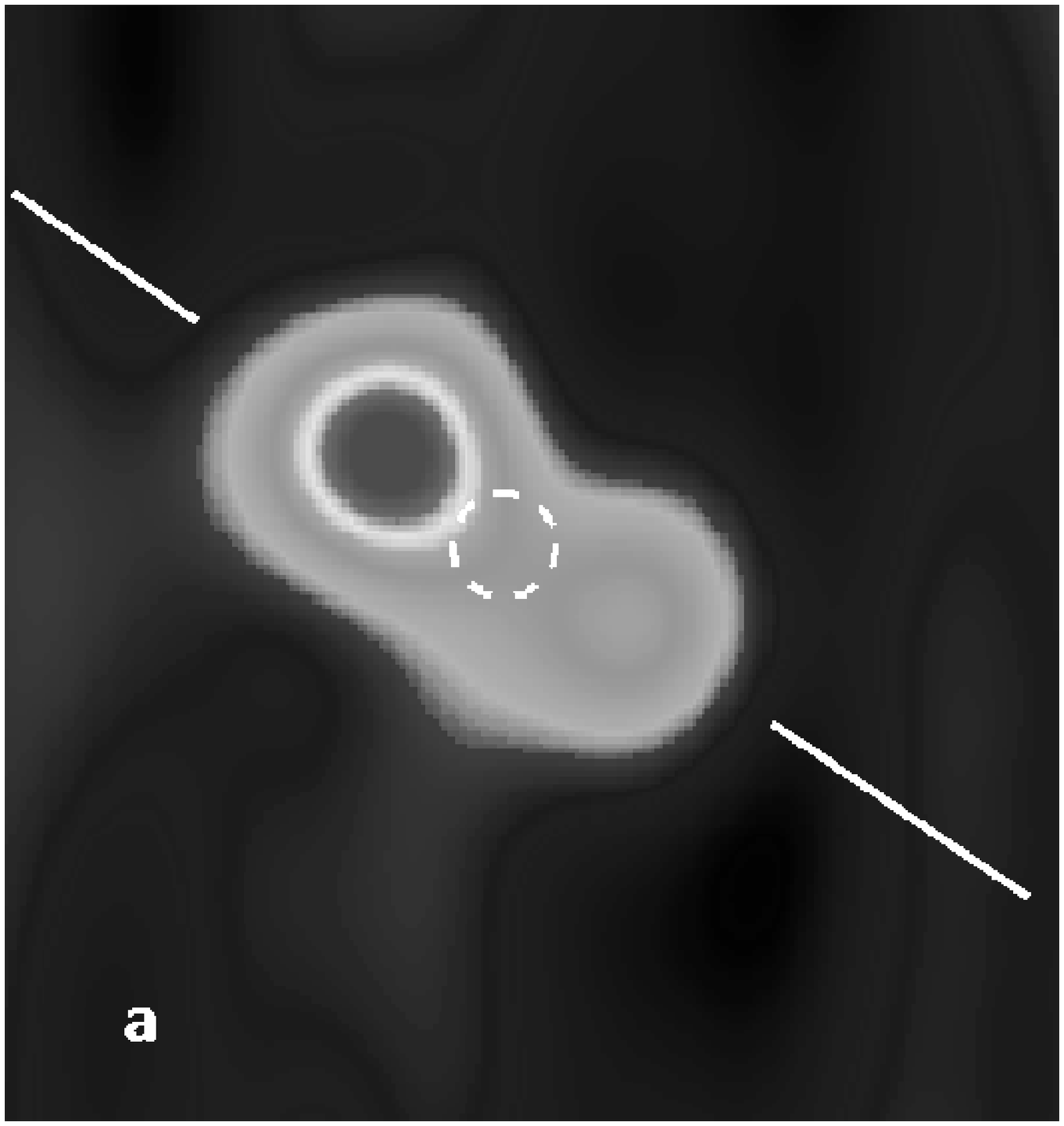}}
\hskip 0.40truecm\resizebox{0.477\textwidth}{!}{\includegraphics{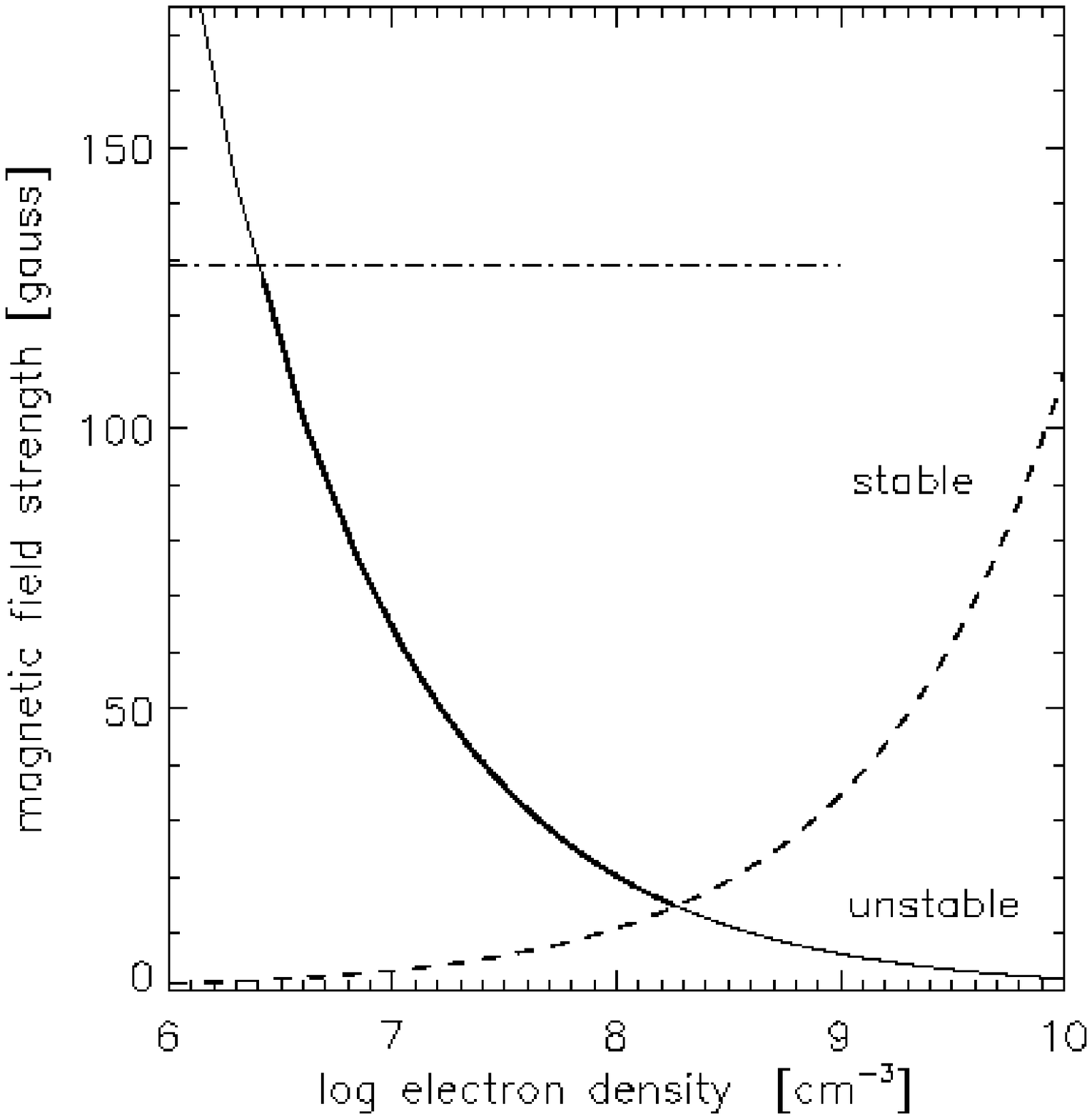}}
}
\caption{{\it (a)} VLBA image of the dMe star UV Cet; the two radio lobes
are separated by about 1.4~mas, while the best angular resolution reaches 
0.7~mas. The straight line shows the orientation of  the putative rotation 
axis, assumed to be parallel to the axis of the orbit of UV Cet around the 
nearby Gl 65~A. The small circle gives the photospheric diameter to size,
although the precise position is unknown (after Benz et al 1998). {\it (b)}
Estimates of the magnetic field and the nonthermal electron density for the same
observation (see text for details).\label{vlba} }
\end{figure}

The dMe star UV Cet was found to be surrounded by a pair of giant synchrotron lobes, with sizes up to 
$2.4\times 10^{10}$~cm and a separation of 4$-$5 stellar radii along the 
putative rotation axis of the star, suggesting very extended magnetic structures above the 
magnetic poles (Fig.~\ref{vlba}a), perhaps arranged in a global dipole as sketched 
in Fig.~\ref{models}b. I discuss this observation is some
detail, following \citet{benz98}, to demonstrate the procedures with which we can characterize 
the magnetic field structure. Throughout, we assume, as detailed further above, that the 
emission is optically thin gyrosynchrotron emission from a power-law population of accelerated electrons with
a number density distribution in energy $\epsilon$
\begin{equation}\label{powerlawradio}
n(\epsilon) = N(\delta-1)\epsilon_0^{\delta-1}\epsilon^{-\delta} 
\quad \mathrm{[cm^{-3}erg^{-1}]}
\end{equation}
where $\epsilon = (\gamma-1)m_ec^2$ is the kinetic particle energy,
$\gamma$ is the Lorentz factor,  and $\delta > 1$ has been assumed 
so that $N$ is the total non-thermal electron number density 
above $\epsilon_0$. 

We first need an expression for the emissivity:
For isotropic pitch angle electron distributions according to (\ref{powerlawradio}) 
with $2 \la \delta  \la 7$, 
for harmonics $10 \la s \la 100$, and for the x-mode \citep{dulk85}
\begin{equation}
\eta_{\nu}     \approx 3.3\times 10^{-24}BN~10^{-0.52\delta}(\sin\theta)^{-0.43+0.65\delta}\left({\nu\over \nu_c}\right)^{1.22-0.90\delta}
\label{gyrosynchrotron_eta}
\end{equation}
(in erg~s$^{-1}$~cm$^{-3}$~Hz$^{-1}$~sterad$^{-1}$). Here, $\theta$ denotes the emission angle to the magnetic field, and $\nu_c = eB/m_ec$ is the
electron gyrofrequency. The coefficient of this expression is applicable if a power-law cutoff 
has been set at 10~keV. This cut-off itself is of little relevance for gyrosynchrotron emission 
since electrons radiate little at such low energies. 

We assume $\theta = \pi/2$ and $\delta \approx 2.5$ (the latter from modeling of 
M dwarf spectra, Sect.~\ref{gyrosynch}). Adding o- and x-mode (which have similar emissivities),
the intensity is
\begin{equation}\label{intensity}
I \approx 1.8\eta_{\nu} D 
\end{equation}
which is to be compared with the observed intensity of $3.9\times 
10^{-8}$~erg$^{-1}$ cm$^{-2}$ Hz$^{-1}$ sterad$^{-1}$. Here, the fitted diameter
of one of the radio blobs (0.3~milliarcsec or $D = 1.2\times 10^{10}$~cm)  
was used. Combining (\ref{gyrosynchrotron_eta})--(\ref{intensity}), we find
\begin{equation}
B \approx 2\times 10^5 N^{-1/2}.
\end{equation} 
This relation is shown in Fig.~\ref{vlba}b together with the condition that the
particle pressure in the magnetic loops is less than the magnetic pressure,
\begin{equation}
N\bar{\epsilon} \le {B^2\over 8\pi}.
\end{equation}
We find a lower limit to $B$ of approximately 15~G and an upper limit to
$N$ of about $2\times 10^8$~cm$^{-3}$.

At least part of the observed emission was slowly decaying, with a 
decay timescale of $\tau = 6650$~s
 (perhaps from a flare
that filled the magnetospheric volume with electrons). Given the large size
of the structure, the dominant energy decay process is likely to be due to
synchrotron radiation loss (\ref{synchloss}).
The average frequency of synchrotron emission can be expressed as
\begin{equation}
\bar{\nu} = 1.3\times 10^{6}B\gamma^2\quad {\rm [Hz]} 
\end{equation}
\citep{melrose80}. Using the observing frequency (8.4~GHz in this case) for
$\bar{\nu}$, we find the decay time
\begin{equation}
\tau = {\gamma\over \dot{\gamma}} \approx 8\times 10^{6}B^{-3/2}\quad {\rm [s]} 
\end{equation}
and therefore $B = 113$~G. This defines
an upper limit to the magnetic field strength because we have ignored other energy losses that
might be present. The upper limit is also drawn in Fig.~\ref{vlba}b. We have thus 
confined the magnetic field strength in the source to 15--130~G.

\begin{figure}[t!]
\centerline{\resizebox{0.46\textwidth}{!}{\includegraphics{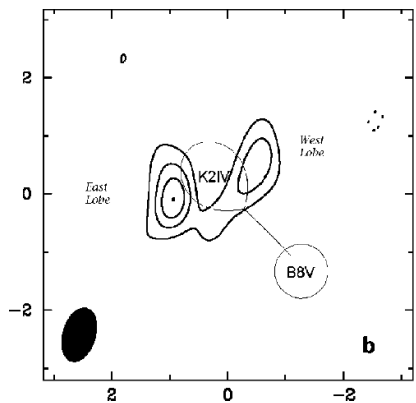}}}
\caption{VLBA observation of Algol at 8.4~GHz, resolving two lobes around the binary. The most
likely configuration of the binary components is also drawn (after \citealt{mutel98}).
\label{vlbaalgol} }
\end{figure}

VLBA imaging and polarimetry of Algol reveals a similar picture with two oppositely 
polarized radio lobes separated along a line perpendicular to the 
orbital plane by more than the diameter of the K star (\citealt{mutel98}, 
Fig.~\ref{vlbaalgol} and Fig.~\ref{models}b). Large-scale  polarization structure further supports models that
assume globally organized magnetic fields around active binary stars \citep{beasley00}.

An important, early VLBI result for RS CVn and Algol-like binaries is evidence  
for a compact core plus extended halo radio structure of a total size that is comparable
to the binary system size  \citep{mutel85}. The basic idea here is the following:
During quiescence, an optically thin, very large magnetosphere is filled with a power-law
distribution of electrons. The emission is essentially optically thin, with a flat radio spectrum. 
During an outburst, an active region injects a larger electron population into magnetic loops. 
The compact source (the ``core''), unresolved by VLBI, becomes optically thick due to synchrotron
self-absorption, and the radio spectral index consequently becomes positive, and the brightness
temperature is thus equal to the effective electron temperature. \citet{mutel85} measured
a few times $10^{10}$~K for outbursts on RS CVn-type binaries. Once the electron injection
has ceased, the lifetime of the energetic electron population is determined essentially by 
the synchrotron loss time. The final phase consists of magnetic-loop expansion owing to
buoyancy, expanding the source until it merges with the pre-flare ``halo'' component. At the same
time, the source becomes progressively more optically thin, developing a negative radio 
spectral index and mild circular polarization. Given the large size, the strength of the 
extended magnetic field  must be rather moderate, of order 10~G. To emit radiation at
frequency $\nu$ in the microwave range, high Lorentz factors are required. Because the
spectral power for synchrotron emission is predominantly emitted at a harmonic of
order $\gamma^3$ of the relativistic gyrofrequency 
\begin{equation}
\nu_c = {eB\over m_e\gamma c} = 2.8\times 10^6 {B\over \gamma},
\end{equation}
we require $\gamma \approx  6\times 10^{-4} (\nu/B)^{1/2}$. For an observing frequency of
5~GHz thus, $\gamma \approx 10$. A model calculation of the spectral development of the
microwave radiation was outlined in Sect.~\ref{gyrosynch}.

\subsection{Radio magnetospheric models}\label{magnetosphere}

\begin{figure}[t!]
\hbox{
\resizebox{0.54\textwidth}{!}{\includegraphics{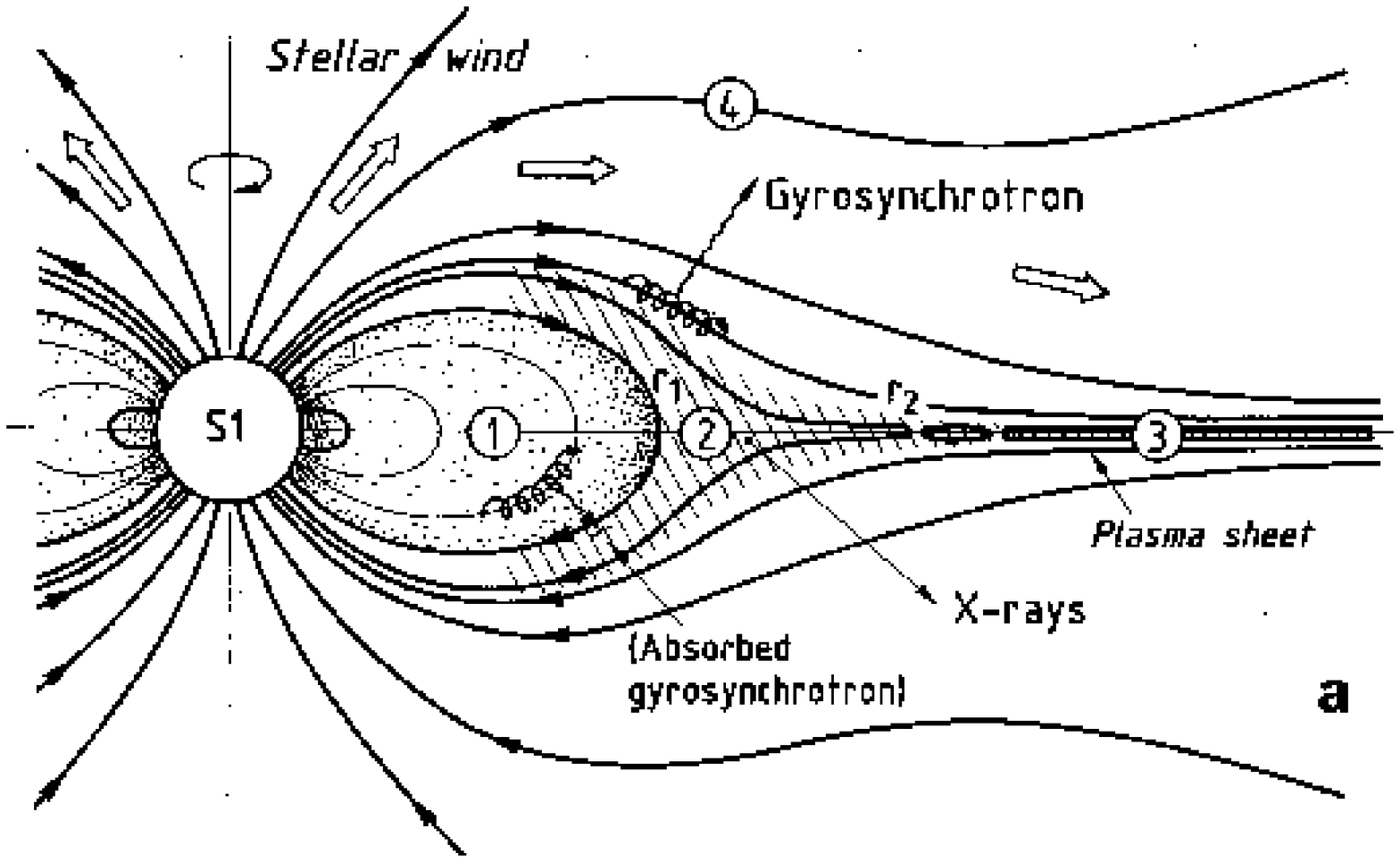}}
\hskip 0.5truecm\resizebox{0.35\textwidth}{!}{\includegraphics{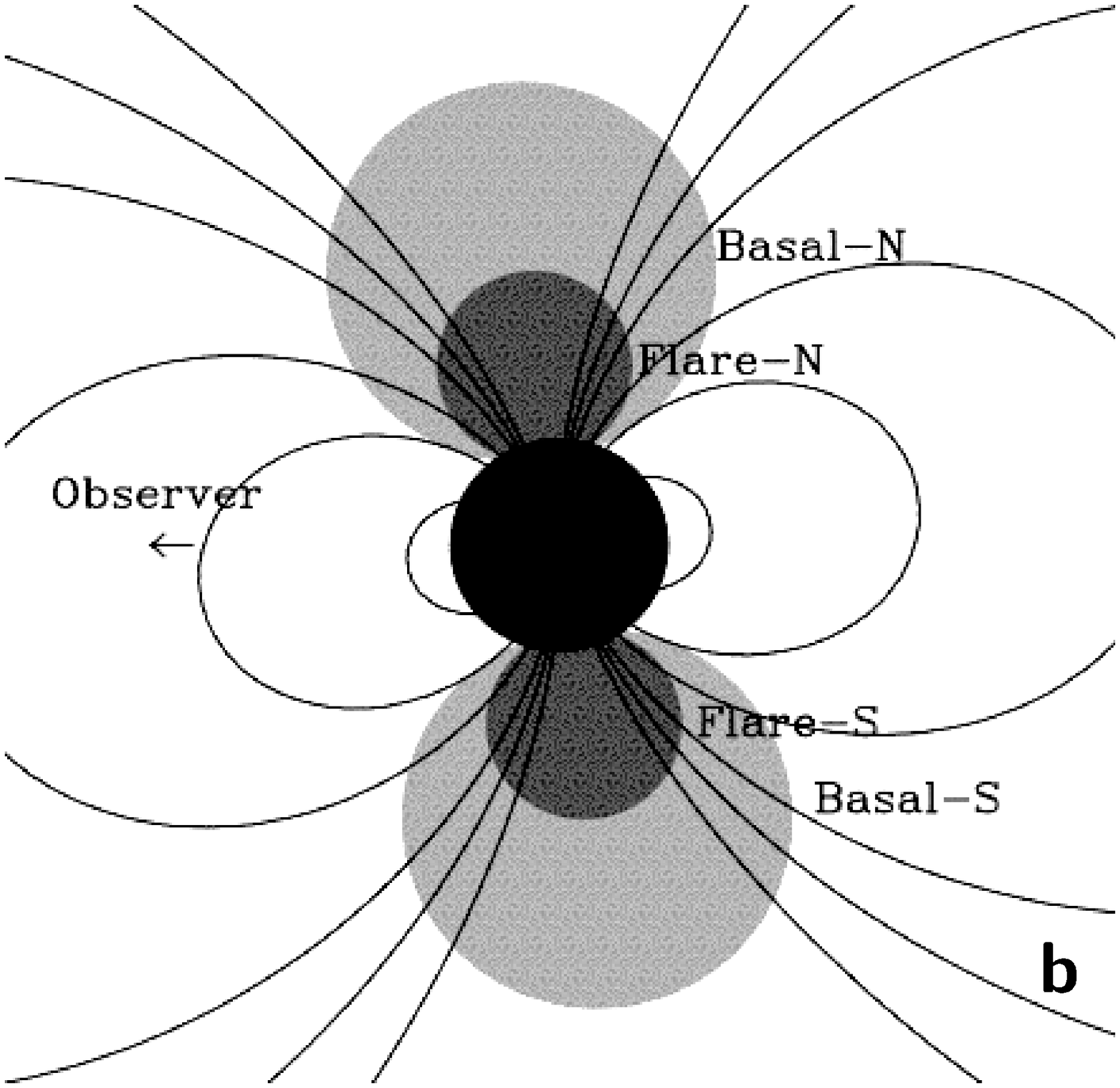}}
}
\caption{{\it (a)} Equatorial model for the magnetosphere of the young B star S1 in $\rho$ Oph
         \citep{andre88}. {\it (b)} Sketch for radio emission from a global dipole
	  consistent with the VLBI observation shown in Fig.~\ref{vlba}b 
	  (\citealt{mutel98}; reproduced with permission of the AAS.)\label{models}}
\end{figure}

VLBI observations of  RS CVn and Algol binaries, T Tau stars, and magnetic Bp/Ap stars
have shown some perplexing structures with sizes at least as large as the binary system,
with polarization properties that suggest that the magnetic fields are
globally ordered. This has led to a series of large-scale magnetic models for such systems.
They have in common  a global, dipole-like structure somewhat
resembling the Earth's Van Allen belts (Fig.~\ref{models}a).  
Stellar winds escaping  along magnetic fields draw the field
lines into a current sheet configuration in the equatorial plane.  
Particles can be accelerated in that region. They subsequently
travel back to become trapped in the dipolar-like, equatorial magnetospheric 
cavity. Variants of this radiation-belt model, partly based on theoretical work of 
\citet{havnes84}, have been applied to RS  CVn binaries (e.g.,  \citealt{morris90}),
in an optically thick version to Bp/Ap stars \citep{drake87, 
linsky92} and in an optically thin version to a young B star  \citep{andre88}.

Polarization observations of RS CVn binaries support these models. The polarization degree
seems to be anticorrelated with the luminosity for any given system, but the sense of polarization
changes between lower and higher frequencies. For the entire binary sample,  the polarization
degree is inversely 
correlated with the stellar inclination angle such that low-inclination (``pole-on'') 
systems show the strongest polarization degrees, as expected from such global systems
 \citep{mutel87, mutel98, morris90}.

Flares still appear to originate in compact sources within such magnetospheres,
probably close to the star. The   ``core plus halo'' model then
correctly describes the radio spectral properties - the halo corresponds to the extended
magnetosphere. This is also supported by magnetic fields inferred to be stronger (80$-$200~G) 
in the flaring core  and weaker (10$-$30~G) in the halo 
\citep{umana93, umana99, mutel98, trigilio01}.  
Because the optical depth is  frequency-dependent, the source is small at high frequencies
(with a size $\approx R_*$, above 10~GHz) and large at small frequencies (with a size comparable
to the binary system size at 1.4~GHz; \citealt{klein87b, jones95}). This effect  
explains the relatively flat, optically thick radio spectra seen during flares.

\subsection{Extended or compact coronae?}\label{extendedcompact}

As the previous discussions imply, we are confronted with mixed evidence for predominantly extended 
(source height $>R_*$) and  predominantly compact ($\ll R_*$) coronal structures or a 
mixture thereof, results that variably come from radio or from X-ray astronomical
observations. There does not seem to be unequivocal agreement on the type of structure
that generally prevails. Several trends can be recognized, however, as summarized below. 

{\it Compact coronal structure.} Steep (portions of the) ingress and egress
light curves or prominent rotational modulation unambiguously argue in favor of short 
scale lengths perpendicular to the line of sight. 
Common to all are relatively high inferred densities ($\approx 10^{10}$~cm$^{-3}$). The pressures of
such active regions may exceed pressures of non-flaring solar active regions by up to two orders 
of magnitude.  Spectroscopic observations of high densities and loop modeling
add further evidence for the presence of some rather compact sources. 
Flare modeling also provides modest sizes,
often of order 0.1--1$R_*$, for the involved magnetic loops (Sect.~\ref{flares}).

{\it Extended structure.} Here, the arguments are less direct and are usually
based on the absence of deep eclipses, or very shallow ingress and egress curves.
Caution is in order in
cases where the sources may be located near one of the polar regions; in those cases, eclipses and
rotational modulation may also be absent regardless of the source size. Complementary
information is available from flare analysis (see Sect.~\ref{flares}) that in some cases 
does suggest quite large loops. The caveat here is that simple single-loop models may not
apply to such flares.
Clear evidence is available from radio interferometry that proves the
presence of large-scale, globally ordered magnetic fields. 
The existence of prominent extended, closed magnetic fields on scales $>R_*$ is therefore 
{\it also} beyond doubt for several active stars.

The most likely answer to the question on coronal structure size is 
therefore an equivocal one: Coronal magnetic structures follow a size distribution from
very compact to extended ($\ga R_*$) with various characteristic densities,
temperatures, non-thermal electron densities, and surface locations. Quite different
loop systems may be responsible when measuring cool X-ray lines or hot lines, or when
observing in the microwave range. This
is no different from what we see on the Sun even though various features observed on magnetically
active stars stretch the comparison perhaps rather too far for comfort.

\section{Stellar radio flares}

Flares arise as a consequence of a sudden energy release and relaxation process of the magnetic field in 
solar and stellar coronae. Present-day models assume that the energy is accumulated and
stored in non-potential magnetic fields prior to an instability that most likely implies
reconnection of neighboring antiparallel magnetic fields. The energy is brought
into the corona by turbulent footpoint motions that tangle the field lines at larger heights.
The explosive energy release becomes measurable across the electromagnetic spectrum
and, in the solar case, as high-energy particles in interplanetary space as well. 

Flares are ubiquitous among coronal stars of all types, with very few exceptions. 
They have, of course, prominently figured in solar studies, and it
is once again solar physics that has paved the way to the interpretation of stellar 
flares, even if not all features are fully understood yet. The complexity that flares reveal to
the solar astronomer is inaccessible in stellar flares, especially in the absence of
spatially resolved observations. Simplified concepts, perhaps tested for solar examples, 
must suffice. The following sections summarize the ``stellar astronomer's way'' of looking at flares.

\subsection{Incoherent radio bursts}

Active stars reveal two principal flare types at radio wavelengths, similar to
the solar case:

Incoherent flares evolve on time scales of minutes to hours,
they show broad-band spectra and moderate degrees of polarization. These bursts
are fully equivalent to solar microwave bursts.
Like the latter, they show evidence for the presence of mildly 
relativistic electrons. The emission is therefore interpreted as gyrosynchrotron 
radiation from coronal magnetic fields.
Many flares on single F/G/K stars are of this type, 
as are almost all radio flares on M dwarfs above 5~GHz \citep{bastian90},
or on RS CVn binaries \citep{mutel85}.
      
\subsection{Coherent radio bursts}

The stellar equivalents to coherent solar radio bursts (showing  high brightness
temperature, short durations, perhaps small bandwidth, and perhaps high polarization degree)
have been observed already in the early days of radio astronomy, given their sometimes extremely
high fluxes. Like for the Sun, they come in a bewildering variety which has made a clear identification
of the ultimate cause difficult
\citep{bastian98}. Coherent bursts are frequent on late-type main-sequence stars, but 
have also been reported from  RS CVn binaries 
\citep{white95}.

Such bursts carry important information in high-time resolution light curves. 
Radio ``spike'' rise times as short as 5$-$20~ms have been reported, requiring, from
the light-time argument,  source sizes of $R< c\Delta t \approx  1500$-$6000$~km. With
\begin{equation}\label{rayleigh2}
S = {2kT_b\nu^2\over c^2}{\pi R^2\over d^2} 
\end{equation}
and measured fluxes up to 1~mJy for sources at a distance of a few parsec, we derive brightness
temperatures of  order  $T_b \approx 10^{16}$~K, a clear proof of the 
presence of a coherent mechanism \citep{lang83, lang86, guedel89a, bastianea90}.

\subsection{Radio dynamic spectra}

The standard means to study solar coherent bursts are ``dynamic spectra'' (flux vs. frequency 
and time). If the elementary frequencies relevant for the coherent emission process,
$\nu_p$ and $\nu_c$, evolve in the source, or if the radiating 
source itself  travels across density or magnetic field gradients, the emission  
leaves characteristic traces on the dynamic spectrum
\citep{bastian98}. The study of drifts, decay times, harmonic structures etc. may
then help identify the emission process and thus infer magnetic field strengths, electron
densities, electron energies and beam velocities etc.

Applying the same technology to stellar observations has turned out to be
extremely challenging, and only few successful dynamic spectra have been recorded to
date. A rich phenomenology  has been uncovered, including: a)
short, highly polarized bursts with structures as narrow as $\Delta\nu/\nu = 0.2$\%
suggesting either plasma emission from a source of size $\sim 3\times 10^8$~cm, or a cyclotron 
maser in magnetic fields of $\sim 250$~G \citep{bastian87, guedel89a, bastianea90}.
b) Evidence for spectral structure with positive
drift rates of 2 MHz~s$^{-1}$ around 20~cm wavelength, taken as evidence
for a disturbance propagating ``downward'' in the plasma emission interpretation
\citep{jackson87}; and c) in solar terminology,
rapid broadband pulsations, ``sudden (flux) reductions'', and 
positive and negative drift rates of 250$-$1000~MHz~s$^{-1}$
(\citealt{bastianea90, abada94, abada97}).  

\begin{figure}[t!] 
\centerline{\psfig{file=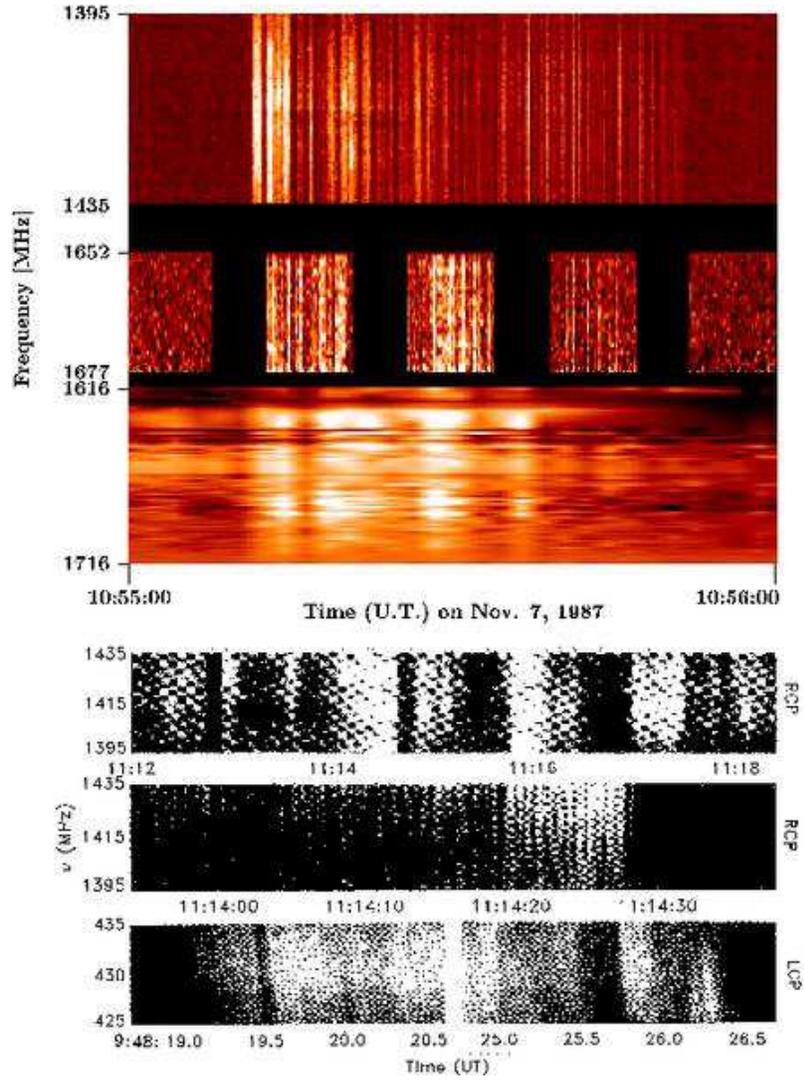,width=10.7truecm}}
\caption{Radio dynamic spectra of M dwarf flares. Upper three panels show
a  flare on  AD Leo, recorded with the Arecibo (top), Effelsberg (middle) and 
Jodrell Bank (bottom) telescopes in different wavelength ranges
(see also \citealt{guedel89a}). Bottom three panels show flares on
AD Leo (top and middle) and YZ CMi (bottom) observed at Arecibo
(after \citealt{bastianea90}, reproduced with permission of the AAS).\label{dyn}}
\end{figure}

Spectral bandwidths as  1\%  of the emitting frequency were found for some  bursts.  
If we conservatively  assume a magnetic scale height of $L_B = 1R_*$ (assuming that 
the emission is gyromagnetic), the source size 
can be estimated to be 
\begin{equation}
r \approx {\Delta\nu\over\nu}L_B
\end{equation}  
which is $\approx$ a few 1000~km for $L_B = R_*$ of an M dwarf, 
again implying very high $T_b$, compatible with the
above light time argument \citep{lang88, guedel89a, bastianea90}.

\begin{figure} 
\hbox{
\hskip -0.8truecm\resizebox{0.55\textwidth}{!}{\includegraphics{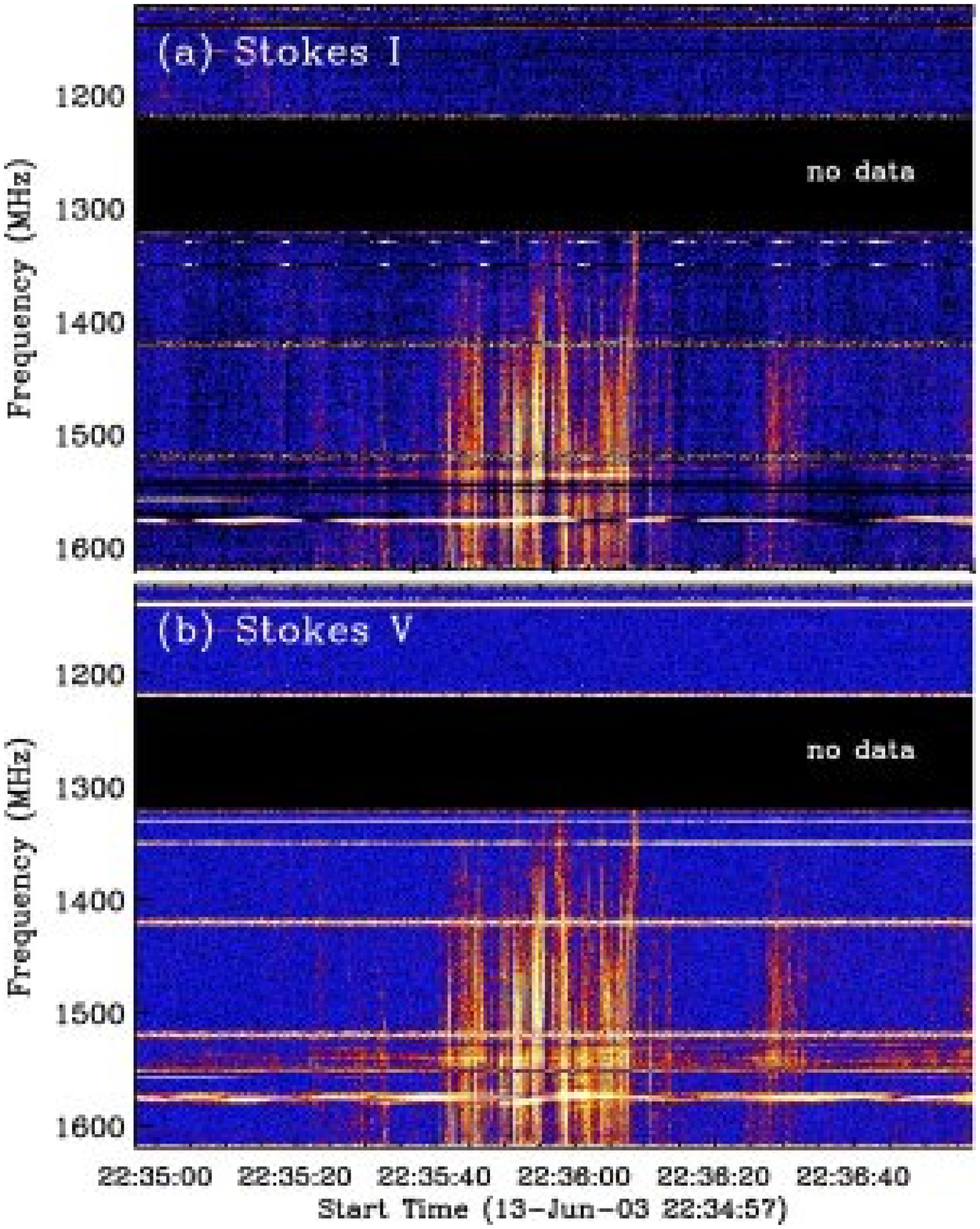}}
\resizebox{0.5\textwidth}{!}{\includegraphics{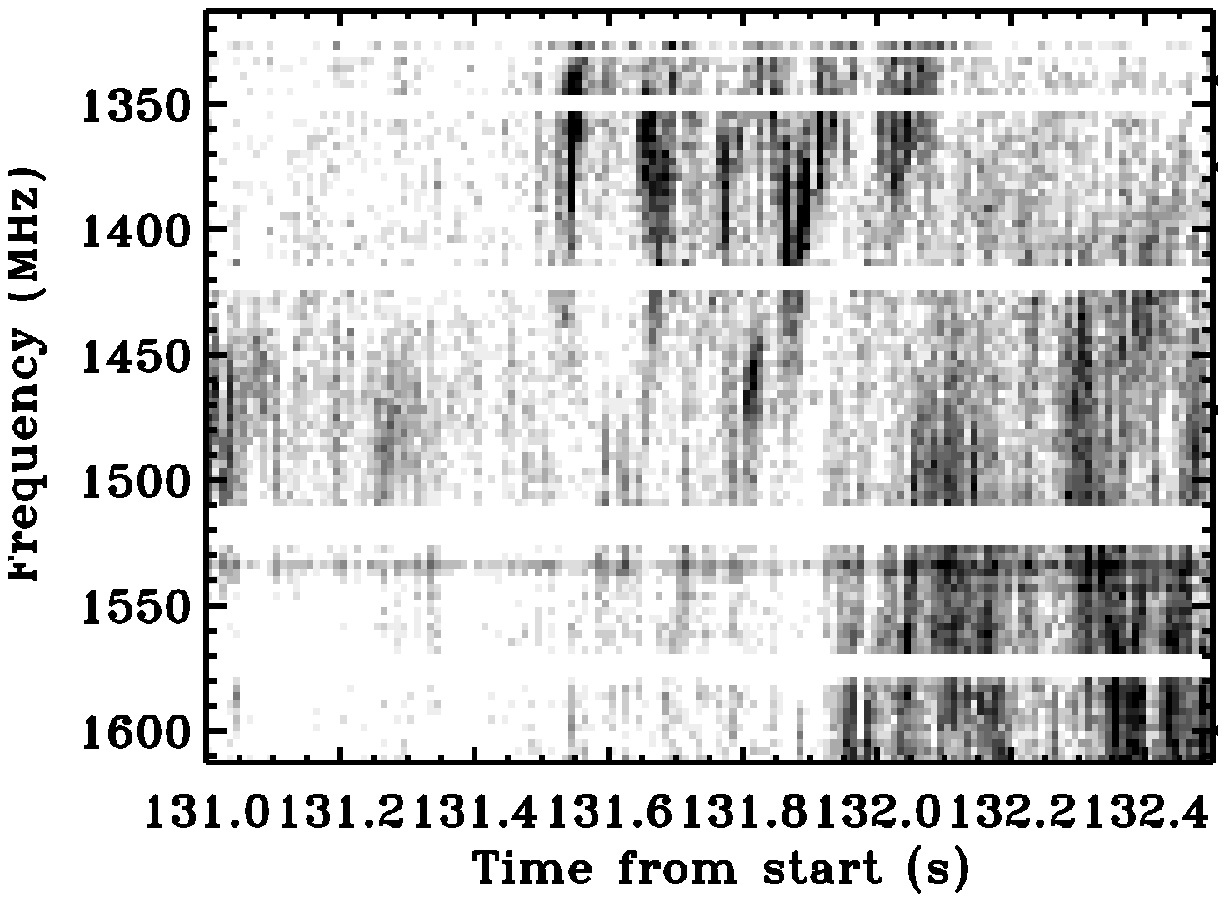}}
}
\vskip -0.3truecm
\caption{Dynamic spectra of radio bursts observed in AD Leo \citep{osten06}.
The left figure shows the entire event, the right figure shows an extract 
revealing frequency drifts and spectral structure.\label{dynosten}}
\end{figure}

Recent developments have permitted recording quite broad regions of the stellar burst
spectrum, allowing for a much better characterization of the burst types. \citet{osten06}
obtained burst spectra from AD Leo recorded in the 1120--1620~MHz range with a time
resolution of 10~ms and a spectral resolution of 0.78~MHz. They find significant frequency-time
drifts in short subbursts (Fig.~\ref{dynosten}). The main characteristics of these
structures are (i) durations of about 30~ms, (ii) high polarization ($> 90$\%), frequency bandwidths
of $\Delta\nu/\nu \approx 5$\%, and inverse drift rates (time interval per drift in frequency)
that are symmetric around zero, with
a characteristic width corresponding to a frequency drift of 2.2~GHz~s$^{-1}$. These characteristics are 
very close to those of narrow-band {\it solar} decimetric ``spike bursts'' but set them
apart from type III bursts associated with electron beams in the solar corona (type III bursts
have considerably longer durations, occupy broader frequency intervals, and are at best
moderately polarized). 

There is one significant difference between the stellar and solar
bursts: the duration of the solar examples is about three times shorter. We can relate this
to Coulomb collisional damping \citep{guedel90, osten06} which is determined by the
electron-ion collision time $\tau_{ei}$, from which we find
\begin{equation}
T = 8100 \nu_{\rm MHz}^{4/3}\tau^{2/3}.
\end{equation}

We now identify $\tau$ 
with the observed characteristic event duration, and $\nu_{\rm MHz}$ with the
radio frequency in MHz. For solar events at 1.5~GHz, $T \approx 5$~MK \citep{guedel90},
but for the longer events on AD Leo, $T \approx 13$~MK \citep{osten06}, which can be explained
by the higher coronal temperature of AD Leo \citep{guedel03a}. Conversely, the
higher temperature of AD Leo's corona increases the gyroresonance  absorption coefficient
for underlying electron cyclotron maser emission significantly:
\begin{equation}
\kappa_{\rm gr} \propto T^{s-1}
\end{equation}
where $s = 2,3$ is the harmonic of the absorbing layer, making the escape of the preferred
harmonics of the maser difficult. This is not the case for plasma radiation for which 
free-free absorption is relevant:
\begin{equation}
\kappa_{\rm ff} \propto T^{-3/2},
\end{equation}
implying that the higher temperature of AD Leo's corona clearly favors a plasma emission
process as long as $\omega_{p} > \Omega_{c}$. The latter condition is easily met as
it requires $B < 500$~G. This is a nice example where a stellar observation  
helps identify the relevant emission mechanism that remains equivocal under solar conditions.

\section{Stellar X-ray flares}\label{flares}

\subsection{Cooling physics}

Flares cool through radiative, conductive, and possibly also
volume expansion processes. We define the flare decay phase as the episode when
the net energy loss by cooling exceeds the energy gain by heating, and
the total thermal energy of the flare plasma  decreases. The thermal energy decay
time scale $\tau_{\rm th}$ is defined as
\begin{equation}\label{decay}
\tau_{\rm th} = {E\over \dot{E}}
\end{equation}
where $E \approx 3n_e kT$  (for $n_e \approx n_{\rm H}$) is the total thermal energy density
in the flaring plasma of electron density $n_e$ and temperature $T$,
and $\dot{E}$  is the volumetric cooling loss rate (in erg~cm$^{-3}$~s$^{-1}$). 
For conduction across temperature gradients in parallel magnetic fields, the
mean loss rate per unit volume is
\begin{equation}\label{condloss}
\dot{E_c} = {1\over L}\kappa_0 T^{5/2}{dT\over ds} \approx {4\over 7L^2}\kappa_0 T^{7/2} 
\end{equation}
where $s$ is the coordinate along the field lines, and the term $\kappa_0 T^{5/2}dT/ds$ 
is the conductive flux in the approximation of \citet{spitzer62}, to be evaluated near the loop 
footpoint where $T$ drops below $10^6$~K, with 
$\kappa_0 \approx 9\times 10^{-7}$~erg~cm$^{-1}$s$^{-1}$K$^{-7/2}$. 
Equations~(\ref{decay}, \ref{condloss}) define the {\it conductive time scale} $\tau_{\rm th} \equiv \tau_c$.
The second equation in (\ref{condloss}) should be used only as an approximation for non-radiating
loops with a constant cross section  down to the loss region
and with uniform heating (or for time-dependent cooling of a constant-pressure loop without
heating; for the factor of 4/7, see \citealt{dowdy85, kopp93}).
We have used $L$ for the characteristic dimension of the source along the
magnetic field lines, for example the half-length of a magnetic loop. 
Strictly speaking, energy is not lost by conduction but is
redistributed within the source; however, we consider energy lost when it is conducted
to a region that is below X-ray emitting temperatures, e.g., the 
transition region/chromosphere at the magnetic loop footpoints.

Radiative losses are by bremsstrahlung (dominant for $T \ga 20$~MK), 2-photon continuum,  
bound-free, and line radiation.  We note that the plasma composition in terms
of element abundances can modify the cooling function $\Lambda(T)$, but the correction 
is of minor importance because stellar flares are usually rather hot. At relevant
temperatures, the dominant radiative losses are by bremsstrahlung, which is little
sensitive to modifications of the heavy-element abundances. The energy loss rate is
\begin{equation}
\dot{E_r} =  n_en_H\Lambda(T)
\label{radloss}
\end{equation}
 For $T \ge 20$~MK,  $\Lambda(T) = \Lambda_0T^{\gamma} \approx 10^{-24.66}T^{1/4}$~erg~cm$^3$~s$^{-1}$ 
(after \citealt{vdoord89} and \citealt{mewe85}). Equations~(\ref{decay}, \ref{radloss}) 
define the {\it radiative time scale} $\tau_{\rm th} \equiv \tau_r$.

\subsection{Interpretation of the decay time}\label{decaytime}

Equations~(\ref{decay}), (\ref{condloss}), and (\ref{radloss}) describe the decay of the thermal
energy, which in flare plasma is primarily due to the decay of  temperature
(with a time scale $\tau_T$) and density. In contrast, the observed light curve decays 
(with a time
scale $\tau_d$ for the {\it luminosity}) primarily due to the decreasing EM and, to a 
lesser extent, due to the decrease  of $\Lambda(T)$ with decreasing temperature above $\approx 15$~MK. 
From the energy equation, the thermal energy decay time scale $\tau_{\rm th}$ is found to be
\begin{equation}\label{decaylc}
{1\over \tau_{\rm th}} = \left(1-{\gamma\over 2}\right){1\over \tau_T} + {1\over 2\tau_d}
\end{equation}
where the right-hand side is usually known from the observations (see \citealt{vdoord88} for
a derivation). The decay time scale
of the EM then follows as $1/\tau_{\rm EM} = 1/\tau_d - \gamma/\tau_T$.
\citet{pan97} derived somewhat different coefficients in (\ref{decaylc}) for the assumption of  
constant volume or constant mass, including the enthalpy flux.
In the absence of  measurements of $\tau_T$, it is often assumed that $\tau_{\rm th} = \tau_d$ 
although this is an inaccurate approximation. 

In (\ref{decaylc}), $\tau_{\rm th}$ is usually set to be $\tau_r$ or $\tau_c$ or,
if both loss terms are significant, $(\tau_r^{-1} + \tau_c^{-1})^{-1}$, taken at the beginning
of the flare decay (note again that a simple identification of $\tau_r$ with $\tau_d$ is not accurate). 
If radiative losses dominate, the density immediately follows
from (\ref{decay}, \ref{radloss})
\begin{equation}\label{cooltime}
\tau_{\rm th} \approx {3kT\over n_e\Lambda(T)}
\end{equation}
and the characteristic size scale $\ell$ of the flaring plasma or the flare-loop
semi-length $L$ for a sample of $\mathcal{N}$ identical loops follow from
\begin{equation}
{\rm EM} = n_en_H (\Gamma + 1)\pi\alpha^2 \mathcal{N} L^3 \approx n^2\ell^3 
\end{equation}
where $\alpha$ is the loop aspect ratio (ratio between loop cross sectional diameter at the base  
and total length $2L$) and $\Gamma$ is the  loop expansion factor.
The loop height for the important case of dominant radiative losses follows 
to be \citep{white86, vdoord88} 
\begin{equation}
H = \left({8\over 9\pi^4}{\Lambda_0^2\over k^2}\right)^{1/3}\left({{\rm EM}\over T^{3/2}}\tau_r^2\right)^{1/3}
   \left(\mathcal{N}\alpha^2\right)^{-1/3}(\Gamma+1)^{-1/3}.
\end{equation}
A lower limit to $H$ is found for  $\tau_r \approx \tau_c$ in the same treatment:
\begin{equation}
H_{\rm min} = {\Lambda_0 \over \kappa_0\pi^2}{{\rm EM}\over T^{3.25}}\left(\mathcal{N}\alpha^2\right)^{-1}.
\end{equation}
$\mathcal{N}$, $\alpha$, and $\Gamma$ are usually unknown and treated as free parameters within
reasonable bounds. Generally, a small $\mathcal{N}$ is compatible with dominant radiative cooling.

\subsection{Quasi-static cooling loops}

\citet{vdoord89} derived the energy equation of a cooling magnetic loop
in such a way that it is formally identical to a static loop \citep{rosner78},
by introducing a slowly varying flare heating rate that balances the total
energy loss, and a possible constant heating rate during the flare decay. This specific
solution thus proceeds through a sequence of different (quasi-)static loops
with decreasing temperature.

The general treatment  involves continued heating that keeps the cooling loop 
at coronal temperatures. If this constant heating term is zero, one finds
for free quasi-static cooling
\begin{eqnarray}
T(t) &=& T_0(1 + t/3\tau_{r,0})^{-8/7}   \\
L_r(t) &=& L_{r,0}(1 + t/3\tau_{r,0})^{-4}  \\
n_e(t) &=& n_{e,0}(1 + t/3\tau_{r,0})^{-13/7}   
\end{eqnarray}
 where $L_r$ is the total radiative loss rate, and $\tau_{r,0}$  is the radiative 
loss time scale (\ref{cooltime})  at the beginning of the flare decay.

This prescription is equivalent to requiring a constant ratio  between
radiative and conductive loss times, i.e., in the approximation of $T \ga 20$~MK 
($\Lambda \propto T^{1/4}$)
\begin{equation}\label{decayratio}
{\tau_r \over \tau_c} = {\rm const}{T^{13/4}\over {\rm EM}} \approx 0.18.
\end{equation}
Accordingly, the applicability of the quasi-static cooling approach can be supported
or rejected based on the run of $T$ and EM during the decay phase. Note, however, 
that a constant ratio (\ref{decayratio}) is not a sufficient condition to fully justify this 
approach.

\subsection{Cooling loops with continued heating}\label{contheat}

\begin{figure} 
\centerline{\resizebox{0.75\textwidth}{!}{\rotatebox{270}{\includegraphics{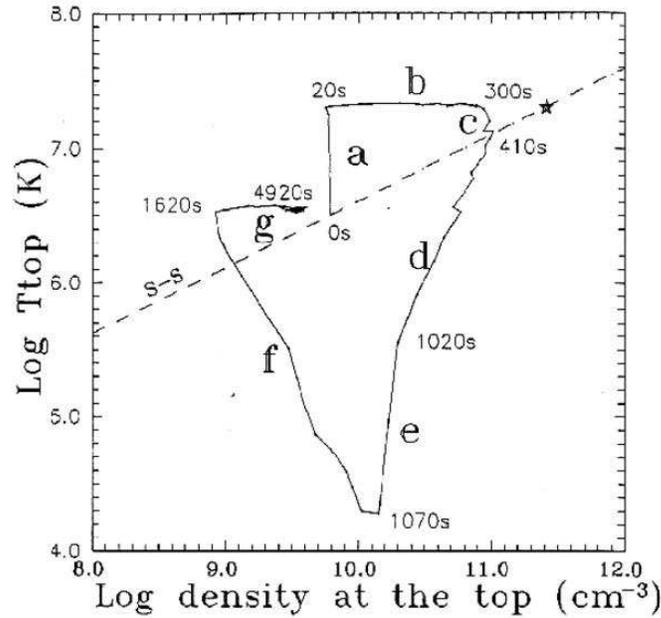}}}}
\caption{Density-temperature diagram of a hydrodynamically simulated flare.
           The flare loop starts from an equilibrium (S-S, steady-state loop according to
	   \citealt{rosner78}); (a) and (b) refer to the heating phase; at (c), the heating 
	   is abruptly turned off, after which the loop cools rapidly (d, e), and only slowly
	   recovers toward a new equilibrium solution (f, g) due to
	   constant background heating (from \citealt{jakimiec92}).}\label{jakimiec}
\end{figure}

Whether or not flaring loops indeed follow a quasi-static cooling path
is best studied on a density-temperature diagram (Fig.~\ref{jakimiec}). Usually, characteristic
values $T = T_{\rm a}$ and $n_e = n_{e, {\rm a}}$ as measured
at the loop apex are used as diagnostics. For a magnetic loop in hydrostatic
equilibrium, with constant cross section assumed, the RTV scaling law~(\ref{RTV})
requires stable solutions $(T, n_e)$ to be located 
where $T^2 \approx 7.6\times 10^{-7} n_e L$ (for $n_e = n_i$). On a diagram of 
log\,$T$ vs. log\,$n_e$, all solutions
are therefore located on a straight line with slope $\zeta = 0.5$. Figure~\ref{jakimiec}
shows the path of a hydrodynamically simulated flare.
The initial rapid heating leads to a rapid increase of
$T$, inducing increased losses by conduction. As chromospheric evaporation 
grows, radiation helps to balance the heating energy input. 
The flare decay sets in once the heating rate drops. At this moment, 
depending on the amount of ongoing heating, the magnetic loop is too dense to 
be in equilibrium, and the radiative losses exceed the heating rate, resulting 
in a thermal instability. In the limit of no heating during the decay, that is, an abrupt 
turn-off of the heating at the flare peak, the slope of the path becomes
\begin{equation}
\zeta \equiv {d{\rm ln}~T\over d{\rm ln}~n_e} \equiv {\tau_n\over \tau_T} = 2  
\end{equation}
implying $T(t) \propto n_e^{\zeta}(t) =  n_e^2(t)$ (see \citealt{serio91} for further discussion).
Here, $\tau_T$ and $\tau_n$ are the e-folding decay times of the temperature and 
the electron density, respectively, under the assumption of exponential decay laws.
Only for a non-vanishing  heating rate 
does the loop slowly recover and eventually settle on a new equilibrium
locus (Fig.~\ref{jakimiec}). In contrast, if heating continues and is
very gradually  reduced, the loop decays along the static solutions ($\zeta = 0.5$). 
Observationally, this path is 
often followed by large solar flares \citep{jakimiec92}. 

Applying this concept to stellar astronomy, we replace $n_e$ by the 
observable $\sqrt{\rm EM}$ and thus assume a constant flare volume, and further 
introduce the following generalization. In the freely cooling case after an abrupt 
heating turnoff, the entropy per particle at the loop apex  decays on the thermodynamic 
decay time
\begin{equation}
\tau_{\rm td} = 3.7\times 10^{-4} {L\over T_0^{1/2}} \quad {\rm [s]}
\end{equation}
where $T_0$ is the flare temperature at the beginning of the decay \citep{serio91}.
When heating is present, we introduce a correction term $F(\zeta) \equiv \tau_{\rm LC}/\tau_{\rm td}$ 
(\citealt{reale97}; $\tau_{\rm LC}$ is the observed light curve decay time)
\begin{equation}\label{taulc}
\tau_{\rm LC} = 3.7\times 10^{-4} {L\over T_0^{1/2}}F(\zeta) \quad {\rm [s]}
 \end{equation}
$F(\zeta)$ is therefore to be numerically calibrated for each X-ray telescope.
With known $F$, (\ref{taulc}) can be solved for $L$. This scheme thus offers
i) an indirect method to study flaring loop geometries ($L$), ii) a way of  
determining the rate and decay time scale of continued heating via $F(\zeta)$
and $\tau_{\rm td}$, and iii) implications for the density decay time via 
$\tau_n = \zeta\tau_T$. Conditions of applicability include $\zeta \ge 0.3$ 
( most values are found in the range of $\sim 0.3 - 1$, \citealt{reale97}) and  a
resulting loop length $L$ of less than one pressure scale height.

\subsection{Two-Ribbon flare models}\label{2R}
 
\begin{figure} 
\centerline{\resizebox{1\textwidth}{!}{\includegraphics{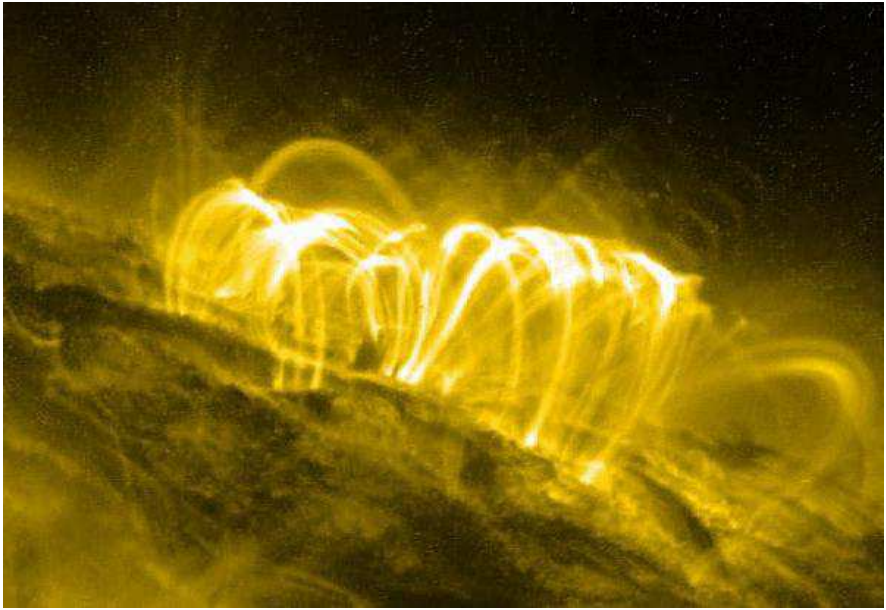}}}
\caption{{\it Trace} image of a flaring magnetic loop arcade.}\label{2R_Trace}
\end{figure}

An approach that is entirely based on continuous heating (as opposed to cooling) 
was developed for the two-ribbon (2-R) class of solar flares. An example of this flare type is 
shown in Fig.~\ref{2R_Trace}.
The 2-R flare model devised initially by \citet{kopp84}
is a parameterized magnetic-energy release model.
The time development of the flare light-curve is completely determined
by the amount of energy available in non-potential magnetic fields, and by
the rate of energy release as a function of time and geometry as the fields
reconnect and relax to a potential-field configuration.
It is assumed that a portion of the total energy is radiated into the observed X-ray band,
while the remaining energy will be lost by other mechanisms.  
2-R flares are well established for the Sun 
(Fig.~\ref{2R_Trace}); they often lead to large,   
long-duration flares that may be accompanied by mass ejections.  
 
The magnetic fields are, for convenience, described along meridional
planes on the star by Legendre polynomials $P_n$ of order $n$, up to the height of the
neutral point; above this level, the field is directed radially, that is, the
field lines are ``open''.  
As time proceeds, field lines nearest to the 
neutral line move inward at coronal levels and reconnect at progressively
larger heights above the neutral line. The reconnection point thus moves upward
as the flare proceeds, leaving closed magnetic-loop systems underneath. 
One loop arcade thus corresponds to one N-S aligned lobe between two zeros of $P_n$ in 
latitude, axisymmetrically continued over some longitude in E--W direction.
The propagation of the neutral point in height, $y(t)$, with a 
time constant $t_0$, 
is prescribed by ($y$ in units of $R_*$, measured from the star's center)
\begin{equation}\label{yt}
y(t) = 1 + {H_{\rm m}\over R_*}\left(1- e^{-t/t_0}\right)
\end{equation}
\begin{equation}\label{ht}
H(t) \equiv [y(t) - 1]R_*
\end{equation}
and the total energy release of the reconnecting arcade per radian in longitude is
equal to the magnetic energy lost by reconnection,
\begin{equation}
{{\rm d}E\over {\rm d}y} = {1\over 8\pi}2n(n+1)(2n+1)^2R_*^3B^2I_{12}(n)
        { y^{2n}(y^{2n+1} - 1)\over [n+(n+1)y^{2n+1}]^3} 
\end{equation}
\begin{equation}\label{dedt}
{dE\over dt} ={dE\over dy}{dy\over dt} 
\end{equation}
\citep{poletto88}. In (\ref{yt}), $H_{\rm m}$ is the maximum height of 
the neutral point for $t\rightarrow \infty$; typically, $H_{\rm m}$  is assumed 
to be equal to the latitudinal extent of the loops, i.e.,
\begin{equation}\label{height}
H_{\rm m} \approx {\pi\over n+1/2}R_*
\end{equation}
for $n > 2$ and $H_{\rm m} = (\pi/2)R_*$ for $n = 2$.
Here, $B$ is the  surface magnetic field strength at the axis of symmetry,
and $R_*$ is the stellar radius. Finally, $I_{12}(n)$ corresponds to 
$\int [P_n({\rm cos}\theta)]^2d({\rm cos}\theta)$ evaluated between the
latitudinal borders  of the lobe (zeros of ${\rm d}P_n/{\rm d}\theta$), 
and $\theta$ is the  co-latitude. 

The free parameters are $B$ and the efficiency of the energy-to-radiation 
conversion, $q$, both of which determine the normalization of the light curve;
the time scale of the reconnection process, $t_0$, and the polynomial degree $n$ determine the 
duration of the flare; and the geometry of the flare is fixed  by 
$n$ and therefore the asymptotic height $H_{\rm m}$ of the reconnection point.
The largest realistic 2-R flare model is based on the Legendre polynomial of degree $n = 2$; 
the loop arcade then stretches out between the equator and the stellar poles. Usually,
solutions can be found for many larger $n$ as well. However, because a larger $n$ requires 
larger surface magnetic field strengths, a natural limit is set to $n$ within
the framework of this model. Once the model solution has been established, further 
parameters, in particular the electron density $n_e$, can be inferred.

\subsection{A  magnetohydrodynamic model}\label{hydrodyn}

Some scaling laws  have been obtained from simulations based on
the full set of magnetohydrodynamic equations \citep{shibata99, shibata02}. 
For the flare peak temperature $T$, the loop 
magnetic field strength $B$, the pre-flare loop electron density $n_0$, and the 
loop semi-length $L$, one finds, under the condition of dominant conductive cooling 
(appropriate for the early phase of a flare),
\begin{equation}\label{shibata1}
T \approx 1.8\times 10^4 B^{6/7}n_0^{-1/7}L^{2/7}~{\rm [K]}.
\end{equation}
 The law follows from the balance between conduction cooling 
($\propto T^{7/2}/L^2$, \ref{condloss}) and magnetic reconnection heating ($\propto B^3/L$).
Assuming loop filling through chromospheric evaporation and balance between
thermal and magnetic pressure in the loop, two further ``pressure-balance scaling laws'' follow:
\begin{eqnarray}\label{shibata2}
{\rm EM}&\approx& 3\times 10^{-17} B^{-5}n_0^{3/2}T^{17/2}~{\rm [cm^{-3}]}\\
{\rm EM}&\approx& 2\times 10^8     L^{5/3}n_0^{2/3}T^{8/3}~{\rm [cm^{-3}]}.
\end{eqnarray}
An alternative scaling law applies if the density development in the initial flare
phase is assumed to follow balance between evaporation enthalpy-flux and conduction 
flux, although the observational support is weaker,
\begin{equation}\label{shibata3}
{\rm EM}\approx 1\times 10^{-5} B^{-3}n_0^{1/2}T^{15/2}~{\rm [cm^{-3}]}.
\end{equation}
And third, a steady solution is found for which the radiative losses balance conductive
losses. This scaling law applies to a steady loop,
\begin{equation} 
{\rm EM} \approx \left\{ 
   \begin{array}{ll}\label{demflaremhd}
       10^{13} T^4L~\quad{\rm [cm^{-3}]}            & \mbox{\quad for\quad  $T < 10^7$~K } \\
       10^{20} T^3L~\quad{\rm [cm^{-3}]}            & \mbox{\quad for\quad  $T > 10^7$~K } 
   \end{array} 
   \right. 
\end{equation} 
and is equivalent to the RTV scaling law (\ref{RTV}).

The advantage of these scaling laws is that they make use exclusively of the flare-peak
parameters $T$, EM, $B$ (and the pre-flare density $n_0$) and do not require knowledge of
the time evolution of these parameters.

\subsection{Observations of stellar X-ray flares}\label{obsflares}

One of the main results that have come from extensive modeling of stellar X-ray flares 
is that extremely large stellar flares require large volumes
under all realistic assumptions for the flare density, i.e., flaring complexes of magnetic loops must
either be high, or must spread across a large surface area. This is because, first, the energy derives 
from the non-potential portion of the magnetic fields that are probably no stronger than a few 
100~G in the corona; and second, small-loop models require higher pressure to produce the observed 
luminosity, hence requiring excessively strong magnetic fields.

There is extensive literature discussing individual flares observed on a variety of
stars. I will not discuss these results individually here. The interested reader
may consult the compilation of results in \citet{guedel04b} and the references given therein.
This section summarizes some systematic trends found in stellar flares.

\begin{figure} 
\hbox{
\resizebox{0.5\textwidth}{!}{\includegraphics{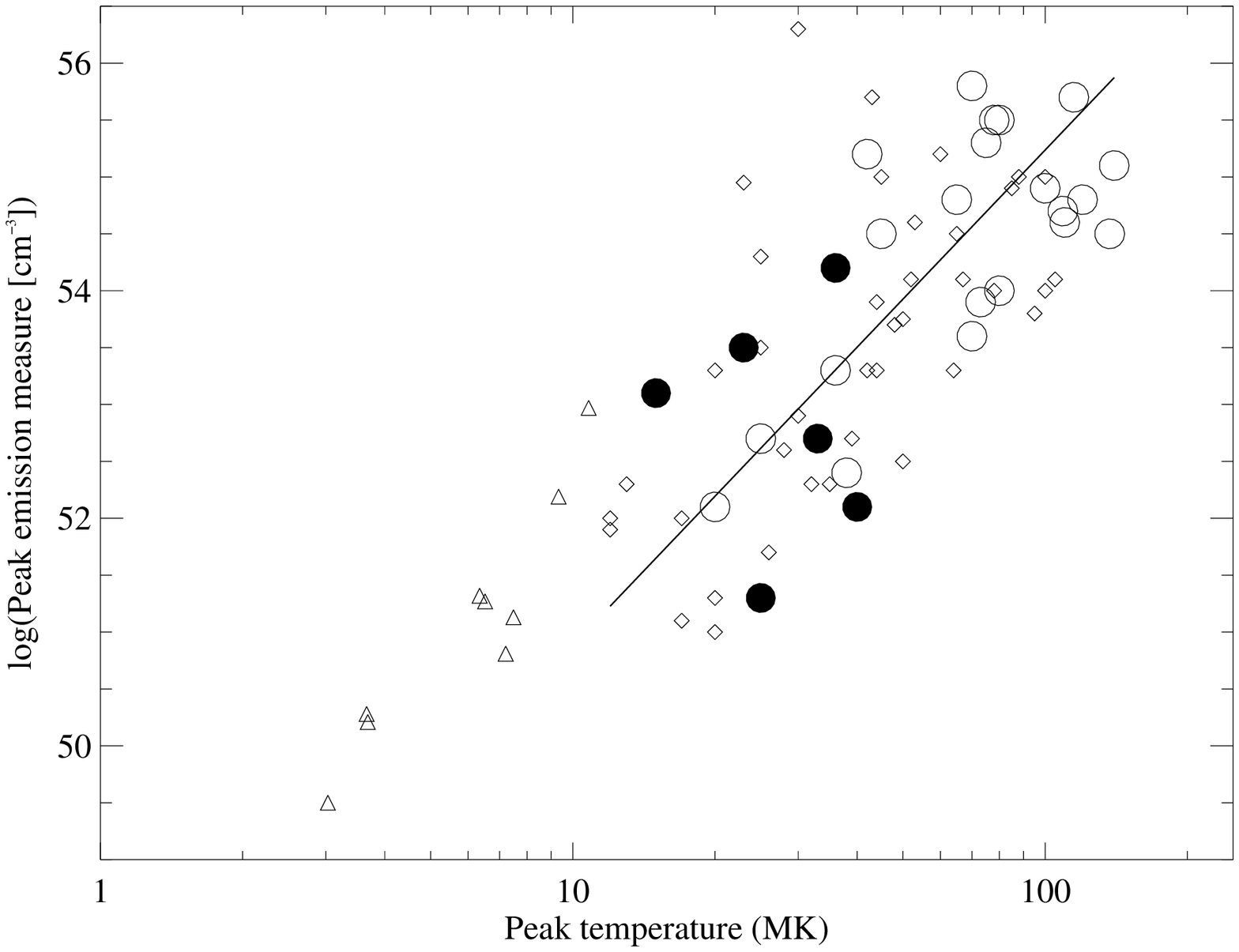}}
\resizebox{0.46\textwidth}{!}{\includegraphics{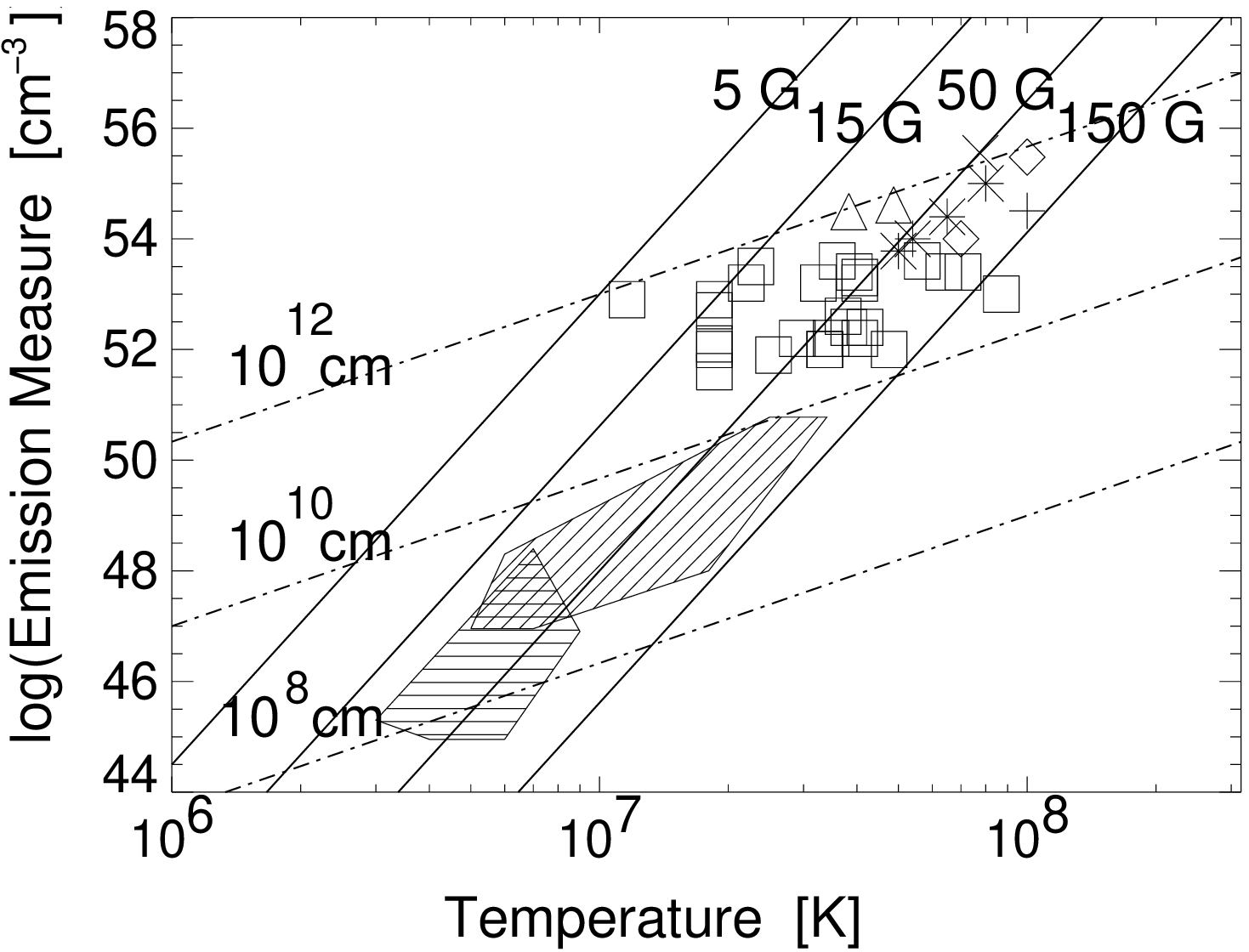}}
}
\caption{{\it Left:} Peak temperatures and EMs of the flares (from \citealt{guedel04b} and references therein). 
Key to the symbols: Filled circles: {\it XMM-Newton} observations. Open circles: {\it ASCA}  or {\it BeppoSAX} observations.
Small diamonds: observations from other satellites. The solid line shows a regression fit (\ref{EM_T}).
Triangles  represent non-flaring parameters of 
G stars, referring to the hotter plasma component in 2-$T$ spectral fits to {\it ROSAT} data.
-- {\it Right:} Theoretical EM-$T$ relations based on the reconnection model by
Shibata \& Yokoyama, showing lines of constant loop length $L$ and
lines of constant magnetic field strength $B$. Hatched areas are loci reported for
solar flares, and other symbols refer to individual stellar flares in star-forming
regions (figure courtesy of K. Shibata and T. Yokoyama, after \citealt{shibata02}).}\label{tem_flare}
\end{figure}

When the flare energy release  evaporates plasma into the corona, heating and cooling
effects compete simultaneously, depending on the density and temperature profiles in a given flare. It is
therefore quite surprising to find a broad correlation between peak temperature $T_p$ and peak emission 
measure EM$_p$, as illustrated in Fig.~\ref{tem_flare} for the sample reported in \citet{guedel04b}. 
A regression fit gives (for 66 entries)

\begin{equation}\label{EM_T}
{\rm EM}_p \propto T_p^{4.30\pm 0.35}.
\end{equation}
The correlation overall indicates that {\it larger flares are hotter.} 
A similar relation was reported previously for solar flares \citep{feldman95}. 

It is interesting to note that  this correlation is similar to the $T-L_X$ correlation for the 
``non-flaring'' coronal stars in Fig.~\ref{temperaturelx} at cooler temperatures. This same sample 
is plotted as triangles in Fig.~\ref{tem_flare}, again only for the hotter plasma component.
The stars follow approximately the same slope as the flares,
albeit at cooler temperatures, and for a given temperature, the EM is higher. This trend may 
suggest that flares systematically contribute to the hot plasma component, although we have not 
temporally averaged the flare temperature  and EM for this
simple comparison. 

In the context of the magnetohydrodynamic scaling laws presented in (\ref{shibata1})--(\ref{shibata2}),
the observed loci of the flares require  loop magnetic field strengths similar to solar flare 
values ($B \approx 10-150$~G) but the loop lengths must increase toward larger flares. This is 
seen in Fig.~\ref{tem_flare} where lines of constant $L$ and $B$ are plotted for this flare model. 
Typical loop lengths are thus of order $L \approx 10^{11}$~cm in this interpretation.

\begin{figure} 
\hbox{\resizebox{1.\textwidth}{!}{\includegraphics{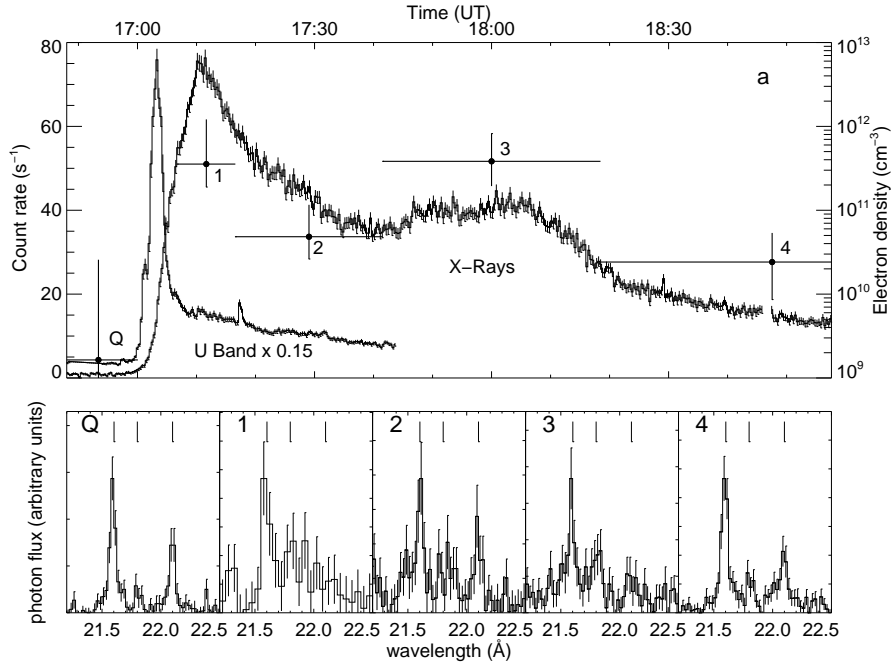}}}
\caption{Flare on Proxima Centauri, observed with {\it XMM-Newton}. The top panel shows
the X-ray light curve and the much shorter  U band flare (around 17 UT). The bottom panel
shows the O\,{\sc vii} He-like triplets observed during various time intervals of the flare. 
The locations of the $r, i$, and $f$ lines are marked by vertical lines. 
The resulting electron densities are given in the top panel by the crosses, where the
horizontal arm lengths indicate the time intervals over which the data were integrated, 
and the right axis gives the logarithmic scale 
(after \citealt{guedel02a}).}\label{proxcenflare}
\end{figure}

\subsection{The ``Neupert Effect''}\label{neuperteffect}

In the framework of chromospheric evaporation, we assume that the emitted radio 
emission or its flux at Earth, $F_{\rm R}(t)$ is at any time proportional to the
deposition rate of kinetic energy by nonthermal electrons into the plasma of the 
chromosphere. This plasma is thereby heated and ``evaporates'' into the 
corona.  We define the conversion factor $\alpha$ as the ratio between the
{\it thermal} energy flux being deposited in the chromosphere by nonthermal
electrons and the observed radio flux density. Conversion
into other forms of energy (mechanical, turbulence, etc.) occurs in parallel. 
Our simplification consists, first, in the assumption that these other energy 
transformations do not interact so that $\alpha$ remains constant over the 
course of the relatively short radio flare; second, we will assume that 
the dominant losses of the {\it thermal} energy occur via radiation; we will 
neglect energy loss into cool, non-X-ray emitting plasma, e.g.  by 
conduction, and cooling by adiabatic expansion. Also, we assume that there is no 
direct heating in parallel to the chromospheric evaporation. 
     Third, we will use one temperature parameter for the flare plasma
     at any given time: The temperature $T$ can be considered as describing
     an isothermal plasma dominating the total losses from the flaring loops. 

The rate of change of the 
total thermal energy $E$ in a plasma of volume $V$ with electron density 
$n_e$ is determined by the influx of kinetic energy and by radiation, hence
the energy conservation equation for the {\it thermal} plasma is
\begin{equation}\label{energy_neupert}
{d\over dt}(3n_ekTV) = \alpha F_{\rm R}(t) - n_e^2V\Lambda(T)  
\end{equation}
where $\Lambda(T)$ is the total luminosity of a plasma with unit 
emission measure (EM) at a temperature of $T$.   Note again that
(\ref{energy_neupert}) does, by definition of $\alpha$, not describe the total
energy budget of the flare, but merely  the energy conversion
of interest here.  
     For large flare temperatures ($\gg 20$~MK), $\Lambda$ is  dominated by 
     bremsstrahlung losses, roughly scaling as $T^{1/2}$. 
For somewhat more moderate temperatures ($\ga  10-20$~MK), losses via 
line emission become important, with the (isothermal) approximation for $\Lambda$
\begin{equation}\label{neupertloss}
\Lambda(T) = 1.86\cdot 10^{-25}T^{1/4}~{\rm [erg~s^{-1}~cm^{3}]}  
\end{equation}
(see after equation~\ref{radloss} - for simplicity, we define here
EM = $n_e^2V$ with $n_H = 0.85n_e$ and a corresponding correction in $\Lambda$).  
In the  absence of  kinetic energy influx $\alpha F_{\rm R}(t)$, the thermal energy  
decays radiatively (neglecting conduction), and hence we define
an e-folding decay time $\tau$ for the thermal energy as a function of the 
     two independent variables temperature $T$ and density $n_e$ at a given instant,
\begin{equation}
\left.{dE(t)\over dt}\right|_{F_{\rm R}=0} = {E(t)\over 
      \tau(n_e(t), T(t))}  
\end{equation}
so that  
\begin{equation}\label{neupert_tau}
\tau(n_e(t), T(t)) = {E(t)\over L_{\rm rad}(t)}  = {3kT\over n_e\Lambda(T)}   
\end{equation}
where $L_{\rm rad}$ is the total luminosity (approximately = X-ray luminosity, $L_X$, but also
contributions from lower temperatures, e.g., in the ultraviolet, which we neglect here). Then,
\begin{equation}\label{neupertdiff}
{dE(t)\over dt} = \alpha F_{\rm R}(t) - L_{\rm rad}(t) = 
   \alpha F_{\rm R}(t) - {E(t) \over \tau(t)}. 
\end{equation}
The general solution of the inhomogeneous, linear differential equation~(\ref{neupertdiff}) 
reads
\begin{equation}
E(t) = e^{-\int_{t_0}^t\tau(y)^{-1}{\rm d}y}
          \left(E_0 +
          \alpha\int_{t_0}^tF_{\rm R}(u)e^{+\int_{t_0}^u\tau(y)^{-1}dy}
            {\rm d}u\right) 
\end{equation}           
where $E_0 = E(t_0)$ for a fixed $t_0$ before the flare start.
The integration over $\tau^{-1}$ is along the time axis.   
If the initial thermal energy content can be neglected ($E_0 = 0$), then 
\begin{equation}
E(t) = \alpha\int_{t_0}^tF_{\rm R}(u)e^{-\int_u^t\tau(y)^{-1}dy}du.
\end{equation}     
We define an average decay constant for any time interval [$u,t$] by
\begin{equation}
{\bar{\tau}}^{-1} = {\bar{\tau}}^{-1}(u, t) = 
   {\int_u^t\tau(y)^{-1}dy \over t-u}.  
\end{equation}
Then   
\begin{equation}\label{generalneupert}
E(t) = \alpha\int_{t_0}^tF_{\rm R}(u)e^{-(t-u)/{\bar\tau(t,u)}}du, 
\end{equation}
i.e., for constant $\tau$ the energy profile is the {\it convolution} 
of the kinetic energy influx with an exponential function.  
Equation~(\ref{generalneupert}) is a generalized form of the 
{\it Neupert effect}. In the limit 
$\tau \rightarrow \infty$,  (\ref{neupertdiff}) and (\ref{generalneupert}) 
become
\begin{equation}\label{neupert1}
{dE(t)\over dt} = \alpha F_{\rm R}(t)  
\end{equation}
\begin{equation}\label{neupert2}
E(t) = \alpha\int_{t_0}^tF_{\rm R}(u)du,  
\end{equation}
i.e., the total energy content of the plasma is the integral of the
kinetic energy influx (and radiation is inhibited). Equations~(\ref{neupert1}) and (\ref{neupert2}) 
remind us   of the classical formulation of the Neupert effect, with
the observed X-ray losses replaced by the thermal energy content of
the plasma. These two equations are applicable  only  for the 
increasing portion of the soft X-ray light curve or, correspondingly, for 
the time interval where $F_{\rm R} \neq 0$.  
On using (\ref{neupert_tau}), we obtain the  generalized Neupert effect for 
the light curve,
\begin{equation}\label{neupertlum}
L_{\rm rad}(t) = {\alpha\over 
       \tau(t)}\int_{t_0}^tF_{\rm R}(u)e^{-(t-u)/{\bar\tau(t,u)}}du. 
\end{equation} 
Note the importance of the thermal {\it energy} decay time
$\tau = 3kT/(n_e\Lambda(T))$.
Serio et al. (1991) and Jakimiec et al. (1992)  find $Tn^{-2} =$ const for the 
radiative cooling phase. If the temperatures  are very high and the 
bremsstrahlung approximation  applies, then  $\tau = T^{1/2}n^{-1}$ 
and thus $\tau = $ const in~(\ref{neupertlum}). However, in the case of more 
moderate temperatures (\ref{neupertloss}),  only for $T^{3/4}n^{-1} =$ const does the 
classical Neupert effect approximately apply to the light curve. The actual
functional dependence of $\psi(T)$ is more complicated.   Thus, in general,  
the relevant parameters to be investigated for the Neupert effect are 
the {\it energies}.   Further,  $L_{\rm rad}(t)$ describes all radiative losses across the 
electromagnetic spectrum; the observed X-ray luminosity $L_{\rm X}$  generally 
depends on the selected bandpass and therefore constitutes only a lower limit 
to $L_{\rm rad}$.

Despite these caveats, it is surprising that the Neupert effect is often well
seen, at least qualitatively, in solar radio and X-ray light curves. Instead of radio emission,
bursts can be monitored in the U band: because they are likely to be a prompt reaction to
the bombardment of the chromosphere by the same electron population that indices
gyrosynchrotron emission, the Neupert effect should hold as well. A stellar example is
shown in Fig.~\ref{proxcenflare}. These observations give clear evidence that at least
some giant flares on active stars are subject to similar evaporation physics as known
from the solar corona.

\section{The statistics of flares}\label{stochasticflares}

The study of coronal structure confronts us with several problems that are difficult
to explain by scaling of solar coronal structure: i) Characteristic 
coronal temperatures increase with increasing magnetic activity. 
ii) Characteristic coronal densities are typically higher in active than in inactive stars, 
and pressures in hot loops can be exceedingly high.
iii) The maximum stellar X-ray luminosities exceed the levels expected from complete 
coverage of the surface with solar-like active regions by up to an order
of magnitude. iv) Radio observations reveal a persistent population 
of non-thermal high-energy electrons in magnetically active stars even if
the lifetime of such a population should only be tens of minutes 
to about an hour  under ideal trapping conditions in coronal loops and 
perhaps much less due to efficient scattering of electrons into the chromosphere.
Several of these features 
are reminiscent of flaring, as are some structural elements in stellar coronae. If flares
are important for any of the above stellar coronal properties indeed, then we must consider
the effects of frequent flares that may be  unresolved in our observations but that may make up
part, if not all, of the ``quiescent'' emission.

\subsection{Stochastic variability - what is ``quiescent emission''?}\label{stochvariability}

The problem has been attacked in several dedicated statistical studies. Early statistical
investigations (fluctuation analysis in light curves) remained ambiguous, reporting a significant
amount of variability, although not necessarily being due to flares, or statistical absence of
low-level flaring within the sensitivity limits (e.g., \citealt{ambruster87, collura88, pallavicini90}).
There are indications in newer data that M dwarfs are continuously  variable on short ($\la$ 1 day)
time scales, and that  the luminosity distribution is very similar to the equivalent
distribution derived for solar flares, which  suggests that the overall stellar light curves of dM
stars are variable in the same way as a statistical sample of solar flares \citep{marino00}.

Very long light curves obtained from the {\it EUVE} satellite reveal an astonishing level of continuous
variability in active M dwarfs. Some of those data can be  used to investigate statistical properties of the 
occurrence rate of flares as a function of total emitted energy.
The increased sensitivity of {\it XMM-Newton} and {\it Chandra} is now revealing extreme levels
of activity. Some X-ray light curves show no steady time interval exceeding a few tens of minutes within
the sensitivity limit. In the day-long light curve in Fig.~\ref{uvcetlight}, no more than 30\%, and probably 
much less, of the average X-ray emission of UV Cet can be attributed to any sort of steady emission, even
outside the obvious, large flares. On the contrary, almost the entire 
light curve is resolved into frequent, stochastically occurring flares of various amplitudes.

\begin{figure} 
\centerline{\resizebox{0.90\textwidth}{!}{\includegraphics{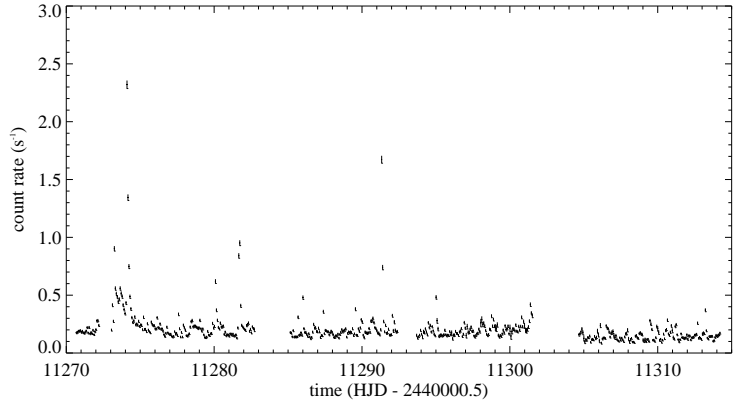}}}
\caption{A long light curve of the dMe star AD Leo, obtained by the DS instrument on {\it EUVE}. Most of the discernible
variability is due to flares (after \citealt{guedel03a}).}\label{adleoeuve}
\end{figure}

\begin{figure} 
\centerline{\resizebox{0.95\textwidth}{!}{\includegraphics{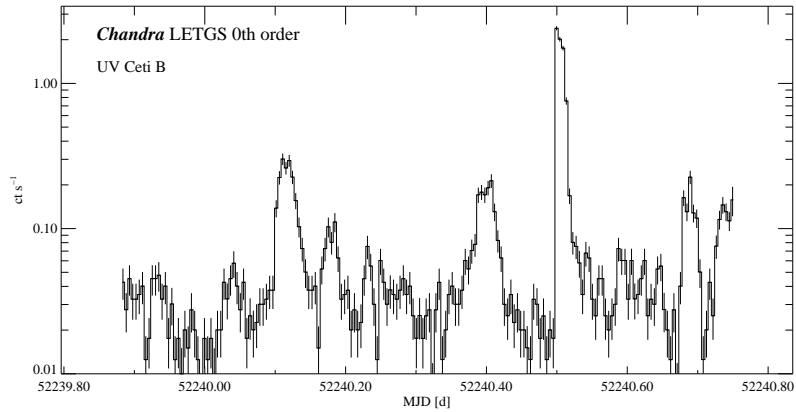}}}
\caption{Light curve of UV Ceti B, observed with the {\it Chandra} LETGS/HRC during about
1 day. Note the logarithmic flux axis (figure courtesy of M. Audard, after \citealt{audard03}).}\label{uvcetlight}
\end{figure}

\begin{figure} 
\centerline{\resizebox{0.95\textwidth}{!}{\includegraphics{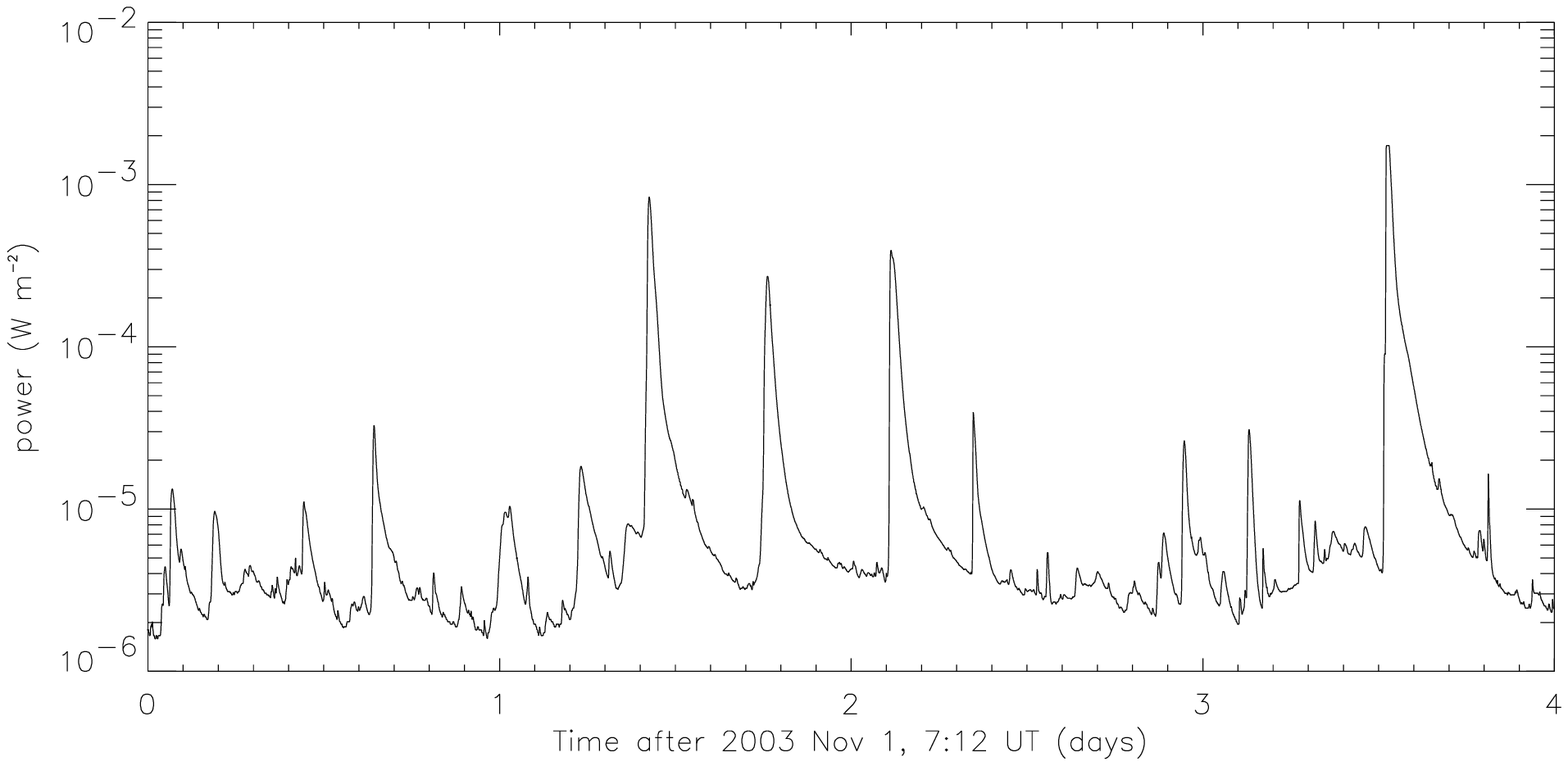}}}
\caption{{\it GOES} full-disk solar X-ray light curve, observed in the 1.5--12~keV band in
        November 2003. The abscissa gives time after 2003 November 1, 7:12 UT in days.}\label{goeslight}
\end{figure}

In the {\it solar} corona,  the flare rate increases steeply toward lower radiative
energies, with no evidence (yet) for a lower threshold (e.g., \citealt{krucker98}). Figure~\ref{goeslight}
shows an example of a {\it GOES} light curve in the 1.5-12~keV range, purposely selected during an extremely
active period in November 2003. While  the {\it GOES} band is harder than typical bands
used for stellar observations, it more clearly reveals the level of the underlying variability (a typical
detector used for stellar observations would see much less contrast).  If the solar
analogy  has any merit in interpreting stellar coronal X-rays, then  
low-level emission in stars that do show flares {\it cannot} be truly quiescent, that is, constant or 
slowly varying exclusively due to long-term evolution of active regions, or due to rotational modulation.
A measure of {\it flare rates} is therefore not meaningful unless it refers to flares above a 
given luminosity or energy threshold.
This is - emphatically - not to say that steady emission is absent in magnetically active stars. 
However, once we accept the solar analogy as a working principle, the question is not so much about
the presence of large numbers of flares, but to what
extent they contribute to the overall X-ray emission from coronae.

\subsection{The flare-energy distribution}\label{energydistribution}

The suggestion that stochastically occurring flares may be largely responsible for
coronal heating is known as the ``microflare'' or ``nanoflare'' hypothesis in solar 
physics \citep{parker88}. Observationally, it is supported by evidence  for the presence of
numerous  small-scale flare events occurring in the solar corona 
at any time (e.g., \citealt{lin84}). Their distribution in energy is  a 
power law,
\begin{equation}\label{powerlaw}
\frac{dN}{dE} = k E^{-\alpha} 
\end{equation}
where $dN$ is the number of flares per unit time with a total 
energy in the  interval [$E,E+dE$], and $k$ is a constant. If $\alpha\ge 2$, then the energy integration (for a given 
time interval) diverges for $E_{\rm min} \rightarrow 0$, that is, by 
extrapolating the power law to sufficiently small flare energies, {\it any} 
energy release power can be attained. This is not the case for $\alpha <2$.
Solar studies have repeatedly resulted in $\alpha$ values of $1.6-1.8$ for ordinary
solar flares \citep{crosby93}, but some recent studies of low-level  flaring
suggest $\alpha = 2.0 - 2.6$ \citep{krucker98, parnell00}.

\begin{table}[b!] 
\caption{Stellar radiative flare-energy distributions} 
\label{flarestat}       
\begin{tabular}{llllll} 
\hline\noalign{\smallskip}
Star sample	       & Photon energies      & log\,(Flare   &          & $\alpha$	 & References \\
         	       & [keV]                & energies)$^a$ &         &          	 &            \\
\noalign{\smallskip}\hline\noalign{\smallskip}
M dwarfs	       &  0.05--2   	      & $30.6-33.2$ & & 1.52$\pm 0.08$ & \citet{collura88} \\    
M dwarfs	       &  0.05--2   	      & $30.5-34.0$ & & 1.7$\pm 0.1$   & \citet{pallavicini90} \\
RS CVn binaries        & EUV		      & $32.9-34.6$ & & 1.6	     & \citet{osten99} \\
Two G dwarfs	       & EUV		      & $33.5-34.8$ & & 2.0--2.2       & \citet{audard99} \\
F-M dwarfs	       & EUV		      & $30.6-35.0$ & & 1.8--2.3       & \citet{audard00} \\
Three M dwarfs         & EUV		      & $29.0-33.7$ & & 2.2--2.7       & \citet{kashyap02}\\
AD Leo  	       & EUV\&0.1--10         & $31.1-33.7$ & & 2.0--2.5       & \citet{guedel03a}  \\
AD Leo  	       & EUV                  & $31.1-33.7$ & & $2.3\pm 0.1$   & \citet{arzner04} \\
\noalign{\smallskip}\hline
\multicolumn{6}{l}{$^a$Total flare-radiated X-ray energies used for the analysis (in ergs).}
\end{tabular}
\end{table}

Relevant stellar studies have been rare (see Table~\ref{flarestat}). Early investigations lumped several stars
together to produce meaningful statistics.  
Full forward modeling of a superposition of stochastic flares was applied to EUV and
X-ray light curves by \citet{kashyap02} and \citet{guedel03a} based on Monte Carlo simulations, 
and by \citet{arzner04} based on an analytical formulation.
The results of these investigations are in full agreement, converging to $\alpha \approx 2.0 - 2.5$
for M dwarfs (Table~\ref{flarestat}). If the power-law flare energy distribution extends by about 1--2 orders of 
magnitude below the actual detection limit in the light curves, then the {\it entire} emission could be explained
by stochastic flares. The coronal heating process in magnetically active stars would - in this extreme limit - be 
one solely due to {\it time-dependent} heating by flares, or, in other words, the X-ray corona would be an entirely 
hydrodynamic phenomenon rather than an ensemble of hydrostatic loops.

\begin{figure} 
\centerline{
\hbox{\hskip 0.6truecm
\resizebox{0.54\textwidth}{!}{\includegraphics{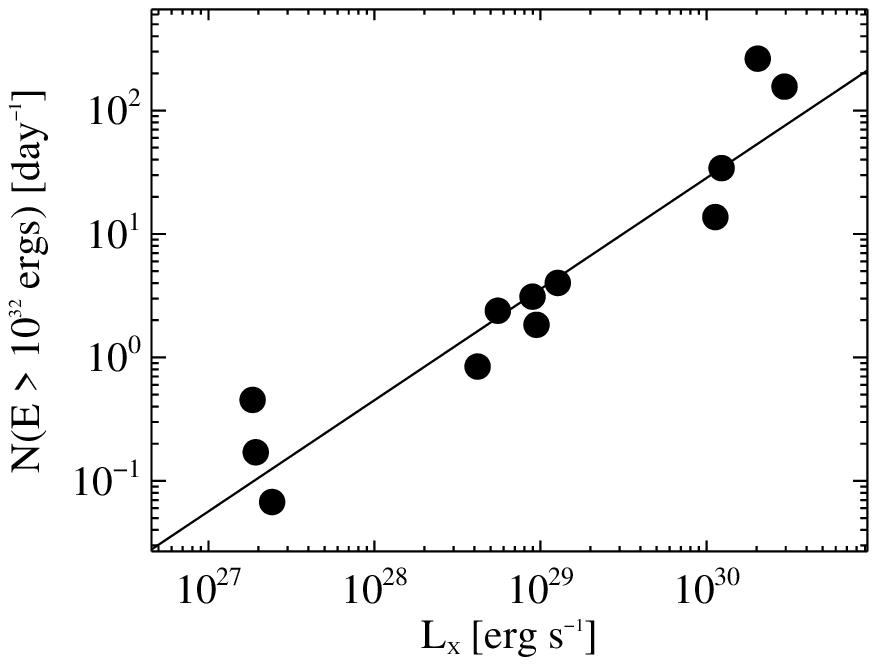}}
\hskip 0.4truecm\resizebox{0.49\textwidth}{!}{\includegraphics{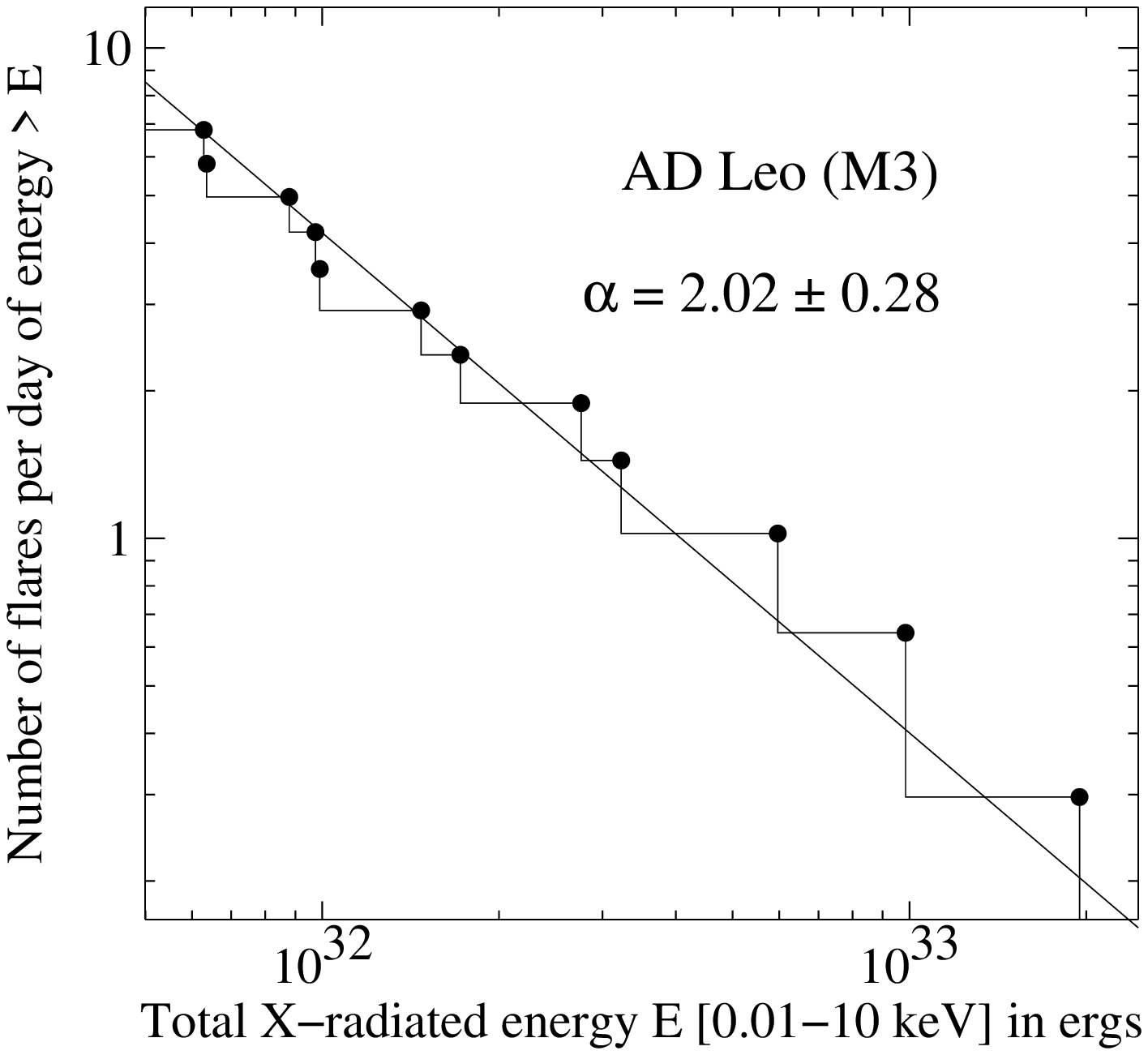}}
}
}
\vskip 0.5truecm
\caption{{\it Left:} The rate of flares above a threshold of $10^{32}$~erg in total
radiated X-ray energy is plotted against the low-level luminosity for several stars, together
with a regression fit.
{\it Right:} Flare energy distribution for AD Leo,
using a flare identification algorithm for an observation with {\it EUVE} (both figures courtesy of M. 
Audard, after \citealt{audard00}).}\label{flaredistrib}
\end{figure}

\subsection{Microflaring at radio wavelengths}

\begin{figure}[t!]
\resizebox{1\textwidth}{!}{\includegraphics{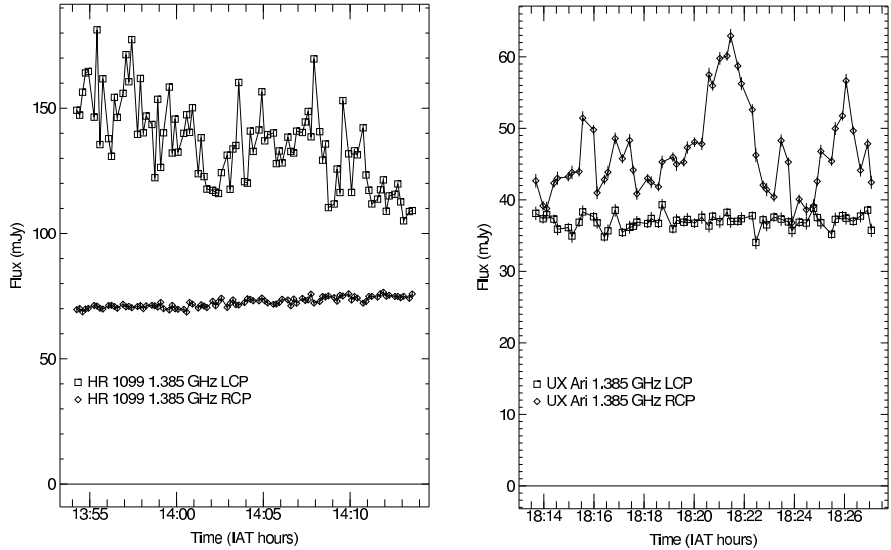}}
\caption{Light curves of HR~1099 ({\it left}) and UX Ari ({\it right}) obtained
in the two senses of circular polarization. The brighter of the two polarized
fluxes varies rapidly and has been interpreted as 100\% polarized 
coherent emission superimposed on a gradually changing gyrosynchrotron component 
(White \& Franciosini 1995; figures from S.~M. White).\label{light}}
\end{figure}

Quiescent radio emission can apparently persist for quite long periods.
Losses by collisions  (\ref{collloss}) require a very low ambient electron density to maintain
the electron population. Alternatively, electrons could be frequently injected 
at many coronal sites. Based on spectral observations, \citet{white95} 
suggest that the emission around 1.4~GHz shown in Fig.~\ref{light} is composed of a steady, weakly polarized 
broad-band gyrosynchrotron component plus superimposed, strongly and oppositely 
polarized, fluctuating plasma emission that is perceived as quasi-steady but that may
occasionally evolve into strong, polarized flare emission. Continual flaring may thus
also reflect in radio light curves.

\section{A flare-heating approach}

The quiescent emission of synchrotron radiation requires the continuous presence
of nonthermal electrons, while X-rays are emitted by the thermal bulk
of typically $10^6$ to a few times $10^7$ K. For
optically thin radio emission ($\nu \ga 5$ GHz), 
the radio luminosity   is
\begin{equation}\label{radiolum}
L_R = 4\pi\eta_{\nu} V_R \ \ [{\rm erg}~{\rm s}^{-1}~{\rm Hz}^{-1}] \ , 
\end{equation}
where $V_R$ is the radio source volume.  The density of nonthermal electrons 
will be assumed to  be distributed in energy according to a power law,
\begin{equation}\label{uvpowerlaw}
n(\varepsilon)={(\delta-1)\over \varepsilon_0}~N~
\left({\varepsilon\over \varepsilon_0}\right)^{-\delta} \ \ 
[{\rm cm}^{-3}~{\rm erg}^{-1}].
\end{equation}
 lower cutoff at $\varepsilon_0 \approx 10$ keV = $1.6\cdot 10^{-8}$ erg
is compatible with most acceleration
processes proposed for stellar coronae and with solar flare observations.
We keep $\varepsilon_0$ fixed at 10 keV in  the following 
considerations. It is not a sensitive value for synchrotron 
emission since the latter becomes appreciable only at higher energies. 
We assume a homogeneous source for simplicity. The gyrosynchrotron 
emissivity  is approximately given by \citet{dulk85} (sum of x and o mode, see
equation~\ref{intensity})

\begin{eqnarray}
\eta &\approx& 1.8\cdot 3.3\times 10^{-24}~BN\ 10^{-0.52\delta}
(\sin\theta)^{-0.43+0.65\delta}
{\left({\nu\over\nu_B}\right)}^{1.22-0.90\delta} \\
   &\equiv& 1.8\cdot \vartheta(B,\delta,\nu,\theta)~N\quad\quad\quad\quad
[{\rm erg}~{\rm s}^{-1}{\rm Hz}^{-1}~{\rm cm}^{-3}~{\rm sterad}^{-1}]\ , 
\end{eqnarray}
where $\nu_B$ is the electron 
gyrofrequency and $\theta$ is the angle between the line of sight and
the magnetic field. 
In a steady-state situation, the density of nonthermal electrons is given by
\begin{equation}\label{electrondens}
n(\varepsilon)={\dot n_{\rm in}(\varepsilon)\over V_R}\tau(\varepsilon)
\end{equation}
where $\dot{n}_{\rm in}(\varepsilon)$ is the total number of electrons 
of energy $\varepsilon$
accelerated per unit time, and $\tau(\varepsilon)$ is the electron lifetime.
Let 
\begin{eqnarray}
\dot n_{\rm in}(\varepsilon) &=& \dot n_{0, in} ~\left({\varepsilon\over 
\varepsilon_0}\right)^{-\kappa}   \\
\tau(\varepsilon) &=& \tau_0 
~\left({\varepsilon\over\varepsilon_0}\right)^{\alpha}
\end{eqnarray}
With (\ref{electrondens}), the power law index of the electrons (\ref{uvpowerlaw}) 
is $\delta \approx \kappa - \alpha$. 

\begin{figure} 
\centerline{
\resizebox{0.54\textwidth}{!}{\includegraphics{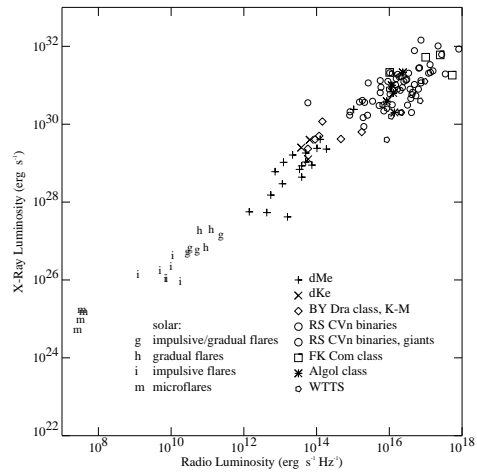}}
}
\caption{Correlation between radio and X-ray luminosities for magnetically active stars
(from \citealt{guedel02d}).}\label{lxlr}
\end{figure}

Let $a$ be the fraction of the energy that goes into accelerated particles, and
$b$ the fraction
of the total coronal energy ultimately radiated into the observed X-ray
band.  Since some of the thermal energy is lost by conduction and other
processes, $b < 1$. Then, 
X-rays are related to the total energy input
\begin{equation}\label{dote} 
\dot E = {1\over a} \int_{\varepsilon_0}^{\infty} \dot n_{\rm in}(\varepsilon) 
\varepsilon ~d\varepsilon = {1\over b} L_X.
\end{equation} 
Using (\ref{radiolum})--(\ref{dote}), the relation between $L_R$ and $L_X$ becomes
\begin{equation}\label{lr1} 
L_R = 1.8\cdot 4\pi \vartheta(B,\nu,\theta,\delta)~{a\over b}
~\varepsilon_0^{-1}\tau_0~\left({\alpha - 1 \over 
\delta - 1} + 1\right)~L_X 
\end{equation}
where we require $\delta > 1$ and $\alpha > 2 - \delta$ for convergence.
Equation~(\ref{lr1}) is general and includes different possible 
scenarios. Let us select typical parameters for stellar observations, viz.
$\delta = 3$ (implying $\alpha > -1$), 
$\nu=5$ GHz, and $\theta=30^\circ$; then
\begin{equation}\label{lr3} 
L_R = 3.5\cdot 10^{-22}B^{2.48}~{a\over b}~\tau_0
(\alpha + 1)~L_X 
\end{equation} 
Late-type main-sequence and subgiant stars appear to follow a linear relation
between $L_R$ and $L_X$: $L_X \approx 10^{15.5} L_R$ (the coefficient is 0.5--1 dex
smaller for RS CVn-type binaries; \citealt{guedel93a}, see Fig.~\ref{lxlr}).
We thus suggest numerical relations between $a$, $b$, $\tau$,
and $B$ for active stars:
\begin{equation}\label{lr2} 
B^{2.48}~{a\over b}~\tau_0(\alpha + 1) \approx \cases{
9.0\cdot 10^5&dMe, dKe, BY Dra\cr
5.4\cdot 10^6&RS CVn, Algols, PTTS, FK Com. \cr}
\end{equation}

Two scenarios are  possible:
If the acceleration efficiency $a$ is close to unity, the nonthermal electrons
first emit a small fraction of the total kinetic energy as synchrotron 
radiation before they lose most of their energy by collisions, thereby heating 
the  X-ray emitting corona (``causal relation'' between nonthermal and 
thermal energy). It is unknown whether fully efficient accelerators are
realized in nature.

The other scenario implies a common
energy release, most of it heating the corona by thermal processes
(e.g. Ohmic heating), and a much smaller fraction $a$ accelerating particles.
Some of the latter energy may also end up as heat, but this is negligible. 
Therefore, the thermal plasma and nonthermal electrons radiate independently
(``common origin'' scenario).
This scenario can explain the relation found by assuming that coronal heating 
in active stars {\it necessarily} implies electron acceleration to relativistic 
energies. The fraction between thermal and nonthermal energy 
release is the crucial, but unknown number. It is unclear whether it can 
be constant over many different types of stars.

The lifetime $\tau$ (or $\tau_0$ and $\alpha$) which is relevant here for
the gyrosynchrotron emitting high-energy particles, depends on the
process that scatters energetic electrons
into the loss-cone of the trapped particle velocity
distribution, such as collisions, whistler wave instability, or cyclotron 
maser action. Their effects on the relation between trapping time and
particle energy are opposite. As an example, let us assume 
$\tau (\varepsilon)= {\rm  const}$ (i.e. $\alpha = 0$) and use solar active region 
and flare
values for $B$ ($\approx 100$ G) and $a/b$ (order of unity; \citealt{dennis88, withbroe77}. 
Equation~(\ref{lr2}) then would 
require $\tau\approx 10$ s for dM(e), dK(e), and BY Dra stars.
The lifetime is an average over the time of flight of the particles that
are lost immediately (moving parallel to the field lines) and the trapped
particles temporarily residing in a lower density plasma. 
The estimated values for $\tau$ are
compatible with the minimum observed time scale of variations in the 
``quiescent'' emission (e.g. \citealt{jackson89}). For the halo
of RS CVn binaries, $B$ can be as low as $\approx 10$ G \citep{mutel85}. Then,
from (\ref{lr2}), $\tau$ is of order of many hours, compatible with 
observed long time scales (e.g., \citealt{massi92}).

\begin{acknowledgement}
It is a pleasure to thank the organizers for inviting me to the Montegufoni 
summer school. I thank several colleagues for providing me figure material, which is
reproduced with permission of the publishers. In particular, some of the material presented here
has been reprinted, with permission, from the {\it Annual Review of Astronomy and Astrophysics},
Volume 40, \copyright\ 2002 by Annual Reviews, www.annualreviews.org (Figures 1b, 2, 13--16, 22, 26, 27; G\"udel 2002).
Some material has been reprinted, with permission, from the  {\it Astronomy and Astrophysics Review} Vol. 12
(Figures 1a, 4, 6--12, 18--25; G\"udel 2004).
Research at PSI has been supported by the Swiss National Science Foundation.
\end{acknowledgement}

\end{document}